\documentclass[10pt,twocolumn,prd,aps,amsmath,amssymb,superscriptaddress,floatfix,nobalancelastpage,
preprintnumbers,reprint,nofootinbib,hyperref]{revtex4-2}

\usepackage{hyperref}

\usepackage{graphicx}      
\usepackage{longtable}     
\usepackage{url}           
\usepackage{bm}            
\usepackage[dvipsnames]{xcolor}
\usepackage{xcolor}
\usepackage{mathrsfs}
\usepackage{blindtext}
\usepackage{float}
\usepackage{bbold}
\usepackage[normalem]{ulem}

\usepackage{setspace}
\usepackage{tabularx}

\newcommand{\vecp}{\mathbf{p}_\nu}
\newcommand{\vecv}{\mathbf{v}_\nu}
\newcommand{\vece}{\mathbf{e}}
\newcommand{\vecF}{\mathbf{F}}
\newcommand{\nue}{\nu_e}
\newcommand{\nuebar}{\bar{\nu}_e}
\newcommand{\nux}{\nu_x}
\newcommand{\nuxbar}{\bar{\nu}_x}
\newcommand{\ff}{\mathfrak{f}}
\newcommand{\ffth}{\mathfrak{f}_{\mathrm{thr}}}
\newcommand{\MBHdot}{\dot{M}_{\mathrm{BH}}}

\newcommand{\dd}{\mathrm{d}}
\newcommand{\yeeq}{Y_e^{\rm eq}}
\newcommand{\yeeqem}{Y_e^{\rm eq,em}}
\newcommand{\yeeqab}{Y_e^{\rm eq,abs}}
\newcommand{\yeeqbar}{\bar{Y}_e^{\rm eq}}
\newcommand{\yeeqembar}{\bar{Y}_e^{\rm eq,em}}

\newcommand{\nfamilye}{n_{\nu_e,q}^{\mathrm{fam}}}
\newcommand{\nfamilyx}{n_{\nu_x,q}^{\mathrm{fam}}}
\newcommand{\neqq}{n_{\mathrm{eq},q}}
\newcommand{\ntot}{n^{\mathrm{tot}}_q}

\newcommand\apjl{ApJL}
\newcommand\apjs{ApJS}
\newcommand\mnras{MNRAS}
\newcommand\jcap{JCAP}
\newcommand\aap{A\&A}
\newcommand\aaps{A\&AS}
\newcommand\pasj{PASJ}

\def\ga{\,\,\raise0.14em\hbox{$>$}\kern-0.76em\lower0.28em\hbox
{$\sim$}\,\,}
\def\la{\,\,\raise0.14em\hbox{$<$}\kern-0.76em\lower0.28em\hbox
{$\sim$}\,\,}
\def\Msun{$M_{\odot}$}

\begin{document}
\title{Fast Neutrino Conversion \\ in Hydrodynamic Simulations of Neutrino-Cooled Accretion Disks}

\author{Oliver Just}
 \email{o.just@gsi.de}
 \affiliation{GSI Helmholtzzentrum f\"ur Schwerionenforschung, Planckstraße 1, D-64291 Darmstadt, Germany}
 \affiliation{Astrophysical Big Bang Laboratory, RIKEN Cluster for Pioneering Research, 2-1 Hirosawa, Wako, Saitama 351-0198, Japan}

 \author{Sajad Abbar}
 \email{abbar@mppmu.mpg.de}
 \affiliation{Max Planck Institut f\"ur Physik (Werner-Heisenberg-Institut), F\"ohringer Ring 6, 80805 M\"unchen, Germany}

 \author{Meng-Ru Wu}
 \email{mwu@gate.sinica.edu.tw}
 \affiliation{Institute of Physics, Academia Sinica, Taipei, 11529, Taiwan}
 \affiliation{Institute of Astronomy and Astrophysics, Academia Sinica, Taipei, 10617, Taiwan}

 \author{Irene Tamborra}
 \email{tamborra@nbi.ku.dk}
 \affiliation{Niels Bohr International Academy and DARK, Niels Bohr Institute,
University of Copenhagen, Blegdamsvej 17, 2100, Copenhagen, Denmark}

 \author{Hans-Thomas Janka}
 \email{thj@mpa-garching.mpg.de}
 \affiliation{Max-Planck-Institut f\"ur Astrophysik, Karl-Schwarzschild-Straße 1, D-85748 Garching, Germany}

 \author{Francesco Capozzi}
 \email{fcapozzi@ific.uv.es}
 \affiliation{Instituto de F\'isica Corpuscular, Universidad de Valencia \& CSIC, Edificio Institutos de Investigaci\'on, Calle Catedr\'atico Jos\'e Beltr\'an 2, 46980 Paterna, Spain}

\date{\today}

\begin{abstract}
 The outflows from neutrino-cooled black-hole accretion disks formed in neutron-star mergers or cores of collapsing stars are expected to be neutron-rich enough to explain a large fraction of elements created by the rapid neutron-capture process, but their precise chemical composition remains elusive. Here, we investigate the role of fast neutrino flavor conversion, motivated by the findings of our post-processing analysis that shows evidence of electron-neutrino lepton-number crossings deep inside the disk, hence suggesting possibly non-trivial effects due to neutrino flavor mixing. We implement a parametric, dynamically self-consistent treatment of fast conversion in time-dependent simulations and examine the impact on the disk and its outflows. By activating the, otherwise inefficient, emission of heavy-lepton neutrinos, fast conversions enhance the disk cooling rates and reduce the absorption rates of electron-type neutrinos, causing a reduction of the electron fraction in the disk by $0.03-0.06$ and in the ejected material by $0.01-0.03$. The rapid neutron-capture process yields are enhanced by typically no more than a factor of two, rendering the overall impact of fast conversions modest. The kilonova is prolonged as a net result of increased lanthanide opacities and enhanced radioactive heating rates. We observe only mild sensitivity to the disk mass, the condition for the onset of flavor conversion, and to the considered cases of flavor mixing. Remarkably, parametric models of flavor mixing that conserve the lepton numbers per family result in an overall smaller impact than models invoking three-flavor equipartition, often assumed in previous works.
\end{abstract}

\maketitle

\section{Introduction}\label{sec:introduction}

Disks accreting material onto stellar mass black holes (BHs) with a rate higher than $\sim 10^{-3}$--$10^{-2}\,M_\odot\,$s$^{-1}$, which are formed after the merger of two neutron stars (NSs) or a NS with a BH \cite{Ruffert1996a, Rosswog2003a, Bauswein2013, Baiotti2017a, Bernuzzi2020b, Ruiz2021d, Radice2018b} or during the collapse of a fast rotating star \cite{MacFadyen1999, Surman2006, Nagataki2007, Siegel2019b, Obergaulinger2021e}, are called neutrino-cooled disks, or neutrino-dominated accretion flows (NDAFs), because weak interactions take place on timescales comparable to the dynamical timescales, enabling neutrinos to radiate away most of the heat generated in the turbulent flow \cite{Ruffert1997, Popham1999, Kohri2002, Beloborodov2003, Chen2007, Metzger2009b, Fernandez2013b, Just2015a, Siegel2018a, Janiuk2019a, Fujibayashi2020a, Hayashi2021a, Most2021q, Murguia-Berthier2021u}. The high neutrino fluxes, and therefore high neutrino pair-annihilation rates, produced in these disks have been suggested to power ultrarelativistic jets of short gamma-ray bursts \cite{Eichler1989, Ruffert1999a, Just2016}. Moreover, the massive, non-relativistic outflows launched from neutrino-cooled disks are believed to represent a rich, possibly even the dominant, source of heavy elements \cite{Surman2008b, Wanajo2012, Just2015a, Wu2016a, Kasen2017a, Siegel2018a, Miller2019a} created through the rapid neutron-capture process (see~\cite{Arnould2020f, Cowan2021g} for recent reviews). A high neutron-richness, or low electron (or proton-to-baryon) fraction, $Y_e$, is a necessary requirement for an efficient r-process, where $Y_e$ is regulated in neutrino-cooled disks by the competition between the $\beta$-processes,
\begin{subequations}\label{eq:beta}
  \begin{align}
    e^-+p\longleftrightarrow n + \nue \, , \\
    e^++n\longleftrightarrow p + \nuebar \, ,
  \end{align}
\end{subequations}
i.e.~electron captures on protons, positron captures on neutrons, and the corresponding inverse reactions. Most existing models of neutrino-cooled disks agree that the material inside the disk is generically very neutron rich, $Y_e< 0.25$, as a consequence of the mild degeneracy of electrons, which suppresses the abundance of positrons \cite{Chen2007, Siegel2018c, Just2021i}. The question of exactly how neutron rich the outflows of neutrino-cooled disks are, and therefore how important these disks are for galactic chemical evolution \cite{Cote2019m, Banerjee:2020eak, van-de-Voort2020r}, is still a matter of active debate and extremely challenging to address with numerical simulations because of the complicated physics connected to neutrino transport, magnetohydrodynamic (MHD) turbulence, and general relativity \cite{Siegel2019r, Fernandez2019b, Fujibayashi2020a, Miller2020a, Just2021i}. Observational support for the viability of BH-torus systems as r-process sites comes from the late, red component of the electromagnetic counterpart, called kilonova \cite{Metzger2010c, Roberts2011, Goriely2011, Kasen2017a, Tanaka2017t,Metzger2019a}, that accompanied the recent gravitational-wave event GW170817. It suggests a significant abundance of lanthanides, which can only be produced in ejecta with $Y_e\la 0.25$--$0.3$ \cite{Hoffman1997, Kasen2015, Lemaitre2021m}. However, whether or not the observed signal truly stems from BH-torus ejecta or from other ejecta components remains to be finally understood (e.g \cite{Kawaguchi2018a}).

Apart from the, already significant, uncertainties connected to neutrino transport that assumes mass-less neutrinos, additional, major questions are connected to the possibility that finite-mass neutrinos could change their flavor during the propagation through and out of the disk. An immediate consequence of flavor conversion, e.g. due to neutrino self-interactions, is that the change in the abundances and mean energies of $\nu_e$ and $\nuebar$ neutrinos would lead to a modified value of the equilibrium electron fraction corresponding to neutrino absorption, $\yeeqab$ \cite{Qian1996, Just2021i}, as well as the timescale over which $Y_e$ approaches $\yeeqab$ in outflows exposed to neutrino irradiation ~\cite{Duan:2010af,Malkus:2012ts,Wu:2014kaa,Pllumbi2015a,Frensel:2016fge,Sasaki:2017jry,Zhu:2016mwa,Wu:2017drk,Deaton2018a,George2020a,Xiong2020c,Myers2021w}.

In a dense neutrino gas, such as the one expected in the innermost regions of neutrino-cooled disks, neutrinos could also experience \textit{fast} flavor conversion~\cite{Sawyer:2005jk, Sawyer:2015dsa}(see also \cite{Capozzi:2022slf, Tamborra:2020cul} and references therein). 
Unlike the traditional collective slow modes~\cite{Duan:2010bg}, which occur on scales on the order of a few km for neutrinos with energies of $\sim 10$\,MeV, fast modes can grow on extremely short scales of just a few cm. Since such scales are shorter than the typical mean-free path between neutrino interactions, fast modes can modify the neutrino transport also at high densities that have long been thought to be the realm of collisional processes (see ~\cite{Martin:2021xyl,Shalgar:2020wcx,Sasaki:2021zld,Capozzi:2018clo,Johns:2021qby,Sigl:2021tmj} for recent work studying the impact of collisions on fast conversions). For instance, if fast conversion affects the diffusion rates out of the disk, neutrino conversion could leave a dynamical imprint on the disk, and among others modify the equilibrium electron fraction corresponding to neutrino emission, $\yeeqem$ (see, e.g.,~\cite{Just2021i}). Needless to say, all the aforementioned effects potentially connected to flavor conversion could have dramatic implications for r-process nucleosynthesis, galactic chemical evolution, and the kilonova signal of neutrino-cooled disks.

It has been demonstrated that fast flavor instabilities in a dense neutrino gas can occur when the difference between the angular distributions of $\nue$ and $\nuebar$ neutrinos changes sign in some direction~\cite{Izaguirre:2016gsx,Capozzi:2017gqd,Yi:2019hrp,Morinaga2021f,Dasgupta:2021gfs}. The search for such zero crossings of the electron-lepton number (ELN), as well as their possible consequences, have since been the subject of numerous studies in the context of core-collapse supernovae (CCSNe) and NS mergers (NSMs)~\cite{Wu2017l, Wu2017a, Tamborra:2017ubu, Abbar:2018shq, DelfanAzari:2019epo, Morinaga:2019wsv,DelfanAzari:2019tez, Nagakura:2019sig, Abbar2019a, Glas:2019ijo, Abbar2021r, Capozzi2021b, Nagakura2021r, Harada2021h, Li2021g, George:2020veu, Xiong2020c}. Since a NSM remnant is born neutron rich, and subsequently heats up and protonizes with time, the emission of $\nuebar$ neutrinos tends to dominate that of $\nue$ neutrinos (at least globally; see, e.g.,~\cite{Wu2017a}), rendering ELN crossings more likely to appear in these environments compared to CCSNe. Indeed, recent work on the occurrence of fast instabilities in post-merger BH disks based on analytical models, which assumed $\nue$ and $\nuebar$ neutrinos to be emitted from two separate sharp neutrinospheres, suggested fast instabilities to occur almost everywhere above the neutrinospheres \cite{Wu2017l, Wu2017a}. While models using sharp neutrinospheres provide a basic understanding of the conditions under which fast instabilities can occur, their reliability is limited. Moreover, these models cannot make any predictions regarding the existence of ELN crossings below the neutrinospheres. Finally, since those studies, like in fact most existing literature on fast pairwise flavor instabilities using data from hydrodynamic simulations, are entirely based  on post-processing approaches, they can hardly make faithful predictions about the dynamical consequences of these phenomena.

In this study, we investigate the existence of fast instabilities in neutrino-cooled accretion disks adopting the recently developed method of Ref.~\cite{Abbar2020m}. The latter does not depend on sharp neutrinospheres but uses as input directly the angular moments of the neutrino distribution function. We apply this method to simulations taken from Ref.~\cite{Just2015a}, which were performed with the neutrino-hydrodynamics code ALCAR \cite{Obergaulinger2008a,Just2015b}, making use of the M1 approximation of neutrino transport. Guided by the results of this crossings-analysis, we perform simulations of disks with an approximate, but dynamically self-consistent, inclusion of fast flavor mixing through a parametric approach. This approach assumes that fast modes, where present according to a varied threshold criterion, lead to an asymptotic equilibrium state in flavor space that can be expressed in terms of the evolved neutrino quantities. Implementing fast conversions in time-dependent simulations is an important step forward compared to previous post-processing studies, because it allows to study the dynamical consequences of fast conversions on the disk evolution, neutrino emission, outflow properties, and nucleosynthesis. Our study expands on the recent work of Ref.~\cite{Li2021g}, which was the first to include fast flavor conversions self-consistently in a dynamical simulation. However, compared to Ref.~\cite{Li2021g}, where only a single model was discussed, here we perform a detailed analysis of the various effects of fast conversions on the weak-interaction equilibria and the corresponding timescales, and we test the sensitivity to the main input assumptions, such as the type of flavor mixing and the condition for the onset of fast conversions.

This paper is structured as follows: In Sec.~\ref{sec:occurr-fast-conv} we present the setup and results of our analysis to spot and characterize the locations of ELN crossings in neutrino-cooled disks, which potentially lead to flavor conversions. The results of this analysis serve as the motivation for the implementation of fast conversion with a simple, parametric criterion in hydrodynamic simulations, as discussed in Sec.~\ref{sec:simul-setup-invest}. We first explore the impact of fast conversions for a fiducial model in Sec.~\ref{sec:results-fiduc-model}, and in Sec.~\ref{sec:model-dependence} we examine the model dependence. We summarize and conclude in Sec.~\ref{sec:summary-conclusions}. Throughout this paper, $h$, $c$, and $G_{\rm F}$ denote the Planck constant, the speed of light, and the Fermi constant, respectively.

\section{Occurrence of fast flavor instabilities in BH tori}\label{sec:occurr-fast-conv}

Before attempting to implement fast conversions in time-dependent numerical simulations, we analyze whether and where favorable conditions for the development of fast flavor instabilities exist in neutrino-cooled BH tori using individual snapshots from previous neutrino-hydrodynamical models. The aim of this section is to find an approximate criterion for the occurrence of fast instabilities, which is simple enough to be evaluated in each time step in an ongoing simulation. In Sec.~\ref{sec:angul-moment-meth} we review the method used to analyze the snapshots, and in Sec.~\ref{sec:occurr-eln-cross} we discuss the results.

\subsection{Angular moment method for detecting fast pairwise instabilities}\label{sec:angul-moment-meth}

As discussed previously, fast instabilities  are thought to occur if, and only if, the angular distributions of $\nu_e$ and $\bar\nu_e$ cross each other, i.e. when the neutrino ELN,
\begin{equation}\label{eq:Gdef}
  G_{\vecv} =
  \sqrt{2} \frac{G_{\mathrm{F}}}{\hbar c}
  \int_0^\infty \frac{E_\nu^2 \mathrm{d} E_\nu}{(hc)^3}
        [f_{\nu_e}(\vecp)- f_{\bar\nu_e}(\vecp)] \, ,
\end{equation}
changes its sign for different unit vectors of the neutrino momentum, $\vecp/|\vecp|$. Here, $E_\nu$ and $\vecv=\vecp c^2/E_\nu$ are the neutrino energy and velocity, respectively, and the $f_\nu$'s are the neutrino occupation numbers. We consider here only crossings in the electron-neutrino sector and implicitly assume that $f_{\nu_{\mu}}=f_{\nu_{\tau}} = f_{\bar\nu_{\mu}}=f_{\bar{\nu}_{\tau}}$, but see Refs.~\cite{Chakraborty:2019wxe, Shalgar:2021wlj,Capozzi:2020kge, Capozzi2021b} for dedicated analyses on three flavor effects. In contrast to the case of (proto- or hypermassive) NSs, the densities and temperatures in neutrino-cooled disks are typically not high enough to produce heavy-lepton neutrinos in large amounts via pair processes or charged-current reactions with muons (see e.g.~Ref.~\cite{Bollig2017a}). We note that \emph{after} the eventual effects of fast flavor mixing, heavy-lepton neutrinos can indeed reach number densities comparable to those of electron-type neutrinos (as will be demonstrated in Sec.~\ref{sec:results-fiduc-model}).

Although the neutrino angular distribution can in general be asymmetric in the azimuthal angle, $\phi_\nu$, we here consider the $\phi_\nu$-integrated distribution of the ELN,
\begin{equation}
  G(\mu) = \int_0^{2\pi}G_{\vecv} \mathrm{d} \phi_\nu \, ,
\end{equation}
and its first $n$ angular moments,
\begin{equation}\label{moments}
I_n = \int_{-1}^{1} \mathrm{d}\mu\ \mu^n\ G(\mu) \, ,
\end{equation}
where $n=0,1,2, 3$. Here, $\mu=\cos\theta_\nu=\vece_r\cdot \vecp/|\vecp|$ is the cosine of the zenith angle (with the unit vector in radial direction in position space, $\vece_r$), and $\phi_\nu$ runs around the axis defined by $\vece_r$. The moments $I_n$ are therefore defined by the radial components of the, generally multidimensional, angular moment tensors of the distribution functions. While in principle we could also use other tensor components in our method, we restrict ourselves to the radial components, as this method already captures the majority of crossings. It is important to bear in mind that although the integration over $\phi_\nu$ might erase some of the ELN crossings, it will not generate any fake crossings in our analysis.

In neutrino-hydrodynamic simulations adopting the so-called M1 scheme, such as in the ones considered in this study, the 0th and 1st angular moments of the neutrino distribution function are evolved, and the 2nd and 3rd moments are derived from the evolved moments by using an approximate closure relation. Thus, one has access to the moments $I_0, I_1, I_2$, and $I_3$. Although a significant amount of information about the neutrino distribution is lost by using an analytic closure, one can already derive sufficient conditions for the occurrence of ELN crossings in a dense neutrino gas on the basis of the available moments. As discussed in Ref.~\cite{Abbar2020m} (see also \cite{Dasgupta:2018ulw,Nagakura2021a,Nagakura:2021suv,Nagakura2021r,Johns2021w} for other methods), ELN crossings are guaranteed to occur if at a given location in space and time a positive function $\mathcal{F}(\mu)$ exists, for which
\begin{equation}\label{eq:crossing_condition}
  I_{\mathcal{F}}I_0<0 \, ,
\end{equation}
where
\begin{equation}\label{IF}
  I_\mathcal{F} = \int_{-1}^{1} \mathrm{d}\mu\ \mathcal{F}(\mu)\  G(\mu) \, .
\end{equation}
Since the neutrino angular information is provided in terms of the $I_n$'s, we choose $\mathcal{F}(\mu)$ to be a polynomial in $\mu$,
\begin{equation} \label{eq:F}
  \mathcal{F}(\mu) =\sum_{n=0}^N a_n \mu^n,
\end{equation}
where $N$ is the highest neutrino angular moment available and the $a_n$'s are arbitrary coefficients for which $\mathcal{F}(\mu)>0$. The quantity $I_\mathcal{F}$ can then be written in terms of the $I_n$'s as
\begin{equation}\label{eq:I_F}
  I_\mathcal{F}=a_0I_0+a_1I_1+\dots+ a_NI_N \, .
\end{equation}
It should be noted that in the M1 scheme $\mathcal{F}$ can be a linear ($a_{1}\neq0, a_{2,3}=0$), a quadratic ($a_{2}\neq 0, a_{3}=0$), or a cubic ($a_{3}\neq0$) polynomial. While the ELN crossings captured by the linear function are the most reliable ones, since $I_0$ and $I_1$ are evolved independently of each other, the cubic function is able to detect a larger number of crossings. This can be understood by noting that higher moments can capture more details of the neutrino angular distribution. Indeed, the narrower the crossings are, the higher is the order of the moments required to capture them. Hence, given the maximum rank of $N=3$ in our case, our method is not guaranteed to capture all the ELN crossings.

\subsection{Occurrence of ELN crossings in snapshots of BH tori} \label{sec:occurr-eln-cross}

\begin{figure*} [htb!]
 \centering
\includegraphics*[width=.9\textwidth, trim=70 85 50 115, clip]{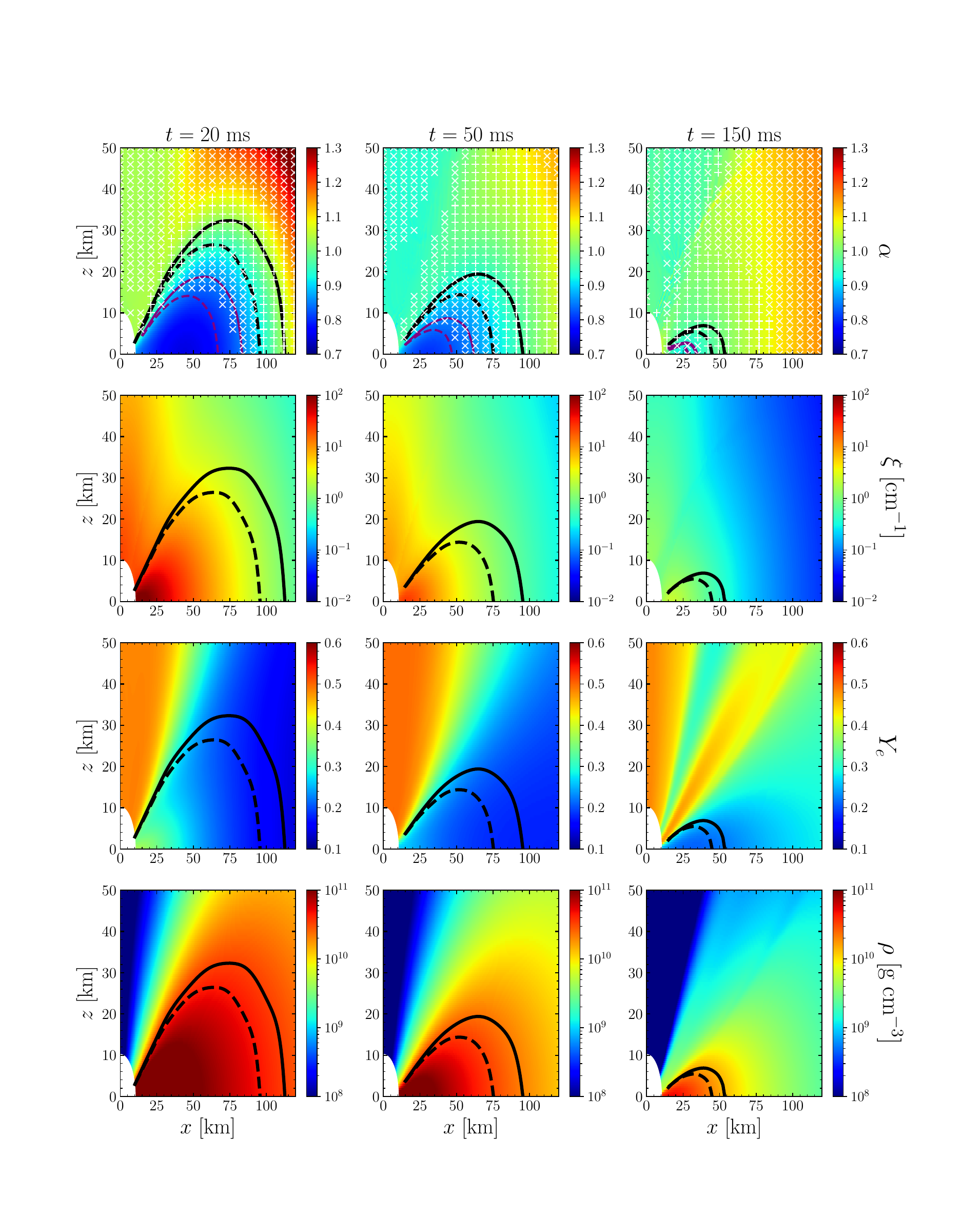}
\caption{Cuts through the polar plane of an axisymmetric, viscous BH-torus simulation (model M3A8m3a5 from \cite{Just2015a}) at times $t=20\,$ms (left column), 50\,ms (middle column), and 150\,ms (right column) showing properties relevant to characterize the occurrence of fast flavor mixing. The torus rotates around the $z$ axis and $x$ denotes the cylindrical radius. The color maps show, from top to bottom, the $\nue-\nuebar$ asymmetry parameter, $\alpha$ (cf. Eq.~(\ref{eq:alpha})), weak interaction strength, $\xi$ (cf. Eq.~(\ref{eq:xi})), electron fraction, $Y_e$, and matter density, $\rho$. Solid (dashed) black lines denote the neutrinospheres of $\nue$ ($\nuebar$) (cf. Eq.~(\ref{eq:ffdef})). Solid (dashed) purple lines in the top panels show the locations where the energy-integrated flux factor of electron antineutrinos $\ff_{\nuebar}=0.18$ (0.14). The white crosses (plus signs) in the top panels indicate regions where ELN crossings, and therefore possibly fast flavor instabilities, exist as predicted by Eq.~(\ref{eq:crossing_condition}) using quadratic/cubic (linear) polynomials for $\mathcal{F}(\mu)$. Note that ELN crossings are more likely to occur where $\alpha$ is close to unity. Since the matter density decreases with time due to mass accretion and viscous spreading, neutrinos decouple at smaller radii at late times, enabling ELN crossings nearly everywhere within the torus.}
\label{fig:1}
\end{figure*}

The formalism outlined above will now be used to investigate the occurrence of ELN crossings in neutrino-cooled BH accretion disks. We analyze numerical simulations obtained with the neutrino-hydrodynamics code AENUS-ALCAR \cite{Obergaulinger2008a,Just2015b}, which adopts an energy-dependent M1 scheme to describe neutrino transport. The scheme evolves the angular moments of the distribution function (energy density and flux-density vector) in a number of energy bins, $N_{\epsilon}$, where typically $N_{\epsilon}\approx 12-20$. For the analysis in this section we adopt the axisymmetric model M3A8m3a5, which was described in Ref.~\cite{Just2015a}. This same model has been previously used also by Ref.~\cite{Wu:2017drk} for a first investigation on the occurrence of fast flavor instabilities, which assumed $\nu_e$ and $\bar\nu_e$ neutrinos to be emitted from sharp neutrinospheres. The specific model considered in this section assumes masses of 3\,\Msun and 0.3\,\Msun for the central BH and initial torus, respectively, and a BH spin parameter of 0.8. We confirmed that the basic results depend only weakly on the particular choice of the model parameters so that our results can be used as a guideline for the approximate implementation of flavor conversions in time-dependent, hydrodynamical simulations. We constrain ourselves here to the case of axisymmetric viscous models. However, we expect that the main result of this exercise (namely that ELN crossings appear in large regions throughout the disk) would also be obtained when using 3D MHD models, which exhibit a similar structure as viscous disks, at least in a time-averaged sense \cite{Siegel2018c, Fernandez2019b, Just2021i, Hayashi2021a}.

In order to check for the presence of ELN crossings in our BH-torus simulation data, we go through a large number of points in the polar ($x-z$) plane and, at each point, scan all appropriate linear, quadratic, and cubic polynomials representing $\mathcal F(\mu)$ using the $I_n$ moments provided by the simulation~\cite{Abbar2021r}.

In the top row of Fig.~\ref{fig:1}, we indicate for different simulation times (20, 50, and 150\,ms after initialization, from left to right) the locations where ELN crossings exist, i.e. where the condition in Eq.~(\ref{eq:crossing_condition}) is satisfied. Here, the crosses denote points where ELN crossings are found by a quadratic or cubic form of $\mathcal F(\mu)$, while the plus signs show locations where the ELN crossings are captured already by linear polynomials of $\mathcal F(\mu)$ (in addition to quadratic/cubic ones).

An intriguing result of this analysis is that fast instabilities in NSM remnant disks may occur already deep within the torus in regions well below the neutrinospheres of $\nu_e$ and $\bar\nu_e$. Note that crossings at such high optical depths cannot be explored by models that only investigate the neutrino field outside of the neutrinospheres, such as employed in Ref.~\cite{Wu:2017drk}. Following Ref.~\cite{Wu:2017drk}, we approximately assume that the (energy-averaged) neutrinosphere is given by the surface at which the (energy-averaged) flux factor
\begin{equation}\label{eq:ffdef}
  \ff_\nu = \frac{|\vecF_{\nu}|}{c n_\nu} = \frac{1}{3} \, ,
\end{equation}
where the neutrino number density and number flux density are respectively defined as
\begin{subequations}
  \begin{align}
 n_{\nu} = \int \frac{ \mathrm{d}^3 p}{h^3} f_{\nu}(\vecp) \, , \label{eq:numdef} \\
 \vecF_{\nu} = \int \frac{ \mathrm{d}^3 p}{h^3} f_{\nu}(\vecp) \vecv \, . 
  \end{align}
\end{subequations}
Our analysis suggests that ELN crossings appear as soon as $\ff_\nu \simeq 0.15-0.2$.

The existence of ELN crossings so deep within the disk can be understood from the following considerations. As a result of the neutron-richness in the disk (with typical values of $Y_e\sim 0.1-0.3$) -- which itself is mainly a consequence of electron-degeneracy (see, e.g., Ref.~\cite{Just2021i} for a detailed discussion of $Y_e$ equilibria in neutrino-cooled disks) -- the absorption opacity of $\nuebar$ ($\propto n_p$) is smaller than that of $\nue$ ($\propto n_n$, with $n_{p/n}$ being the number density of protons/neutrons). This means that $\nuebar$ neutrinos diffuse out of the torus more readily than $\nue$ neutrinos and, therefore, 
\begin{equation}\label{eq:ffgtone}
  \frac{\ff_{\nuebar}}{\ff_{\nue}} > 1
\end{equation}
in most regions of the torus. The condition of Eq.~(\ref{eq:ffgtone}) is realized in the disk (i.e. roughly in the region enclosed by the neutrinospheres) with an asymmetry parameter of
\begin{equation}\label{eq:alpha}
\alpha = \frac{n_{\nuebar}}{n_{\nue}}  \la 1 \, ,
\end{equation}
namely close to, but slightly below, unity (cf. top row in Fig.~\ref{fig:1}), while at the same time the number loss rates by which neutrinos leave the torus, $L_{N,\nu}$, fulfill
\begin{equation}\label{eq:Lratio}
\frac{L_{N,\nuebar}}{L_{N,\nue}} \sim  \frac{|\vecF_{\nuebar}|}{|\vecF_{\nue}|} \ga 1 \, ,
\end{equation}
because the expanding torus slowly protonizes with time (see, e.g., Refs.~\cite{Wu:2017drk, Just2021i}). In regions where the number densities of both species are similar but the angular distribution of one species is more forward-peaked than that of the other species, ELN crossings are likely to occur. We note that similar conditions were found to exist also in the neutrino decoupling region above the proto-NS in CCSNe~\cite{Abbar2019a,Nagakura:2019sig}. Going even deeper into the torus, $\alpha$ becomes very small and the angular distribution of both species too similar (i.e. isotropic) for ELN crossings to develop.

As shown by the crosses in Fig.~\ref{fig:1}, ubiquitous ELN crossings (i.e. crossings captured already by linear polynomials) develop only in the vicinity of the neutrinospheres, where $\alpha$ happens to be close to unity. Further away from the torus we again find ELN crossings captured by quadratic/cubic polynomials. Eventually, a band appears near the polar regions where our method cannot capture any crossings (see, e.g., top middle panel in Fig.~\ref{fig:1}). However, this band gets narrower at later times and finally disappears. In this band both the number density and angular distribution of $\nue$ and $\nuebar$ are very similar and thus do not exhibit ELN crossings. Since in this region $\alpha$ is close to unity, it is quite likely that here ELN crossings can only be captured with higher neutrino angular moments.

At later times ($t>70-80$ ms for this model), fast instabilities can appear almost everywhere inside the disk. One reason for that is simply the decreasing optical depth of the disintegrating torus, which causes the neutrinospheres to shrink with time. Another reason is that due to the lower levels of electron degeneracy at later times, the abundance ratio $\alpha$ becomes closer to unity. Since both of these effects are sensitive mainly to the mass of the disk, ELN crossings appear inside the disk earlier (later) in models with lower (higher) torus mass.

While $\alpha$ is a measure of the relative asymmetry between both species, the quantity
\begin{equation}\label{eq:xi}
  \xi = \sqrt{2}\frac{G_{\rm F}}{\hbar c} n_{\nu_e} \, ,
\end{equation}
plotted in the second row of Fig.~\ref{fig:1}, is a measure of the absolute strength of weak interactions at a given point. Even at late times, e.g. $t=150\,$ms in the considered model, and with correspondingly low densities in the torus, the neutrino density is still relatively high, $\xi\sim 10^5\,$km$^{-1}$, meaning that corresponding fast modes would occur on length scales of $\sim 1$\,cm.

In summary, our analysis predicts that fast instabilities are a generic feature of neutrino-cooled BH disks formed after NSMs, both deep within the disk as well as above the disk.

\section{Impact of neutrino fast conversion in simulations of BH tori}\label{sec:impact-fast-conv}

After having gained insight about the appearance of ELN crossings in the disk, we now adopt this knowledge to design simulations that test the possible ramifications of fast-flavor conversions for the dynamics, composition, and outflows in neutrino-cooled BH tori. For simplicity, we assume that ELN crossings always lead to fast-flavor conversions, and that flavor conversions take place in such a way that the angular moments of the neutrino distribution functions after conversion are simple, i.e. time and location independent, algebraic functions of the moments before the flavor conversion (see, e.g., Refs.~\cite{Bhattacharyya:2020jpj,Wu2021q, Richers2021j, Xiong2021o, Padilla-Gay:2021haz,Padilla-Gay2021y,Duan:2021woc} for recent studies challenging the simplified mixing scenarios assumed here, outlined in Sec.~\ref{sec:simul-setup-invest}).

The disk model, the adopted simulation code, as well as all numerical methods are identical to those used in Refs.~\cite{Just2015a,Just2021i} except for aspects explicitly mentioned in the following. We refer to the aforementioned references for detailed descriptions of the simulation setup. In contrast to most models of \cite{Just2021i} we include weak magnetism corrections in the charged-current and neutral-current neutrino-nucleon interactions \cite{Horowitz2002a} in all of the present models, and we evolve the two species $\nu_x$ and $\bar\nu_x$ collectively representing $\nu_{\mu/\tau}$ neutrinos and $\bar\nu_{\mu/\tau}$ neutrinos, respectively\footnote{\label{footnote:species} We note that in the adopted set of neutrino interactions the only difference between the interaction rates of $\nux$ and $\nuxbar$ exists in the weak-magnetism correction factors \cite{Horowitz2002a}. Since these differences are too small to create a significant asymmetry between $\nux$ and $\nuxbar$ neutrinos, we neglect them in our simulations and use the same weak-magnetism correction factors for both $\nux$ and $\nuxbar$ neutrinos (obtained by averaging). Nevertheless, we distinguish between $\nux$ and $\nuxbar$ neutrinos in our simulations (instead of evolving a single heavy-lepton species and assuming $f_{\nux}=f_{\nuxbar}$), because one of our considered flavor mixing scenarios (case ``mix2'') explicitly breaks the symmetry between $\nux$ and $\nuxbar$ and, therefore, requires an independent evolution of both species.}. In our simulations heavy-lepton neutrinos can interact via isoenergetic scattering off nucleons \cite{Bruenn1985} as well as thermal pair annihilation \cite{Pons1998} and bremsstrahlung \cite{Hannestad1998} using the effective treatment by Ref.~\cite{OConnor2015a}. However, compared to fast flavor conversions, these ``conventional'' reactions only play a sub-dominant role as a source or sink of heavy-lepton neutrinos in our models due to the relatively low densities ($\la 10^{12}\,$g\,cm$^{-3}$) and temperatures ($\la 5\,$MeV) in the disk. In the remainder of this paper, we  use the symbol $\nux$ ($\nuxbar$) to refer to \emph{one} of the species $\nu_\mu$ or $\nu_\tau$ ($\bar\nu_\mu$ or $\bar\nu_\tau$), e.g. $L_{\nux}$ refers to the luminosity of a single species and not the summed luminosities of several species.

\subsection{Implementation and investigated models} \label{sec:simul-setup-invest}

\newlength{\mytabcolsep}
\setlength{\tabcolsep}{2pt}
\begin{table*}
  \centering
  \caption{Summary of numerical simulations performed in this study and properties of ejected material. From left to right: Model name, inclusion of flavor conversion (FC), initial torus mass, adopted flavor mixing scheme, treatment of turbulent viscosity (either using $\alpha$-viscosity or MHD), dimensionality, final simulation time, ejecta mass, ejecta velocity at extraction radius of $r=10^4\,$km ($3000\,$km) for viscous (MHD) simulations, average electron fraction and entropy per baryon of the ejecta at $T=5\,$GK, total mass fractions of 2nd-peak elements (with atomic mass numbers fulfilling $119 \leq A \leq 138$), of lanthanides (with atomic charge numbers of $57\leq Z \leq 70$) plus actinides ($89\leq Z \leq 102$), and of 3rd-peak elements (with $185 \leq A \leq 204$) synthesized in the material that becomes ejected during the simulated times.}
  \begin{center}
    \begin{tabular}{lcccccccccccccc}
      \hline \hline
      model      & FC       & $m_{\rm tor}^0$ & mixing & $\ffth$ & turbul.       & dimensi- & $t_{\rm fin}$ & $M_{\rm ej}$      & $\langle v \rangle_{\rm ej}$ & $\langle Y_e\rangle_{\rm ej}^{\rm 5GK}$ & $\langle s \rangle_{\rm ej}^{\rm 5GK}$ & $X_{\rm 2nd}$ & $X_{\rm LA}$  & $X_{\rm 3rd}$ \\
                 & enabled? & [$M_\odot$]     & scheme &                          & viscosity     & onality  & [s]           & [$m_{\rm tor}^0$] & [$c$]                        &                                         & [$k_{\rm B}$]             & [$10^{-1}$] & [$10^{-2}$] & [$10^{-2}$] \\
      \hline                                                                                                                                                                                                                                                   
      m1         & no       & 0.1             & -      & -                        & $\alpha$-vis. & 2D       & 10            & 0.204             & 0.058                        & 0.262                                   & 21.68                                  & 3.32        & 1.77        & 3.35 \\ 
      m1mix1     & yes      & 0.1             & mix1   & 0.175                    & $\alpha$-vis. & 2D       & 10            & 0.186             & 0.051                        & 0.246                                   & 21.23                                  & 4.69        & 3.45        & 3.51 \\ 
      m1mix1f    & yes      & 0.1             & mix1f  & 0.175                    & $\alpha$-vis. & 2D       & 10            & 0.189             & 0.051                        & 0.252                                   & 20.49                                  & 4.37        & 2.49        & 3.48 \\ 
      m1mix2     & yes      & 0.1             & mix2   & 0.175                    & $\alpha$-vis. & 2D       & 10            & 0.179             & 0.054                        & 0.240                                   & 21.60                                  & 4.74        & 4.83        & 4.59 \\ 
      m1mix3     & yes      & 0.1             & mix3   & 0.175                    & $\alpha$-vis. & 2D       & 10            & 0.186             & 0.050                        & 0.255                                   & 20.92                                  & 4.62        & 1.55        & 2.43 \\ 
      m1f4       & yes      & 0.1             & mix1   & 0.4                      & $\alpha$-vis. & 2D       & 10            & 0.182             & 0.052                        & 0.236                                   & 20.88                                  & 5.10        & 4.60        & 4.94 \\ 
      m1f1       & yes      & 0.1             & mix1   & 0.1                      & $\alpha$-vis. & 2D       & 10            & 0.186             & 0.047                        & 0.243                                   & 20.85                                  & 5.23        & 4.00        & 3.86 \\ 
      m1f0       & yes      & 0.1             & mix1   & 0                        & $\alpha$-vis. & 2D       & 10            & 0.187             & 0.049                        & 0.240                                   & 20.53                                  & 5.16        & 3.86        & 4.18 \\ 
      m01        & no       & 0.01            & -      & -                        & $\alpha$-vis. & 2D       & 10            & 0.220             & 0.053                        & 0.236                                   & 27.72                                  & 4.98        & 5.96        & 10.1 \\ 
      m01mix1    & yes      & 0.01            & mix1   & 0.175                    & $\alpha$-vis. & 2D       & 10            & 0.218             & 0.053                        & 0.220                                   & 27.92                                  & 4.95        & 6.36        & 13.1 \\ 
      m3         & no       & 0.3             & -      & -                        & $\alpha$-vis. & 2D       & 10            & 0.203             & 0.053                        & 0.278                                   & 17.99                                  & 3.04        & 1.41        & 1.91 \\ 
      m3mix1     & yes      & 0.3             & mix1   & 0.175                    & $\alpha$-vis. & 2D       & 10            & 0.190             & 0.051                        & 0.256                                   & 18.44                                  & 4.43        & 2.43        & 2.32 \\ 
      m1mag      & no       & 0.1             & -      & -                        & MHD           & 3D       & 0.5           & 0.152             & 0.186                        & 0.283                                   & 29.23                                  & 4.11        & 1.81        & 3.20 \\ 
      m1magmix1f & yes      & 0.1             & mix1f  & 0.175                    & MHD           & 3D       & 0.5           & 0.085             & 0.194                        & 0.253                                   & 26.30                                  & 5.79        & 2.31        & 5.08 \\ 
      \hline \hline
    \end{tabular}
  \end{center}
  \label{tab:prop}
\end{table*}

Fast flavor conversions can be implemented in a two-moment scheme, such as ours \cite{Just2015b}, in a relatively straightforward manner if one assumes that flavor instabilities produce a new steady-state equilibrium in flavor space on timescales much shorter than any other dynamical timescale~\cite{Shalgar:2019qwg}. Since the fast-conversion timescales (given roughly by $(c\xi)^{-1}\sim 10^{-11}$\,s) are shorter than the time step used to integrate the neutrino-hydrodynamics equations (typically $10^{-7}$\,s), we simply switch to the equilibrium state from one time step to another without having to follow the detailed evolution leading to that state, which would require to solve a significantly more complicated set of equations~\cite{Vlasenko:2013fja, Volpe:2015rla}. The treatment of fast conversions in our simulations therefore only depends on two ingredients: First, a local condition for the occurrence of fast conversions that can be evaluated on-the-fly during the simulation, and second a prescription for the flavor mixing characterizing the steady state after the operation of the flavor instability. In the following we outline the treatment of both ingredients. We note that a similar treatment of fast conversions was recently also applied in a 3D general relativistic MHD model in Ref.~\cite{Li2021g} 

In order to detect occurrences of fast conversions, we cannot employ the detailed scheme of Sec.~\ref{sec:occurr-fast-conv}, which is too expensive to be evaluated at each time step in an ongoing simulation. Instead, motivated by the findings of Sec.~\ref{sec:occurr-fast-conv}, we assume that flavor conversion develops where
\begin{equation}\label{eq:fccond}
  \ff_{\nuebar} > \ffth  \, ,
\end{equation}
i.e., where energy-averaged flux factor of $\nuebar$ neutrinos, $\ff_{\nuebar}$ (cf. Eq.~(\ref{eq:ffdef})), rises above some critical threshold value, $\ffth$, which lies broadly within 0.15-0.2. We choose $\ffth=0.175$ in the fiducial case, but we also perform simulations with values of $\ffth=0, 0.1$, and~$0.4$ to test the sensitivity. We ignore flavor conversions further away from the torus in regions where the density drops below $10^6\,$g\,cm$^{-3}$, mostly because otherwise we occasionally encountered stability problems with the numerical scheme. However, this should be of minor relevance for the evolution, because the neutrino field arriving in regions of such low densities was already subject to flavor conversions during its previous propagation out of the disk.

The second component of our implementation consists of the prescription to mimic flavor mixing at points where the instability condition, Eq.~(\ref{eq:fccond}), is satisfied. We stress again that the outcome of fast conversions for arbitrary initial neutrino distributions is far from trivial and an actively pursued research area on its own. Therefore, the prescriptions used here are likely to carry poorly known, but potentially significant, uncertainties. 

We consider three different prescriptions to express the moments of the flavor-mixed steady state (no superscript) as functions of the moments before fast conversions (superscript ``0''), which differ in the degree of mixing and in the respected conservation laws. In what follows, we first outline the detailed mixing treatment of the number densities, $n_{\nu,q}$ (having units of cm$^{-3}$), for neutrino species $\nu\in\{\nue,\nuebar,\nux,\nuxbar\}$ in energy bin $q= 1,\ldots,N_\epsilon$ centered around energy $\epsilon_q$, and we subsequently comment on the treatment of the corresponding 1st moments, $\vecF_{\nu,q}$. In all three cases, the post-conversion number densities are given as a linear superposition of the pre-conversion number densities, i.e.
\begin{align}\label{eq:superpos}
  n_{\nu,q}  = \sum_{\tilde{\nu}} c_{\nu\tilde{\nu}} n_{\tilde{\nu},q}^0 \, ,
\end{align}
where $\tilde{\nu}$ runs through $\{\nue,\nuebar,\nux,\nuxbar\}$ and the mixing coefficients $c_{\nu\tilde{\nu}}$ are constants that are defined by the constraints that lepton number conservation is assumed to be fulfilled (or not) in the respective case of flavor mixing. Since we deal with four neutrino species, we need altogether four conditions to define $c_{\nu\tilde{\nu}}$ for a given species. One of these conditions, shared by all mixing cases, is the conservation of the total number of neutrinos, i.e. 
\begin{align}\label{eq:totlepcon}
  \sum_\nu n_{\nu,q}  = \sum_{\tilde{\nu}} n_{\nu,q}^0 \, .
\end{align}

\begin{itemize}

\item \label{case1} \textbf{Case ``mix1'': lepton number per family conserved.} This case assumes that both 
  \begin{align}\label{eq:nfame}
    \nfamilye = n_{\nue,q} - n_{\nuebar,q} 
  \end{align}
  and
  \begin{align}\label{eq:nfamx}
    \nfamilyx = n_{\nux,q} - n_{\nuxbar,q} 
  \end{align}
  are conserved. Among the three cases of flavor mixing considered here, this is the case most compatible with Standard Model physics, but also the most restrictive one concerning the possible degree of flavor redistribution. The remaining condition needed to fix the mixing coefficients, $c_{\nu\tilde{\nu}}$, is that number equipartition shall be achieved among the species with subdominant number densities, i.e.
  \begin{align}\label{eq:neq0}
    \min\{n_{\nue,q},n_{\nuebar,q}\} & = \min\{n_{\nux,q},n_{\nuxbar,q}\}\nonumber \\
    & = \neqq \, ,
  \end{align}
which defines the quantity $\neqq$. The mixing equations result as follows:
  \begin{subequations}
      \begin{align}\label{eq:mix1}
        n_{\nue,q}      & = \neqq +\max\{\nfamilye,0\} \, , \\
        n_{\nuebar,q}    & = \neqq -\min\{\nfamilye,0\} \, , \\
        n_{\nux,q}      & = \neqq +\max\{\nfamilyx,0\} \, , \\
        n_{\nuxbar,q}    & = \neqq -\min\{\nfamilyx,0\} 
      \end{align}
      \end{subequations}
  with
  \begin{align}\label{eq:neq}
    \qquad \neqq = \frac{1}{3} \min\{n_{\nue,q}^0,n_{\nuebar,q}^0\} + \frac{2}{3} \min\{n_{\nux,q}^0,n_{\nuxbar,q}^0\} \, .
  \end{align}

\item \label{case2} \textbf{Case ``mix2'': total lepton number conserved.} This case relaxes
  the conditions of $\nfamilye$- and $\nfamilyx$-conservation underlying Eqs.~(\ref{eq:mix1}) and assumes that neutrinos can mix across different neutrino families while still conserving the total lepton number:
  \begin{equation}\label{eq:ntot}
    \ntot=n_{\nue,q}-n_{\nuebar,q} + 2\left( n_{\nux,q}-n_{\nuxbar,q}\right) \, .
  \end{equation}
  The resulting mixing equations are
  \begin{subequations}
      \begin{align}\label{eq:mix2}
        n_{\nu,q}       & = \frac{1}{3}\left(n_{\nue,q}^0+2n_{\nu_x,q}^0\right) \, , \\
        n_{\bar\nu,q}   & = \frac{1}{3}\left(n_{\nuebar,q}^0+2n_{\nuxbar,q}^0\right) 
      \end{align}
      \end{subequations}
  for $\nu\in\{\nue,\nux\}$ and $\bar\nu\in\{\nuebar,\nuxbar\}$. This approach to mimic the effect of
  flavor mixing, which was assumed in Refs.~\cite{Wu2017a,Li2021g}, leads to a potentially higher degree of equipartition compared to case ``mix1''. Note that this scenario violates family lepton number conservation and is therefore inconsistent with the Standard Model of particle physics.

\item \label{case3} \textbf{Case ``mix3'': total equipartition.} Finally, the case with the largest degree of flavor redistribution is the one of complete equipartition among all six neutrino species, i.e.
  \begin{equation}\label{eq:mix3}
    n_\nu=\frac{1}{6}\left(n_{\nue,q}^0+n_{\nuebar,q}^0+2n_{\nux,q}^0+2 n_{\nuxbar,q}^0\right) 
  \end{equation}
  for $\nu\in\{\nue,\nuebar,\nux,\nuxbar\}$. Here, neutrinos not only can mix across families but also with their anti-particles. Such a case is exotic but could possibly be realized for Majorana neutrinos in the presence of strong magnetic fields and in beyond-Standard-Model scenarios for the neutrino magnetic moments~\cite{Abbar:2020ggq, Sasaki:2021bvu, Kharlanov:2020cti}. We include this scenario in order to explore the maximal impact of flavor mixing on the disk and its composition.
\end{itemize}

We remark that (see also Footnote~\ref{footnote:species}) the above case ``mix2'', and only this case, breaks the symmetry between heavy-lepton neutrinos and their antiparticles. As a consequence, all neutrino properties, such as luminosities and mean energies, may differ between $\nux$ and $\nuxbar$ in the model using ``mix2'', whereas they are identical in all other models.

Since our neutrino transport scheme evolves the 1st-moment vector, with components $F_{\nu,q}^i$ (where $i=r,\theta,\phi$)\footnote{We note that the fluxes in azimuthal ($\phi$) direction vanish as a result of the approximation that the $\phi$-velocities entering the neutrino-transport equations are neglected in our simulations (see, e.g., Refs.~\cite{Just2015a,Just2021i} for more details).}, independently from the 0th moments, we also need appropriate mixing relations for the 1st moments. The simplest and most straightforward treatment, which is adopted in the majority of our simulations, consists of using the same mixing coefficients, $c_{\nu\tilde{\nu}}$, as used for the 0th moments (cf. Eq.~(\ref{eq:superpos})) and to compute the flavor-mixed flux densities of any of the four evolved species, $\nu$, as
\begin{equation}\label{eq:fluxmix1}
  F_{\nu,q}^i = \sum_{\tilde{\nu}} c_{\nu\tilde{\nu}} F_{\tilde{\nu},q}^{i,0}
\end{equation}
as functions of the unmixed flux densities $F_{\tilde{\nu},q}^{i,0}$. This case is equivalent to assuming that flavor mixing takes place independently of angle in momentum space. We also consider two models with a ``mix1'' treatment of the 0th moments but a slightly different, non-linear mixing of the 1st moments, which assumes that the flux factor at any given energy remains unchanged, i.e.
\begin{align}\label{eq:fluxmix2}
  F_{\nu,q}^i  & = \frac{F_{\nu,q}^{i,0}}{n_{\nu,q}^0}n_{\nu,q} \, .
\end{align}
The mixing scheme for these models, called ``mix1f'', will be used to test the sensitivity of our results to the mixing treatment of the 1st-moments. As will be seen in Sec.~\ref{sec:diff-prescr-flav}), this sensitivity turns out to be very small.

Table~\ref{tab:prop} provides an overview of all performed simulations. Apart from varying $\ffth$ and the scenario adopted for modeling flavor mixing, we also vary the initial torus mass, $m_{\rm tor}^0$, which in NS mergers depends sensitively on the initial binary masses and the nuclear equation of state. Most simulations are conducted for a time of $t_{\rm fin}=10\,$s in axisymmetry and adopt the $\alpha$-viscosity scheme \cite{Shakura1973} to approximately describe angular momentum transport and dissipation due to turbulent stresses. Explicitly, the dynamic viscosity is assumed to be $\eta_{\rm vis}= \alpha_{\rm vis} P_{\rm gas}/\Omega_{\rm Kep}$ in terms of the gas pressure, $P_{\rm gas}$, and Keplerian angular velocity, $\Omega_{\rm Kep}$, and $\alpha_{\rm vis}=0.06$. 

As suggested by previous studies \cite{Fernandez2013b, Just2015a, Siegel2018c, Fernandez2019b, Just2021i, Zhou2021j}, disk models using the $\alpha$-viscosity scheme reproduce most basic features of the torus evolution and ejecta nucleosynthesis fairly well. Nonetheless, the $\alpha$-viscosity is merely an effective mean-field model mimicking angular momentum transport and dissipation generated by MHD turbulence, which in turn is driven by the magnetorotational instability (MRI). Hence, it carries non-negligible uncertainties related to the poorly constrained functional form and amplitude of the viscous stress tensor (see, e.g., \cite{Just2021i} for a comparison of different parametrization choices) and it is unable to describe magnetically dominated outflows. In order to assess the sensitivity of our results to the treatment of turbulent angular momentum transport, we additionally ran two MHD models, one without flavor conversion and one using the mix1f conversion scheme\footnote{The reason for the MHD model adopting the mix1f scheme, instead of the mix1 scheme, is simply because we conducted this simulation first, and the small differences between mix1 and mix1f observed for the viscous models did not justify the efforts to run an additional expensive MHD model.}. However, in contrast to viscous models\footnote{Unlike MHD models, viscous models do not require the third dimension to capture angular momentum transport. In principle, hydrodynamic disks can also develop instabilities other than the MRI that could trigger non-axisymmetric modes, such as the Papaloizou-Pringle instability \cite{Papaloizou1984d,Xie2020j} or the runaway instability~\cite{Korobkin2012l}. However, so far no study known to us has demonstrated the relevance of these instabilities in realistic post-merger disks, which have a low disk-to-BH mass ratio and an extended, nearly Keplerian angular momentum profile.}, MHD models need to be simulated in 3D and with high numerical resolution in order to capture sustained angular moment transport by the MRI~\cite{Moffatt1978}, which makes them considerably more expensive than axisymmetric, viscous disk models. For this reason, we could only run these models until $t_{\rm fin}=0.5\,$s, i.e. for a much shorter time than the viscous models.

For details regarding the initial models, numerical evolution scheme, adopted numerical resolution, as well as the extraction of ejecta properties we refer to Ref.~\cite{Just2021i}. For viscous (MHD) models we count as ejecta all material that crosses $r=10^4\,$km (3000\,km) until $t=t_{\rm fin}$. For the MHD models we additionally ignore the very early and fast ejecta that reach a radius of 3000\,km before $t=50\,$ms, because this material is to a large extent driven by the strong and transient hydrodynamical adjustment that commences during the initial phase.

\subsection{Results for our fiducial BH-torus model}\label{sec:results-fiduc-model}

In this section we investigate the impact of flavor conversions on the evolution of BH-accretion tori and their outflows by comparing a fiducial model with and without flavor conversions (models m1mix1 and m1, respectively).

\subsubsection{Impact on neutrino emission and torus evolution}\label{sec:impact-neutr-emiss}

\begin{figure*} 
 \centering
\includegraphics*[width=.99\textwidth]{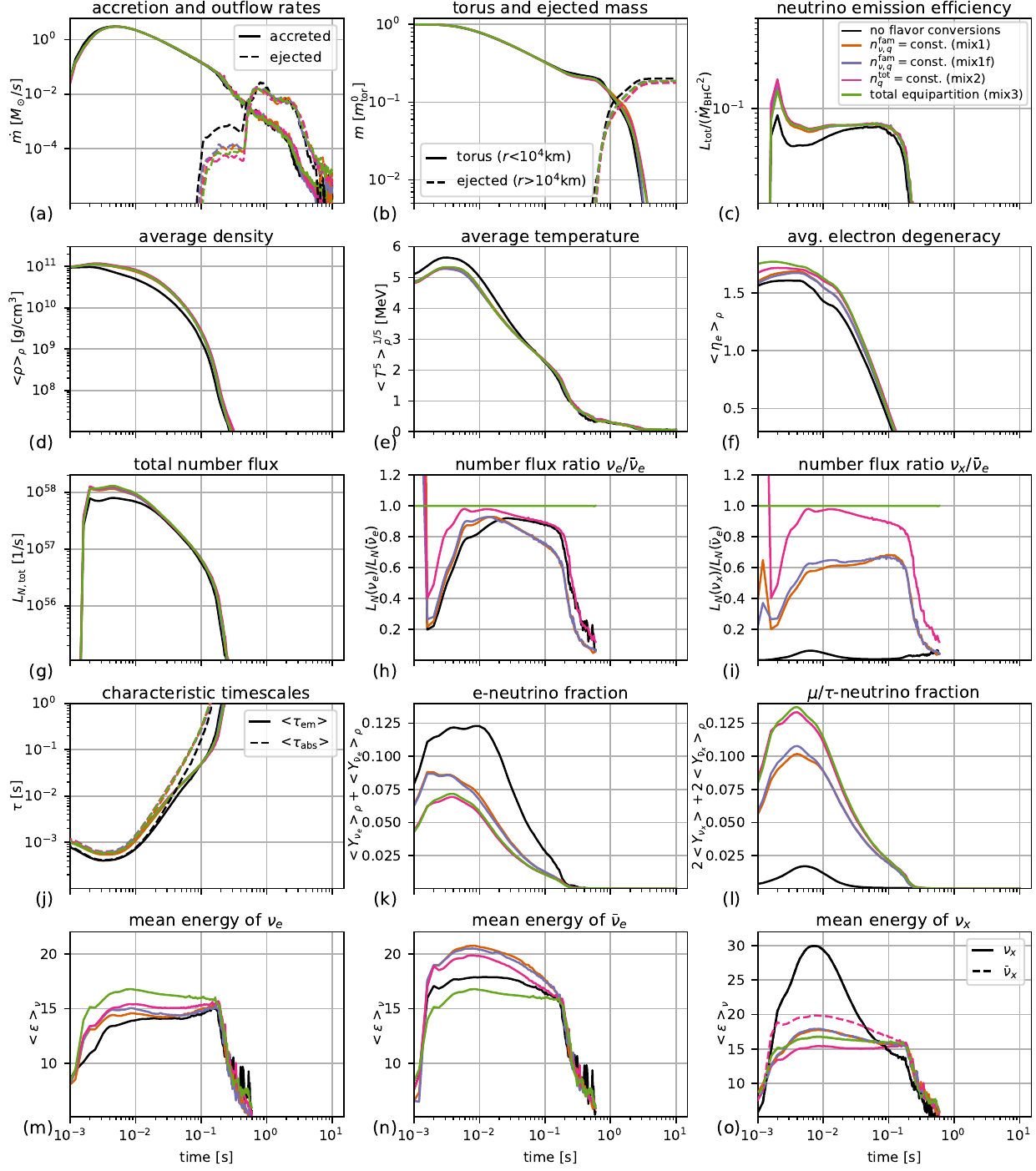}
\caption{Global properties as functions of time for models m1, m1mix1, m1mix1f, m1mix2, and m1mix3 that adopt different treatments of flavor mixing. The panels show: (a) mass accretion rate into the BH, $\dot{M}_{\rm BH}$ (solid), and mass flux through the sphere at a radius of $r=10^4\,$km, (b) torus mass and ejecta mass, (c) neutrino emission efficiency (cf. Eq.~(\ref{eq:eta})), (d)-(f) mass-averaged density, temperature (computed as $\langle T^5\rangle_\rho^{1/5}$ to account for the $T^5$ dependence of the neutrino emission rates), and electron degeneracy, (g) number flux summed over all six neutrino species measured at $r=500\,$km in the laboratory frame, (h) ratio of $\nue$ to $\nuebar$ number fluxes, (i) ratio of $\nux$ (single species) to $\nuebar$ number fluxes, (j) characteristic timescales of emission and absorption computed as in Eq.~(16) and (23) of Ref.~\cite{Just2021i}, (k) average abundance of $\nu_e$ plus $\nuebar$ neutrinos relative to nucleons, (l) average abundance of the four heavy-lepton neutrinos relative to nucleons, (m)-(o) mean energies of $\nue$, $\nuebar$, and $\nux/\nuxbar$ neutrinos, respectively, computed as $L_{\nu}/L_{N,\nu}$ at $r=500\,$km in the laboratory frame. }
\label{globec}
\end{figure*}
\begin{figure} [htb!]
 \centering
\includegraphics*[width=.49\textwidth]{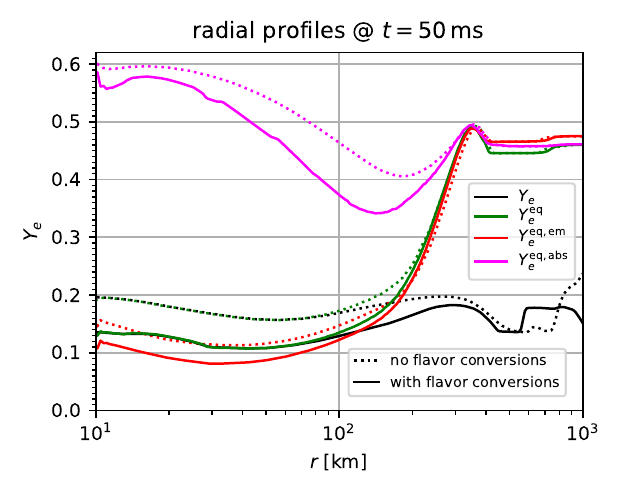}
\caption{Radial profiles of $Y_e$ along the equator for models m1mix1 (solid lines) and m1 (dashed lines) with and without flavor conversions, respectively, as well as the equilibrium values that would result for a constant thermodynamic background and neutrino field, $\yeeq$, for a constant thermodynamic background and vanishing neutrino field, $\yeeqem$, and for vanishing neutrino emission rates and a constant neutrino field, $\yeeqab$. Explicit expressions for $\yeeq, \yeeqem$,~and~$\yeeqab$ are provided in Sec.~2 of Ref.~\cite{Just2021i}.}
\label{yeprof}
\end{figure}
\begin{figure*} [htb!]
 \centering
\includegraphics*[width=.99\textwidth]{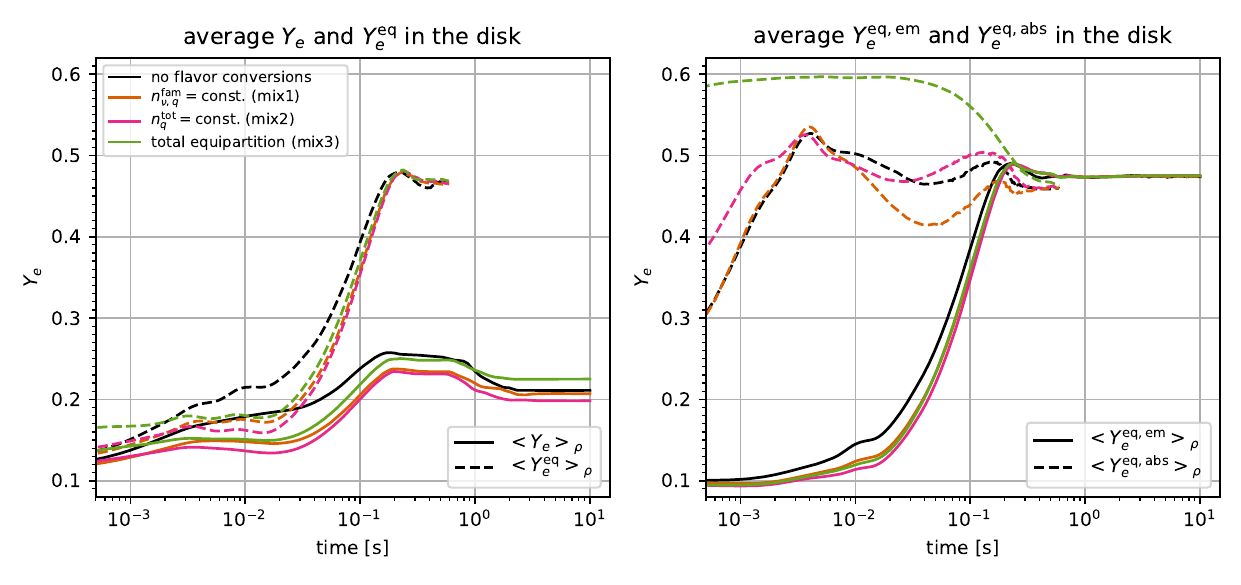}
\caption{Time evolution of mass-averaged values of $Y_e$ and $\yeeq$ (left panel) as well as $\yeeqem$ and $\yeeqab$ (right panel) in the disk (i.e. within radii of $r=10^9\,$cm) for models m1, m1mix1, m1mix2, and m1mix3 that adopt different types of flavor mixing. The combined impact of flavor conversions on neutrino emission and absorption reduces both $\yeeqem$ and $\yeeqab$. The equilibrium value for pure neutrino absorption, $\yeeqab$, is significantly affected only in the case ``mix3'' where neutrinos can mix with with antineutrinos (green dashed lines in right panel).}
\label{yeevo}
\end{figure*}

We summarize global properties characterizing the mass accretion and ejection, as well as the neutrino emission for models with different equilibrium schemes in Fig.~\ref{globec}. The time evolution of post-merger BH tori in multiple dimensions has been investigated in several previous studies using various methods and approximations \cite{Setiawan2006, Fernandez2013b, Just2015a, Siegel2017b, Fujibayashi2020a, Miller2019a, Most2021q, Zhou2021j}. First of all, looking at the rates of mass accretion and matter ejection (cf. panel~(a) of Fig.~\ref{globec}) as well as the torus mass and cumulative ejecta mass (panel~(b)), which are all rather similar, it appears that the flavor conversion does not change the accretion dynamics at a qualitative level. With regard to neutrino effects, however, we do notice significant differences, as we discuss in the following.

The disk traverses three characteristic states while accreting onto the central BH and continuously losing mass: An optically thick and inefficiently cooled, optically thin and efficiently cooled, and optically thin and inefficiently cooled state, and these states are typically encountered for high, medium, and low mass accretion rates onto the black hole, respectively (see, e.g., \cite{Metzger2008c, Just2021i}). Since photons are completely trapped at the considered conditions, the disk is cooled only as a result of neutrino emission. The cooling or neutrino emission efficiency can be measured by the quantity
\begin{align}\label{eq:eta}
  \eta_{\nu} = \frac{\sum_{i} L_{\nu_i}}{\MBHdot c^2} \, ,
\end{align}
i.e. the ratio of the total energy loss rate due to neutrino emission and rest-mass energy accreted onto the BH per unit of time; see panel~(c) of Fig.~\ref{globec}. All neutrino luminosities provided in this study are measured at $r=500\,$km in the laboratory frame. Apart from the brief spike at $t\approx 2\,$ms caused by the transition from an initially non-accreting to an accreting configuration, the cooling efficiency exhibits relatively low values of $\eta_\nu\approx 0.04$ during the early phase for the standard case without flavor conversions (black lines in Fig.~\ref{globec}). During this phase, low values of $\eta_\nu$ are a consequence of finite optical depth effects, i.e. the circumstance that the timescale for neutrinos to diffuse out of the disk is longer than, or comparable to, the timescale over which they are advected into the central BH together with the fluid. On the other hand, considering in Fig.~\ref{globec} any of the models with flavor conversions taken into account, we notice that the early trough in $\eta_\nu$ is basically absent. This indicates that the disk in these models is cooled with high efficiency already in the early phase of evolution. As a consequence of the higher cooling efficiency, disks with active flavor conversions exhibit on average higher densities, lower temperatures, and, as a result, higher electron degeneracies (cf. panels~(d),~(e),~and~(f) of Fig.~\ref{globec}) with respect to the disk without flavor conversions. We compute density-weighted averages of any quantity $X$ as 
\begin{align}\label{eq:rhoavg}
  \langle X\rangle_\rho = \frac{\int_{V_1} \rho X \dd V}{\int_{V_1} \rho \,\dd V} \, ,
\end{align}
where the reference volume, $V_1$, is given by the region below the radius of $10^4\,$km. Although the neutrino emission efficiency is significantly boosted only for the first $\sim 20$\,ms of disk evolution, the impact on the average density and electron degeneracy appears to be long lasting and is visible even at much later times.

What is the reason for the enhanced neutrino emission rates? As can be seen in panel~(i) of Fig.~\ref{globec}, the luminosity ratio of $\nu_x$ neutrinos relative to $\bar\nu_e$ neutrinos grows from $\la 10\,\%$ to $\ga 60\,\%$ when switching on flavor conversions, while the luminosity ratio between $\nue$ and $\nuebar$ does not seem to change significantly from the original values close to unity (cf. panel~(h)). Hence, fast flavor conversions effectively activate the production of heavy-lepton neutrinos in neutrino-cooled disks, which is otherwise strongly subdominant due to the low densities and temperatures in disks compared to proto-NSs or hypermassive NSs. The flavor conversion thus opens up an additional, more efficient channel for disk cooling, because heavy-lepton neutrinos do not (significantly) interact via charged-current (i.e. absorption) reactions with nucleons but only through neutral-current (i.e. scattering) reactions (e.g.~\cite{Bruenn1985}). The optical depth of heavy-lepton neutrinos is therefore reduced roughly by a factor of two compared to that of electron-flavor neutrinos, meaning that they can diffuse more quickly through and out of the disk.

The second important consequence of neutrino flavor conversion, which was already pointed out by Refs.~\cite{Wu:2017drk, Li2021g}, can be grasped by considering panels~(k)~and~(l) of Fig.~\ref{globec}. These panels provide averages of the number fractions of all electron-type neutrinos, $Y_{\nue}+Y_{\nuebar}$, and all heavy-lepton neutrinos, $2Y_{\nux}+2Y_{\nuxbar}$, where $Y_i$ is the particle number of species $i$ relative to the total number of nucleons. Since most of the $\mu/\tau$ neutrinos are produced at the expense of converting electron neutrinos, the abundance of electron-type neutrinos is significantly reduced, namely by a factor of 2--3 for all flavor-mixing cases, compared to the case without flavor conversions.

Since neutrino absorption rates are not only sensitive to the number densities of neutrinos, but also to their mean energies, it is also worth inspecting the change of the neutrino mean energies induced by fast conversions. To this end, we plot in panels~(m),~(n),~and~(o) of Fig.~\ref{globec} the mean energies of radiated neutrinos, computed as
\begin{align}\label{eq:emean}
  \langle \epsilon \rangle_\nu = \frac{L_{\nu}}{L_{N, \nu}}
\end{align}
for the neutrino fluxes measured at $500\,$km in the observer frame. Without flavor conversion, the hierarchy of mean energies is $\langle \epsilon \rangle_{\nue}<\langle \epsilon \rangle_{\nuebar}<\langle \epsilon \rangle_{\nux}$ (and $\langle \epsilon \rangle_{\nux}\approx \langle \epsilon \rangle_{\nuxbar}$), which is typical for neutron-rich disks (e.g. \cite{Just2021i, Miller2020a}) and reflects the circumstances that $\nue$ neutrinos ``see'' more absorption targets (neutrons) than $\nuebar$ neutrinos (protons) and are therefore effectively emitted from a less dense and cooler neutrinosphere. On the other hand, heavy-lepton neutrinos are released from the hot and dense inner region of the disk, because they experience no absorption at all. If flavor conversion via the ``mix1'' prescription is switched on (cf. orange lines of Fig.~\ref{globec}), we observe a swap in the hierarchy of mean energies, i.e. $\langle \epsilon \rangle_{\nux}$ ends up somewhere between $\langle \epsilon \rangle_{\nue}$ and $\langle \epsilon \rangle_{\nuebar}$. This is a consequence of the fact that $n_{\nue,q}>n_{\nuebar,q}$ ($n_{\nue,q}<n_{\nuebar,q}$) in low (high) energy bins $q$, which, according to the ``mix1'' prescription (cf. Eq.~\ref{eq:mix1}) means that the flavor mixing results in $n_{\nux,q}=n_{\nuebar,q}$ for low $q$ and $n_{\nux,q}=n_{\nue,q}$ for high $q$. In other words: Since heavy lepton neutrinos always attain the lower of the two occupation numbers, $n_{\nue,q}$ and $n_{\nuebar,q}$, at a given $q$ their spectrum must peak between the two spectra of the electron-type neutrinos. 

Besides a changed hierarchy, we also observe a spectral hardening (i.e. enhanced mean energies) for both electron-type neutrinos. This result is less straightforward to understand. We suspect it to be a consequence of the possibility that high-energy heavy-lepton neutrinos, which can diffuse more easily out of the disk than electron-type neutrinos, repopulate the high-energy tail of electron-type neutrinos through flavor mixing, or equivalently, that low-energy electron-type neutrinos are converted more effectively into heavy-lepton neutrinos than high-energy electron-type neutrinos. However, we leave a more conclusive analysis to future work.
Finally, we point out that the impact of flavor conversion on the mean energies seen in model m1mix1 and described above is not universal but depends on the cases of flavor mixing; see Sec.~\ref{sec:diff-prescr-flav} for their discussion.

\subsubsection{Impact on the electron fraction in the torus}\label{sec:impact-y_e-torus}

We now examine the consequences of the effects identified in the previous section for the evolution of the electron fraction, $Y_e$, in the disk. A useful proxy for the electron fraction is the equilibrium value $\yeeq$ that would result for $Y_e$ for a fixed density, temperature, and neutrino field \cite{Just2021i}. For the considered conditions in the disk, nuclei are fully dissociated into free nucleons during most of the time such that $\yeeq$ is determined entirely by the $\beta$-processes on free nucleons of Eqs.~(\ref{eq:beta}). In order to gain insight about the individual sensitivities of $\yeeq$ on the thermodynamic state and on the neutrino field, $\yeeq$ can be further decomposed into a value characterizing a pure emission equilibrium, $\yeeqem$, as well as a quantity corresponding to a pure absorption equilibrium, $\yeeqab$. Given the relatively low optical depth of neutrino-cooled disks, $\yeeq$ typically lies close to, but slightly above, $\yeeqem$, while the difference, $\yeeq-\yeeqem$, grows with the abundance of $\nue$ and $\nuebar$ neutrinos relative to nucleons. While $\yeeqab$ depends mainly on the neutrino field, $\yeeqem$ depends solely on the thermodynamic state (namely on $\rho$ and $T$ in nuclear statistical equilibrium). Importantly, the state of pure emission equilibrium characterized by $\yeeqem$ becomes more neutron rich for higher levels of electron degeneracy \cite{Beloborodov2003,Chen2007} because of the concomitant suppression of positrons and correspondingly low rates of $e^+$ captures on neutrons.

Figure~\ref{yeprof} shows radial profiles of $Y_e$, $\yeeq$, $\yeeqem$,~and~$\yeeqab$ along the equator at $t=50\,$ms for two models with (solid lines) and without (dashed lines) fast conversions. In both models the hot and dense part of the disk in the innermost $\sim 100\,$km is close to weak equilibrium, $Y_e\approx \yeeq$, while in the expanding outer layers of the disk $Y_e$ departs from its local equilibrium value and eventually freezes out. In the model with active flavor conversions we observe values of $\yeeq$ (green lines) that are reduced throughout the first $\sim 100\,$km by approximately 0.05--0.07 compared to the model without conversions. This reduction results from the combination of the two effects related to flavor conversions that were previously identified in Sec.~\ref{sec:impact-neutr-emiss}: First, the enhanced cooling rates and higher electron degeneracies cause the neutrino emission rates to favor a more neutron-rich equilibrium, as can be inferred from the fact that $\yeeqem$ (red lines) is lower by about 0.04 when flavor conversions are taken into account. Second, the attenuation of electron-type neutrino abundances in the disk (cf. panel~(k) of Fig.~\ref{globec} and discussion of Sec.~\ref{sec:impact-neutr-emiss}) reduces the tendency of neutrino captures to drive $Y_e$ towards $\yeeqab\sim 0.5$.

In Fig.~\ref{yeevo} the global averages of $Y_e$ and its equilibria are plotted as functions of time. In all models $\langle Y_e\rangle_\rho$ tends to follow $\langle \yeeq \rangle_\rho$ but gradually decouples from it with time, because a growing fraction of the disk expands and cools down to temperatures $T\la 1\,$MeV, where neutrino emission timescales become much longer than the evolution timescales of the disk of $\mathcal{O}(\la 1\,\mathrm{s})$. By $t\sim 200\,$ms weak interactions have ceased completely and $Y_e$ remains constant along all Lagrangian fluid elements. Enabling fast flavor conversions causes a reduction of the global average, $\langle Y_e\rangle_\rho$, by about 0.03--0.06 during the first $\sim 100\,$ms, as a result of the two previously discussed effects related to neutrino emission and absorption.

\subsubsection{Impact on outflow properties}\label{sec:impact-outfl-prop}

\begin{figure*} [htb!]
  \centering
  \includegraphics*[width=.99\textwidth]{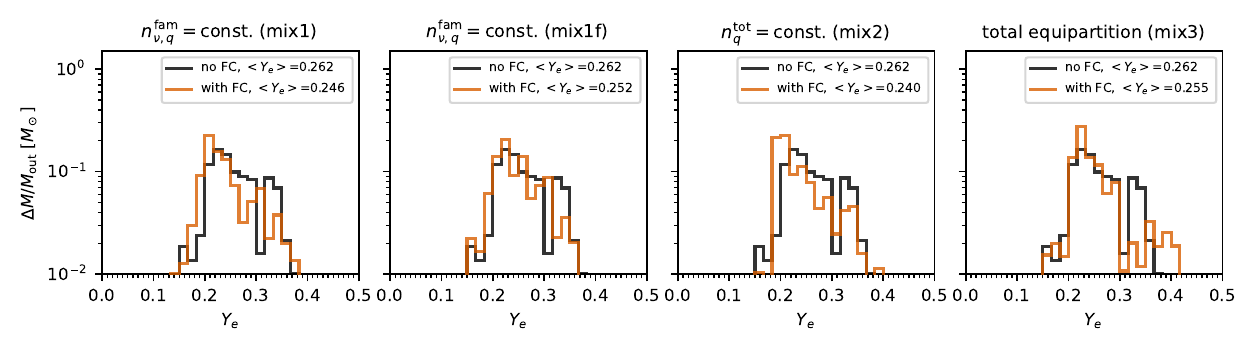}
  \caption{Mass versus $Y_e$ histograms of the ejected material for models m1mix1, m1mix1f, m1mix2, and m1mix3 (orange lines, from left to right) using different schemes of flavor mixing in regions where flavor conversions occur. Black lines denote the reference model without flavor conversions.}
  \label{hist41}
\end{figure*}
\begin{figure*} [htb!]
 \centering
\includegraphics*[width=.99\textwidth]{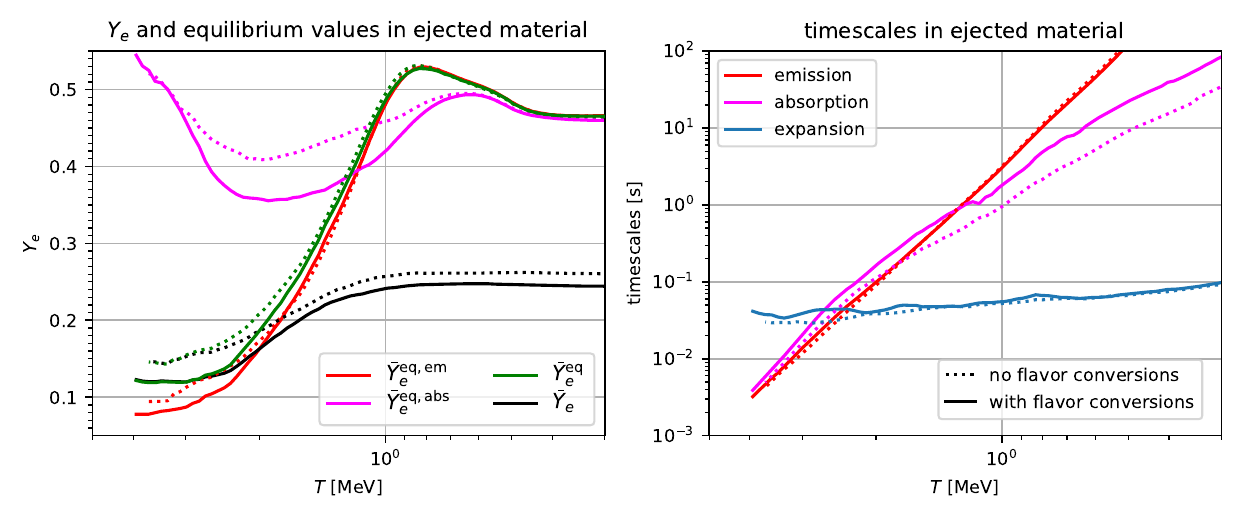}
\caption{Electron fraction and its equilibrium values (left panel) as well as characteristic timescales (right panel) averaged at given temperature during the expansion of ejected material for models m1mix1 (solid lines) and m1 (dashed lines) including (not including) flavor conversions. Note that the temperature decreases from left to right.}
\label{trajavg}
\end{figure*}
Backed by the insights about the disk evolution, we now take a look at the ramifications of flavor conversions for the properties of the ejected material. First, we note that the ejecta masses, bulk velocities, and entropies, which are summarized in Table~\ref{tab:prop}, all seem to exhibit slightly lower values in most of the models with flavor conversions \footnote{A noteworthy exception to this trend is observed for the ejecta velocities in the two MHD models. However, since both models are evolved only until $t=0.5\,$s, the ejecta properties in these models are not finalized and a comparison is therefore inconclusive.}, however, only at the level of a few per cent. A likely explanation for this tendency is the enhanced cooling efficiency during the early phase of disk evolution (cf. Sec.~\ref{sec:impact-neutr-emiss}). Since the mass ejection processes in the considered disks are driven mainly by viscous heating, higher rates of neutrino cooling counteract these processes and may therefore lead to slower, cooler, and less massive ejecta.

Next, we consider the distribution of $Y_e$ in the ejected material. To this end we extract outflow trajectories exactly in the way described in Ref.~\cite{Just2021i} by sampling all material that crosses the sphere at a radius of $10^4\,$km ($3\times 10^3\,$km) in the viscous (MHD) models within the simulated times. The histograms in the left panel of Fig.~\ref{hist41} provide information about the amount of mass that is ejected with given electron fractions. The histograms of both models with and without flavor conversion are rather similar and exhibit a broad peak of width $\Delta Y_e\sim 0.15$ centered around values close to $Y_e\sim 0.25$. For models including flavor conversions the distributions are slightly shifted to lower $Y_e$ by about 0.01--0.03 depending on the model.

Figure~\ref{trajavg} provides more information about the evolution of average properties of the expanding ejecta for two models with and without flavor conversions, namely $Y_e, \yeeq, \yeeqem,$~and~$\yeeqab$ (left panel) as well as the characteristic timescales of emission, absorption, and expansion (right panel), all averaged at given temperatures during the expansion (cf. \cite{Just2021i} for details regarding the computation of these quantities). We first consider the timescales. Since the neutrino emission rates per baryon are mainly a function of the temperature (roughly $\propto T^5$) the emission timescales are basically insensitive to flavor conversions as function of temperature. The absorption timescales, in contrast, become significantly longer in the model with fast conversions owing to the diminished abundance of $\nue$ and $\nuebar$ neutrinos. Finally, the blue lines indicate slightly longer ejection timescales, i.e. smaller outflow velocities, in the model with flavor conversions. These lower outflow velocities are probably a consequence of the enhanced rates of neutrino cooling, which effectively reduce the initial, total energy of fluid elements ending up in the ejecta and thereby reduce their final kinetic energies.

Now looking at the left panel of Fig.~\ref{trajavg}, we observe at high temperatures, which are sampled primarily by material located near the density maximum of the torus, the same features already identified in the previous section, namely a reduction of $\yeeqbar$ and $\bar{Y}_e$ due to the combination of increased electron degeneracies and reduced absorption rates of $\nue$ and $\nuebar$ neutrinos. However, following the material along its expansion to lower temperatures the differences in $\bar{Y}_e$, $\yeeqbar$,~and~$\yeeqembar$ between both models decrease, such that the values of $\bar{Y}_e$ at freeze-out (i.e. at $T\ga 1\,$MeV) lie much closer together than initially in the torus.

The result that the net effect on the ejecta-$Y_e$ is relatively small, while the impact on the torus-$Y_e$ is more significant, is not particularly surprising in light of what is known from previous investigations of neutrino-cooled disks; see, e.g., Ref.~\cite{Just2021i} for a systematic study of the torus-$Y_e$ and the ejecta-$Y_e$. After leaving the hot and dense equilibrium conditions in the early stage of the disk evolution but before entering weak freeze-out -- namely in regions where $2\,\mathrm{MeV}>T>1\,\mathrm{MeV}$ -- the material is still subject to numerous weak interactions. Both emission and absorption reactions tend to increase $Y_e$ for $T\la 2\,$MeV (cf. left panel of Fig.~\ref{trajavg}), and by doing so can partially erase the memory of the original torus-$Y_e$. Hence, the final $Y_e$ in the outflow material does not only depend on $Y_e$ in the bulk of the torus (which lies close to $\yeeq$), but also on the detailed conditions during the expansion, such as the expansion timescale. The fact that material expands on average slower in the models with flavor conversions, as a consequence of enhanced neutrino cooling, may contribute to the explanation for the only modest impact of flavor conversions on the ejecta-$Y_e$. This is because slower ejecta material is able to adapt $Y_e$ for a longer period of time to its local equilibrium value before freezing out. Thus, the effect of enhanced neutrino cooling due to flavor conversions can in that way also act in the opposite direction, i.e. tend to increase $Y_e$, compared to the initial reduction of $\yeeq$ in the torus.

\subsubsection{Impact on nucleosynthesis and electromagnetic counterpart}

\begin{figure*} [htb!]
  \centering
  \raisebox{+0.6cm}{\includegraphics*[width=.49\textwidth]{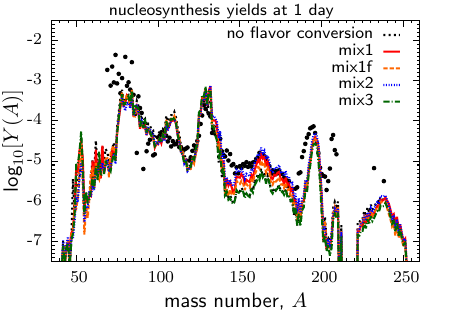}}
  \includegraphics*[width=.46\textwidth]{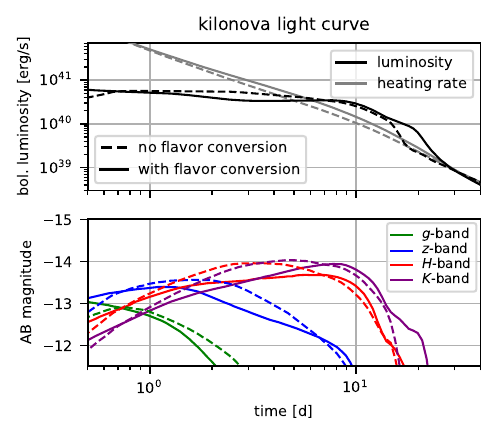}
  \caption{\emph{Left panel:} Abundance distributions as functions of the atomic mass number of elements synthesized in the ejected material in models m1, m1mix1, m1mix1f, m1mix2, and m1mix3 measured at $t=1\,$d after the birth of the disk. The imprint of flavor conversions is most visible in the enhanced abundances of lanthanides. \emph{Right panel:} Kilonova signal powered by radioactive heating of synthesized material for models m1 (dashed lines) and m1mix1 (solid lines) estimated using spherically averaged ejecta properties. The top panel shows the bolometric luminosities (black) and effective heating rates (including thermalization; grey), the bottom panel depicts AB magnitudes in selected bands. Flavor conversions induce more powerful heating but also higher opacities, causing the peak emission to take place with nearly the same luminosity but for an extended period of time.}
  \label{nuckilo}
\end{figure*}

We finally inspect the impact of flavor conversions on the nucleosynthesis yields and kilonova emission. The left panel of Fig.~\ref{nuckilo} shows the elemental abundances as function of atomic mass number, $A$, for the series of models varying the flavor-mixing schemes, while Table~\ref{tab:prop} provides the mass fractions of all 2nd-peak elements, $X_{\rm 2nd}$, lanthanides plus actinides, $X_{\rm LA}$, and 3rd-peak elements, $X_{\rm 3rd}$ synthesized in the ejecta of each model. The trajectories have been post-processed using the same nuclear reaction network as adopted in~\cite{Wu2016a}. We start the network calculations at $T=10\,$GK for the trajectories whose temperatures exceed 10\,GK during the evolution. For a small part of the trajectories for which the temperatures are always below 10\,GK, we start the evolution at the time $t=0$ corresponding to the start of the hydrodynamic simulation. As anticipated from the previously found reduction of $Y_e$ in the ejecta, flavor conversions enhance the production of nearly all r-process elements, while the largest relative increase (of up to a factor of $\sim 2$ depending on the model) is observed for the lanthanides. Not surprisingly, for different models the size of the impact of flavor conversions on the mass fractions scales pretty well with the size of the impact on $Y_e$, i.e. models with smaller reduction of $Y_e$ exhibit a milder increase of $X_{\rm LA}$ etc.

In order to assess the impact on the kilonova light curves, we use the trajectories and results from the nucleosynthesis analysis, assume constant velocities beyond $r=10^9\,$cm, and construct spherically symmetric distributions of mass, heating rates, mass fractions of lanthanides plus actinides, and mean atomic mass numbers as functions of velocity (as was also done in Ref.~\cite{Just2021i}). We then plug these data into the spherically symmetric version of the scheme described in Ref.~\cite{Just2022a}, which solves the radiative transfer equations in the M1 approximation using simplified, parametrized opacities (see \cite{Just2022a} for technical details of the solver). The right panels of Fig.~\ref{nuckilo} provide the results for the two models m1 (dashed lines) and m1mix1 (solid lines), namely the radioactive heating rates powering the light curve and bolometric luminosities (top panel) and the broadband magnitudes for selected frequency bands (bottom panel).

The kilonova is affected in two ways by the modified nucleosynthesis pattern in models with flavor oscillations: First, the radioactive heating rates are boosted at $3\la t\la 20\,$d by several tens of percent mostly as a consequence of the increased abundance of 2nd-peak elements, which dominate the heating rates during this period of time. The second effect is given by the increased opacities, which mainly result from the higher abundance of lanthanides. Since the second effect to some extent counteracts the first effect, the light curve in the model with flavor conversions is barely more luminous until the plateau-like peak epoch at about $t\approx 10\,$d than in the model without conversions. After the plateau the light curve decays more slowly and reaches the asymptotic behavior (given by the radioactive heating rate) several days later. The broadband light curves exhibit similar differences between both models. Overall, the impact of fast flavor conversions on the kilonova predicted by our models is noticeable mostly in the duration of the high-luminosity emission.

\subsection{Model dependence} \label{sec:model-dependence}

In this section we examine the sensitivity of the findings of the previous section to variations of the flavor-mixing prescription, the chosen threshold for the onset of flavor instabilities, the disk mass, and to replacing the $\alpha$-viscosity with an MHD treatment.

\subsubsection{Dependence on flavor mixing scenario}\label{sec:diff-prescr-flav}

Since all of our flavor-mixing treatments are likely to be simplified, we now compare different prescriptions in order to get a feeling for the sensitivity of the results to the final state of the neutrino distributions reached as a result of the flavor instability.

In the fiducial model that we have considered thus far, m1mix1, neutrinos cannot reach full equipartition of all flavors at unstable points, because the instability is required to respect the conservation of lepton number of each family. Relaxing this requirement, as in model m1mix2, leads to a stronger conversion of $\nue$ and $\nuebar$ neutrinos into $\nux$ and $\nuxbar$ neutrinos. As a result, the system reaches an almost complete global flavor equilibrium, in which $L_{N,\nue}=L_{N,\nux}$ and $L_{N,\nuebar}= L_{N,\nuxbar}$ is fulfilled nearly perfectly (not shown explicitly in Fig.~\ref{globec}), $L_{N,\nue}\approx L_{N,\nuebar}$ (cf. purple line in panel~(h) of Fig.~\ref{globec}), as well as $\langle Y_{\nue}\rangle_{\rho} + \langle Y_{\nuebar} \rangle_{\rho} \approx \langle Y_{\nux}\rangle_{\rho}+\langle Y_{\nuxbar}\rangle_{\rho}$ (panels~(k)~and~(l)). The mean energies of emitted neutrinos (cf. panels~(m)-(o) of Fig.~\ref{globec}) retain the $\langle\epsilon\rangle_{\nue}<\langle\epsilon\rangle_{\nuebar}$ behavior but, in contrast to the ``mix1'' case of flavor mixing (cf. Sec.~\ref{sec:impact-neutr-emiss}), now fulfill $\langle\epsilon\rangle_{\nue} \approx \langle\epsilon\rangle_{\nux}$ as well as $\langle\epsilon\rangle_{\nuebar}\approx \langle\epsilon\rangle_{\nuxbar}$. The nearly perfect alignment of all neutrino mean energies and all antineutrino mean energies was to be expected considering the complete (but separate for neutrinos and antineutrinos) equipartition in this scenario. As can be inferred from Figs.~\ref{globec} and~\ref{yeevo} as well as Table~\ref{tab:prop}, all effects identified and attributed to flavor conversions in Sec.~\ref{sec:results-fiduc-model} turn out to be qualitatively similar but quantitatively more pronounced compared to the case with conserved family lepton number. In particular, electrons are slightly more degenerate, electron neutrinos slightly less abundant, and $\yeeqem$ and $\yeeq$ are correspondingly lower. These effects render the overall impact of flavor conversions on the ejected material and the nucleosynthesis yields (cf. Fig.~\ref{nuckilo} and Table~\ref{tab:prop}) somewhat more significant. 

Next, we consider model m1mix3 using the flavor-mixing treatment ``mix3'' (cf. Sec.~\ref{sec:simul-setup-invest}), in which neutrino flavors are mixed even between neutrinos and antineutrinos. As the green lines in panels~(h)~and~(i) of Fig.~\ref{globec} show, this prescription leads to a perfect equipartition of the numbers of neutrinos emitted per time for all six flavors. However, Fig.~\ref{globec} reveals that compared to model m1mix2, where flavor mixing is already almost complete, the quantitative impact remains small in most diagnostic properties. Nevertheless, a remarkable characteristic of model m1mix3 is the complete degeneracy of all mean energies of emitted neutrinos, $\langle\epsilon\rangle_{\nu}$; cf. panels~(m),~(n),~and~(o) in Fig.~\ref{globec}. The consequence of this feature is a larger value of \cite{Qian1996}
\begin{align}\label{eq:yeeqab}
  \yeeqab \approx \left(1+\frac{L_{N,\nuebar}}{L_{N,\nuebar}}\cdot \frac{\langle\epsilon\rangle_{\nuebar}-2.6\,\mathrm{MeV}}{\langle\epsilon\rangle_{\nue}+2.6\,\mathrm{MeV}}\right)^{-1} \, , 
\end{align}
i.e. neutrino absorption tends to drive $Y_e$ to values as high as $\approx 0.6$, instead of $\approx 0.45$--0.5 as in the other models (cf. dashed lines in the right panel of Fig.~\ref{yeevo}). As a result, the values of $\yeeq$ and $Y_e$ are slightly enhanced as well, as are the electron fractions in the ejected material, compared to the other flavor-mixing cases. For this reason, the mix3 case of flavor mixing turns out to have the smallest impact among the considered prescriptions on the composition of the torus and its ejecta.

Models m1mix1, m1mix2, and m1mix3 all assume the same linear flavor-mixing coefficients for the components of the flux density vector, $F^i_{\nu,q}$, as were used for the number densities, $n_{\nu,q}$ (cf. Eq.~(\ref{eq:fluxmix1})). Model m1mix1f instead assumes a different treatment of the flux densities, namely that the flux factor remains unaffected by the flavor instability (cf. Eq.~(\ref{eq:fluxmix2})). This model thus allows an assessment of the sensitivity of our results with respect to a variation of the 1st-moment treatment. As can be seen from comparing the purple lines with the orange lines in Fig.~\ref{globec}, this sensitivity turns out to be very mild, indicating that most of the impact of flavor conversions is connected to the mixing of neutrino numbers and not, at least not in the systems considered here, to changes of the angular distribution induced by flavor conversions.

\subsubsection{Dependence on threshold for the onset of flavor mixing}\label{sec:diff-inst-thresh}
\begin{figure*} [htb!]
 \centering
\includegraphics*[width=.99\textwidth]{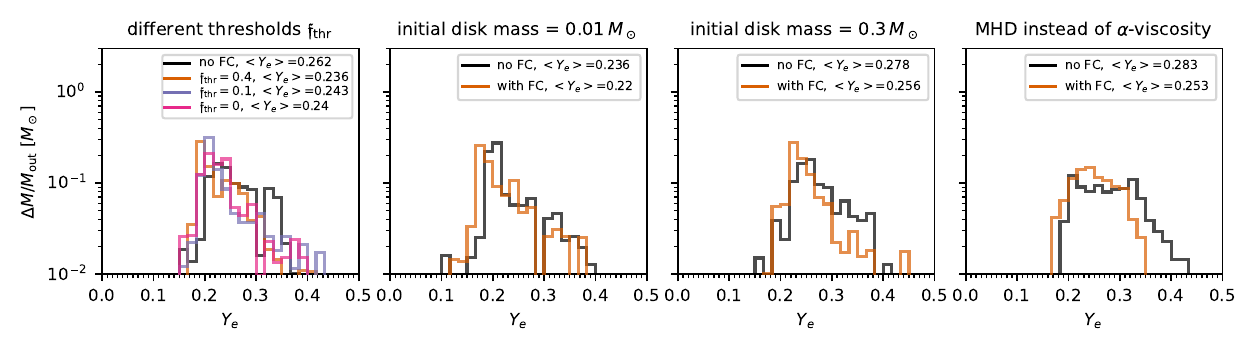}
\caption{Mass versus $Y_e$ histograms of the ejected material for models m1, m1f4, m1f1,~and m1f0 (left panel), models m01 and m01mix1 (second panel from left), models m3 and m3mix1 (third panel from left), as well as models m1mag and m1magmix1f (right panel). The colored lines refer to models including flavor conversion, while the black lines denote the corresponding models without flavor conversion. In the three left panels (right panel), all material outside of a radius of $10^4$\,km ($3000\,$km) at a time of $10\,$s ($0.5\,$s) is counted as ejecta.}
\label{hist42}
\end{figure*}

The second ingredient of our implementation of fast conversions, which was motivated by the search for ELN crossings in disk snapshots discussed in Sec.~\ref{sec:occurr-fast-conv}, is the threshold for the flux factor, $\ffth$, above which we assume fast flavor mixing to occur. Next to the models with $\ffth=0.175$, we ran additional models with $\ffth =0.4, 0.1$,~and $0$ to test the sensitivity to the choice of $\ffth$. The case of $\ffth=0$ is equivalent to assuming flavor conversion to take place everywhere inside of the torus, while the case of $\ffth =0.4$ shifts the surface, outside of which conversions take place, further away from the densest regions in the disk compared to the fiducial model, m1mix1. As revealed by Table~\ref{tab:prop} and the histograms in Fig.~\ref{hist42}, the ejecta properties are not particularly sensitive to $\ffth$, i.e. all models exhibit a similar reduction of the ejecta-$Y_e$ and corresponding enhancement of r-process abundances. The weak dependence of the ejecta properties on $\ffth$ signals saturation of the impact of flavor conversions and is likely related to the circumstance that the surface $\ff=\ffth$, outside of which conversions are activated, shrinks quickly during the first $\sim 100\,$ms. This renders the differences between models with different values of $\ffth$ small already before the majority of ejecta experience weak freeze out. Somewhat surprisingly, the final $Y_e$ in the ejecta varies non-monotonically with $\ffth$ (Table~\ref{tab:prop}). This behavior is possibly related to the competition of the two counteracting consequences of enhanced cooling mentioned in Sec.~\ref{sec:impact-outfl-prop} (reduction of torus $Y_e$ versus reduced expansion velocities, which cause a stronger late-time increase of $Y_e$ in the ejecta).

\subsubsection{Dependence on torus mass}

\begin{figure*} [htb!]
 \centering
\includegraphics*[width=.99\textwidth]{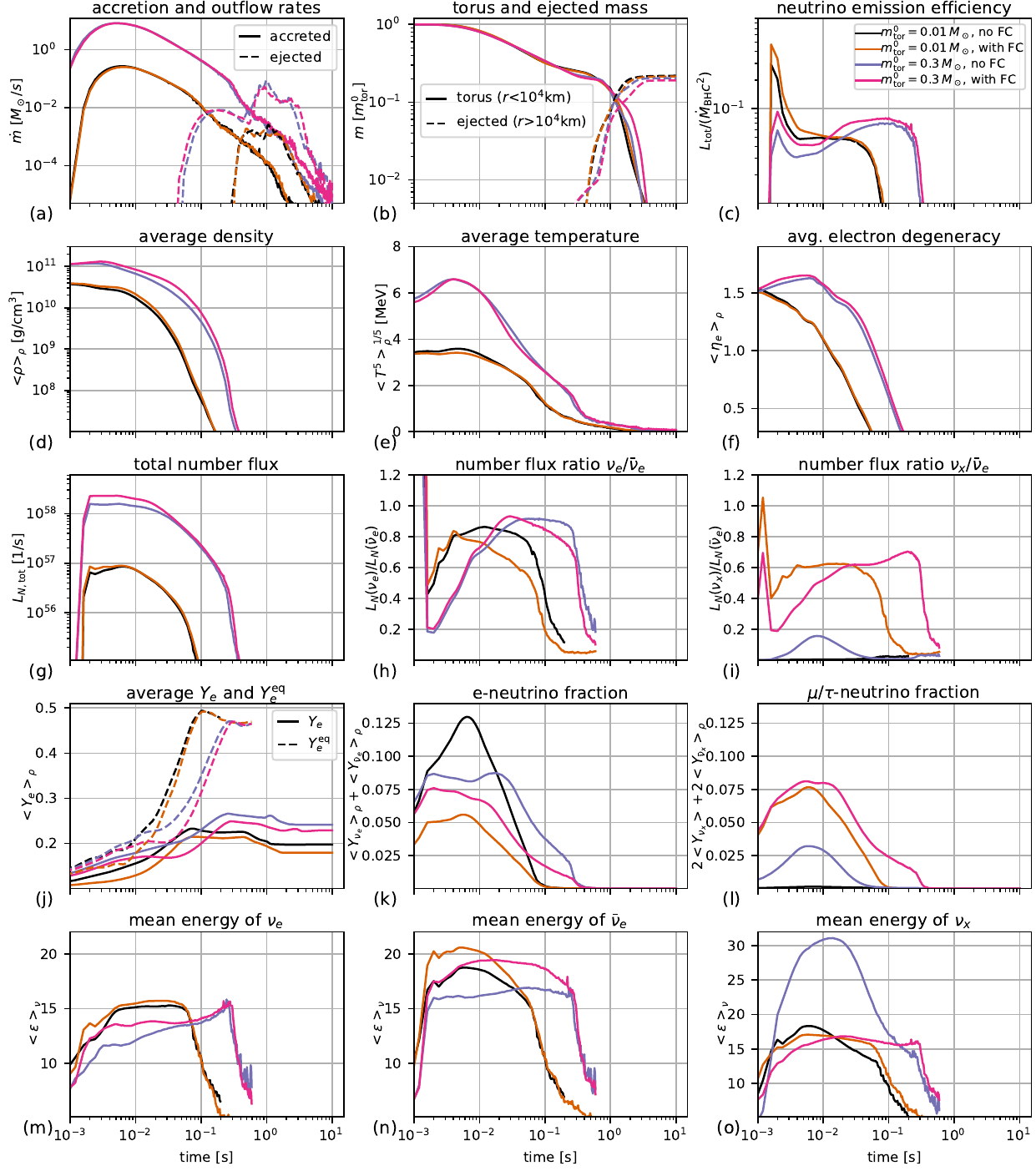}
\caption{Same as Fig.~\ref{globec} but for models m01, m01mix1, m3, and m3mix1 using the different initial disk masses as indicated in the legend of panel~(c). Note that the characteristic timescales of panel~(j) in Fig.~\ref{globec} are replaced here by the averages of the electron fraction and its equilibrium value.}
\label{globmdisk}
\end{figure*}

An important global parameter controlling the weak interaction rates in the torus, and therefore the composition of the outflows, is the disk mass (see, e.g., Refs.\cite{Fernandez2013b, Fujibayashi2020a, De2020a, Just2021i}). Figure~\ref{globmdisk} shows the global properties of models with lower ($0.01\,M_\odot$) and higher ($0.3\,M_\odot$) disk masses compared to the fiducial model ($0.1\,M_\odot$) that was considered in the previous sections. The second and third panels from the left in Fig.~\ref{hist42} additionally provide the corresponding $Y_e$ histograms. As can be seen in these figures, as well as in Table~\ref{tab:prop}, neither of the cases, lower or higher disk mass, leads to more dramatic consequences of flavor conversions. The differences in the average $Y_e$ of ejected material due to flavor conversions are $0.016, 0.016$,~and~$0.018$ for the series of models with torus masses of $0.01, 0.1$,~and~$0.3\,M_\odot$, respectively, i.e. rather insensitive to the disk mass. This seems somewhat surprising at first in view of the fact that the role of neutrino absorption generally grows in more massive disks \cite{Just2021i}. The explanation for this mild dependence is likely connected to the circumstance that the extent of neutrino flavor conversion, by which the number of electron neutrinos is reduced \emph{in the disk}, is smaller in a more massive disk. This can be seen by the high abundances of electron neutrinos in the $0.3\,M_\odot$ disk that survive the flavor instabilities (cf. panel~(k) in Fig.~\ref{globmdisk}), at least during the first tens of milliseconds of evolution. This feature, in turn, is a consequence of the fact that flavor instabilities take place further away from the densest regions in a more massive disk compared to a less massive disk, because a larger fraction of the disk exhibits neutrinos in flavor-stable ($\ff<\ffth$) conditions. Moreover, an additional reason for the reduced conversion ratio is that in the $0.3\,M_\odot$ disk heavy-lepton neutrinos are produced with non-negligible rates already without flavor mixing (cf. purple line in panel~(l) of Fig.~\ref{globmdisk}). In summary, we do not see a strong dependence on the disk mass in our models.

\subsubsection{MHD treatment instead of $\alpha$-viscosity}

\begin{figure*} [htb!]
 \centering
\includegraphics*[width=.99\textwidth]{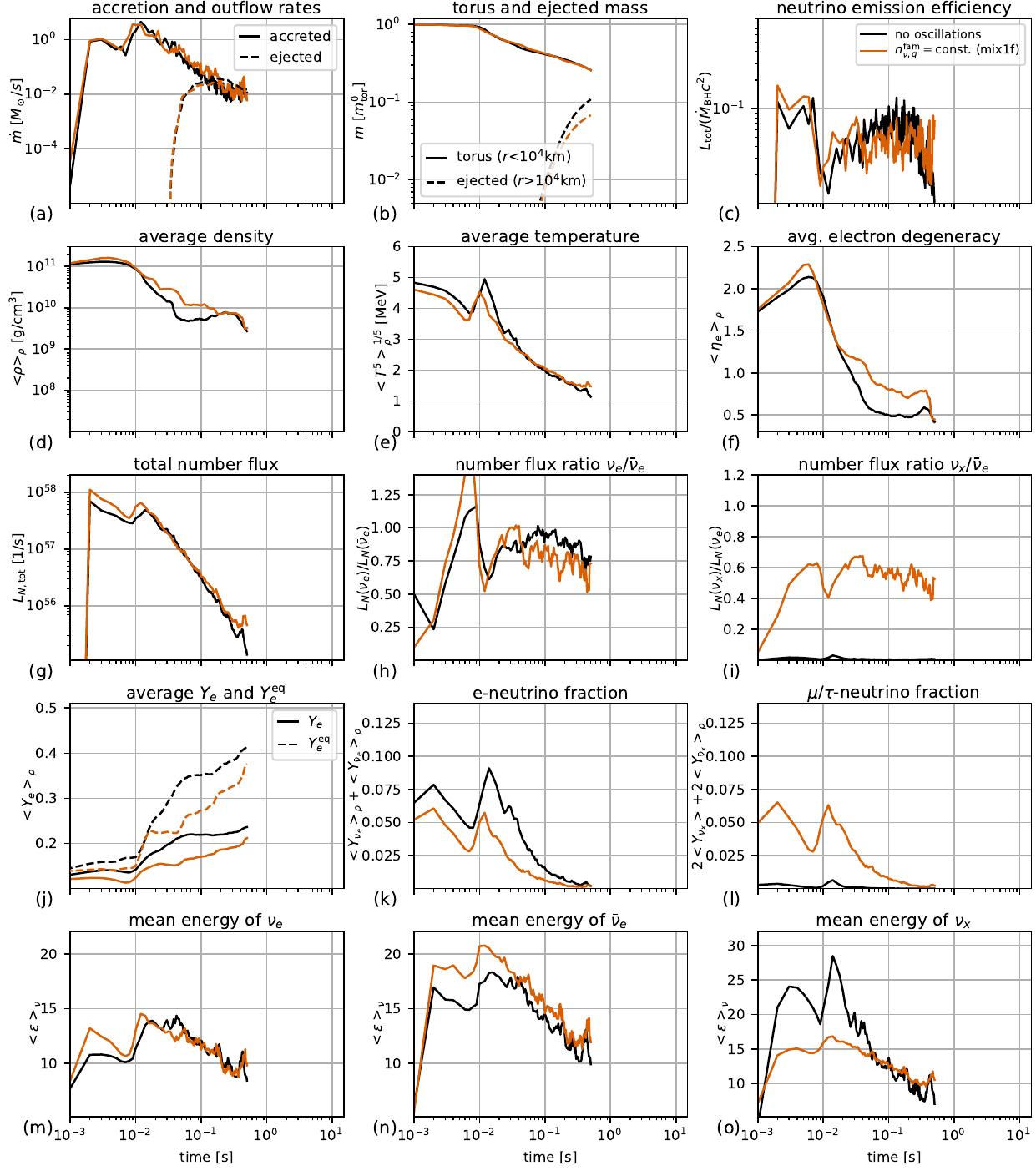}
\caption{Same as Fig.~\ref{globec} but for models m1mag and m1magmix1f using an MHD description instead of $\alpha$-viscosity treatment. Note that the characteristic timescales of panel~(j) in Fig.~\ref{globec} are replaced here by the averages of the electron fraction and its equilibrium value. Also note that panels~(a) and (b) count only material ejected beyond $r=10^4\,$km, whereas for the ejecta analysis of the MHD models we use all material crossing $r=3000\,$km.}
\label{globmhd}
\end{figure*}
All models discussed so far adopt the $\alpha$-viscosity scheme to describe MHD turbulence by an effectively laminar behavior to capture turbulent angular momentum transport and energy dissipation. In order to check the sensitivity of our results to the treatment of turbulent angular momentum transport, we now take a look at the MHD models keeping in mind, however, that these models could only be simulated for a much shorter evolution time of $t_f=0.5\,$s.

The same global properties as discussed before are shown in Fig.~\ref{globmhd}. In contrast to the viscous models, most global properties now carry significant temporal fluctuations as a result of the flow pattern being strongly turbulent at all times in the MHD models -- whereas the viscous models are purely laminar during the neutrino-cooled phase and exhibit convective motions only afterwards. However, considering the time-averaged behavior, Fig.~\ref{globmhd} reveals that fast conversions induce qualitatively, and to some extent quantitatively, the same effects that were observed for the viscous tori, namely higher densities, lower temperatures, and higher electron degeneracies in the disk (cf. panels~(d),~(e),~and~(f)), as well as a significant reduction (increase) of the number fraction of electron ($\mu/\tau$) neutrinos (panels~(k) and (l), respectively). As a result of these effects, the equilibrium and actual electron fractions are again lower than without flavor conversions (panel~(j)). As can be seen in panel~(b) showing the outflow rates across the sphere at $r=10^4\,$km, the matter ejection process is not as powerful and somewhat delayed in the model with flavor conversions. One reason for that are probably the enhanced cooling rates. However, at this point we cannot exclude that another, purely numerical reason might play a role, namely the possibility that the MRI is more poorly resolved in the cooler, and therefore geometrically thinner, torus when flavor conversions are enabled. A similar suppression of the MRI was suspected in Ref.~\cite{Just2021i} to explain the relatively slow evolution in a model neglecting neutrino absorption. Resolving this issue, e.g. by conducting an expensive resolution study, is beyond the scope of the present work.

Before considering the ejecta properties, we stress that due to the shorter evolution times of only $t_{\rm fin}=0.5$\,s (compared to 10\,s for the viscous models), only a fraction of the final amount of ejecta is captured by our models. Therefore, since the matter-ejection processes are time dependent, and this time dependence is non-trivial and may differ between viscous and MHD models~\cite{Fernandez2019b, Just2021i, Li2021g}, a detailed quantitative comparison of the ejecta properties between MHD and viscous models (even if we only considered the material that is ejected beyond $3000$ km until $0.5$ s) is likely to be inconclusive and omitted here. In fact, even the comparison between both MHD models is inconclusive, because, as previously mentioned, mass ejection happens to proceed more slowly in the model with flavor conversions, causing the fraction of the measured amount of ejecta relative to the total amount of ejecta to be significantly smaller in this model compared to the model without conversions. For instance, the large factor of almost two (cf. Table~\ref{tab:prop}) between the ejecta masses of both models obtained until $t=t_{\rm fin}$ may be strongly exaggerated by the ambiguous time of measurement (at $t=t_{\rm fin}$) and would probably be much smaller if we had evolved the disks until complete disintegration.

The $Y_e$ distributions of all material ejected beyond $r=3000\,$km until $t=0.5\,$s is shown in the right panel of Fig.~\ref{hist42}. Both models, with and without flavor conversions, lead to $Y_e$ distributions that are in the same ballpark as for the viscous models. The impact of flavor conversions on $Y_e$ seems to be somewhat more pronounced in the MHD models, where the average $Y_e$ value is reduced from 0.283 to 0.253. This may be connected to the faster expansion speeds compared to the viscous models (cf. Table~\ref{tab:prop}), which allow outflow material to retain a $Y_e$ value that lies closer to the initial value in the torus (recalling that in the torus $Y_e$ is significantly reduced as a result of flavor conversions). Consistent with the results for the viscous models, the nucleosynthesis yields of r-process elements also increase by several tens of percent with flavor conversions being activated.

Finally, we note that our results are in qualitative agreement with the results reported recently by Ref.~\cite{Li2021g} for a 3D general relativistic MHD model that was simulated until $t=0.4\,$s and used a more simplistic scheme than the one described in Sec.~\ref{sec:occurr-fast-conv} for finding ELN-crossings. The authors of Ref.~\cite{Li2021g} find a similar reduction of $Y_e$ and corresponding enhancement of heavy-element abundances, although the quantitative impact appears to be stronger in their study. One reason for the stronger impact with respect to our fiducial model, m1mix1, is the three-flavor mixing (case ``mix2'') assumed in their study (cf. Sec.~\ref{sec:diff-prescr-flav}.  However, a more detailed investigation of additional reason(s) for the discrepancy is not feasible at this point, because many numerical and physics details are treated slightly differently in their model compared to ours, and the quantitative uncertainties associated with these discrepancies are poorly studied so far for neutrino-MHD models \cite{Siegel2018c,Fernandez2019b,Christie2019a,Miller2019a,Just2021i,Zhou2021j}. It is worth to note, however, that, judging from their $Y_e$ histograms, both models in Ref.~\cite{Li2021g}, with and without flavor conversions, seem to produce ejecta with overall lower $Y_e$ values compared to our models.


\section{Summary and conclusions}\label{sec:summary-conclusions}

In this paper we investigated the appearance and consequences of fast neutrino flavor mixing in BH-torus systems that can be formed after the merger of two NSs or a NS and a BH, or in the core of a collapsing star. We first analyzed representative snapshots of neutrino-hydrodynamic models adopted from Ref.~\cite{Just2015a} for the occurrence of ELN crossings by relying on the algorithm developed in Ref.~\cite{Abbar2020m}. The results were then used as a motivation to implement fast flavor conversion through a parametric approach in dynamical simulations using an energy-dependent M1 two-moment neutrino transport scheme. Motivated by the short timescales of the fast flavor instability, we assume that flavor mixing takes place effectively instantaneously after each integration step of the transport scheme. Since we do not solve the quantum kinetic equations, and we therefore cannot predict the detailed amount and momentum-space distribution of flavor conversion, we assume for simplicity that the flavor instability leads to a flavor-mixed steady state that, at a given neutrino energy, is defined entirely in terms of the first two angular moments, namely the local number densities and flux densities. Under these assumptions we investigated a set of 2D axisymmetric simulations using an $\alpha$-viscosity scheme as well as two 3D MHD simulations for the impact of flavor conversions on the disk dynamics, neutrino emission, ejecta electron fraction, and r-process yields. We compared the impact for different assumptions regarding the appearance and outcome of flavor mixing, initial disk masses, and treatments of turbulent viscosity. Our main results are as follows:
\begin{itemize}

\item Using the methodology of Ref.~\cite{Abbar2020m} to spot ELN crossings in snapshots of angular-moment distributions, we confirmed the findings of Refs.~\cite{Wu2017l, Wu:2017drk, Li2021g} in that flavor instabilities are ubiquitous in disk environments. In particular, we found that ELN crossings already start to occur deep within the disk when the neutrino distributions are still close to isotropic and the energy-averaged flux factor of $\nuebar$ neutrinos, $\ff_{\nuebar}$, exceeds $\ffth\approx 0.15-0.2$ (cf. Fig.~\ref{fig:1}).

\item Motivated by these findings, we implemented fast flavor mixing in a parametric fashion and  using the criterion $\ff>\ffth$ to activate flavor conversions in neutrino-hydrodynamic simulations, using different choices of $\ffth$ and assuming different possibilities for the neutrino-flavor steady state that is attained as a result of the flavor mixing. We found that fast conversions induce two main effects in neutrino-cooled disks, which together reduce the equilibrium electron fractions, $\yeeq$, in the disk by about 0.03--0.06 in our models (cf. Figs.~\ref{globec} and~\ref{yeevo}): First, by activating the -- otherwise negligible -- transport of $\mu/\tau$ (anti-)neutrinos, which diffuse out of the disk more quickly than $\nue$ and $\nuebar$ neutrinos, fast conversions enhance the cooling rates of the disk. This leads to a higher electron degeneracy and, in turn, to a more neutron-rich equilibrium state for $e^\pm$ captures on nucleons (i.e. a lower value of $\yeeqem$, cf. Ref.~\cite{Just2021i}). Second, fast flavor conversions reduce the number density of $\nue$ and $\nuebar$ neutrinos in the disk and by that the rate of neutrino absorption reactions onto free nucleons, which tend to increase $\yeeq$ (the equilibrium value of $Y_e$ for all $\beta$-processes for $\nue$ and $\nuebar$ absorption and emission~\cite{Just2021i}). The second effect was already discussed in Refs.~\cite{Wu:2017drk, Li2021g}.

\item The corresponding reduction of $Y_e$ in the ejecta by $\sim 0.01$--$0.03$ is significantly smaller than the reduction in the disk. This is connected to the circumstance that $\yeeq$ increases during the expansion, and both neutrino emission and absorption drive $Y_e$ towards high values, which partially erases the differences in $Y_e$ between models with and without flavor conversions (cf. Fig.~\ref{trajavg}). In addition, the enhanced cooling rates with flavor conversion lead to slightly slower expansion velocities, giving the ejecta more time for $Y_e$ to rise towards $\yeeq$.

\item Flavor conversion leads to an increase of r-process yields by typically tens of percent and at most a factor of two in our models (left panel of Fig.~\ref{nuckilo}). The rate of radioactive heating is boosted by $\sim 20$--$50\,\%$ percent, primarily due to higher abundances of 2nd-peak elements, while at the same time the ejecta become more opaque due to the higher abundance of lanthanides. The resulting kilonova signal therefore peaks roughly at the same time and with the same brightness, but decays over a longer timescale (right panel of Fig.~\ref{nuckilo}).
  
\item While qualitatively the same effects were seen in all models, the quantitative impact of flavor conversions depend on the assumed flavor mixing scenario, which dictates the ratio by which $\nue$ and $\nuebar$ neutrinos can possibly be converted into heavy-lepton neutrinos. Assuming that the number of leptons per family is conserved (case ``mix1'') leads to a ratio of $\nux$ (where $\nux\in\{\nu_\mu,\nu_\tau\}$) to $\nuebar$ neutrino fluxes of $L_{N,\nux}/L_{N,\nuebar}\approx 60\,\%$ (panel (i) of Fig.~\ref{globec}). Allowing all three neutrinos and, separately, all three antineutrinos to reach equipartition -- while violating the conservation of the family lepton number -- increases this value to $\approx 90-95\,\%$. A ratio of exactly $100\,\%$ is obtained in the exotic case that all six neutrinos can reach equipartition. In this case, however, the impact of flavor conversion on $Y_e$ is reduced because of the resulting degeneracy of the mean energies of all species, which leads to a high value of $Y_e$ corresponding to neutrino-capture equilibrium, $\yeeqab$ (cf. right panel of Fig.~\ref{yeevo}).

\item Varying the initial disk mass only leads to modest variations, despite the more important role of neutrino absorption in more massive disks. We attribute the modest impact in more massive disks to the reduced conversion ratio of $\nu_e$ neutrinos into $\nu_x$ neutrinos, which, in turn, results because of a larger fraction of neutrinos being in diffusive conditions with $\ff_{\nuebar}<\ffth$, i.e. being stable with regard to fast flavor conversions. An additional reason is that in more massive disks a larger fraction of $\nu_x$ neutrinos is present already without fast flavor conversions.

\item Replacing the $\alpha$-viscosity treatment with an MHD evolution results in qualitatively similar effects related to flavor conversions as in the viscous models but a slightly more significant reduction of the average $Y_e$ in the ejected material by $\approx 0.03$. However, due to the high computational cost, MHD models could not be followed for long enough time to obtain the final amount and properties of the ejected material, rendering a direct comparison with the viscous models inconclusive. Our results are in broad agreement with the 3D general relativistic MHD results recently reported by Ref.~\cite{Li2021g}.

\end{itemize}
Our study is among the first where neutrino flavor mixing was implemented in time-dependent neutrino-hydrodynamic simulations. In agreement with Ref.~\cite{Li2021g} our results suggest that fast conversions can make neutrino-cooled disks an even more prolific r-process site than previously thought. Even though the observed impact of flavor conversions is not overly dramatic and the remaining modeling uncertainties are still considerable, our results certainly motivate further research concerning the role of fast pairwise oscillations in neutron star mergers. An important point highlighted by our results is that flavor conversions can not only alter the neutrino field surrounding the neutrino-emitting region but also the thermodynamic conditions deep inside the disk. While the former effect can in principle be studied by using post-processing methods based on the output of conventional neutrino-hydrodynamic simulations, the dynamical feedback of flavor conversions can only be assessed by coupling hydrodynamics simulations self-consistently with a neutrino transport scheme including flavor-changing effects.

One reason for the relatively mild impact of fast flavor conversions witnessed in our models with respect to $Y_e$ in the disk and in the ejecta is related to the fact that $\nue$ and $\nuebar$ neutrinos are emitted by similar rates already to begin with, i.e. the ratio of number luminosities $L_{N,\nue}/L_{N,\nuebar}\approx 1$ even without flavor conversions. Fast flavor conversions that lead to equipartition along all directions in the neutrino-momentum space -- as assumed in this study -- will not move the ratio of $L_{N,\nue}/L_{N,\nuebar}$ significantly away from its original value close to unity, and hence, cannot induce a large impact on the neutrino-capture equilibrium value, $\yeeqab$. If, on the other hand, flavor conversion would lead to an appreciable asymmetry between the $\nue$ and $\nuebar$ distributions, even if only along certain momentum directions, the potential impact on $Y_e$ could possibly turn out to be more sizable than the one obtained in our models.

The present study is an early exploration of possible dynamical effects and corresponding sensitivities related to fast flavor conversions in neutrino-cooled disks. Future investigations will have to develop a deeper understanding of the amount of flavor mixing and the oscillation phenomenology in global, time-dependent models. While our study only provides a tentative glimpse of the potential impact, many challenging questions have yet to be tackled in order to assess the true role of the intriguing phenomenon of neutrino oscillations in NS mergers and core-collapse supernovae.

\emph{Data availability:} The data underlying this article will be shared on reasonable request to the corresponding author.


\section*{Acknowledgments}
OJ wishes to thank Andreas Bauswein, Samuel Giuliani, Gabriel Mart{\'\i}nez-Pinedo, Ninoy Rahman, and Zewei Xiong for inspiring discussions. OJ acknowledges support by the European Research Council (ERC) under the European Union's Horizon 2020 research and innovation programme under grant agreement No.~759253 and by the Sonderforschungsbereich SFB~1245 ``Nuclei: From Fundamental Interactions to Structure and Stars'' of the Deutsche Forschungsgemeinschaft (DFG, German Research Foundation). OJ is grateful for computational support by the HOKUSAI computer centre at RIKEN/Japan, Max-Planck Computing and Data Facility (MPCDF) at Garching/Germany, and VIRGO computer cluster at GSI Darmstadt. SA, IT, and HTJ acknowledge support by the DFG through SFB 1258 ``Neutrinos and Dark Matter in Astro- and Particle Physics'' (NDM). MRW acknowledges support by the Ministry of Science and Technology, Taiwan, under Grant No.~110-2112-M-001-050, the Academia Sinica under Project No.~AS-CDA-109-M11, and by the Physics Division of the National Center for Theoretical Sciences, Taiwan. MRW also appreciates the computing resources provided by the Academia Sinica Grid-computing Center. The work of IT is further supported by the Villum Foundation (Projects No.~37358) and the Danmarks Frie Forskningsfonds (Project No.~8049-00038B). HTJ also acknowledges support by the DFG under Germany’s Excellence Strategy through Cluster of Excellence
ORIGINS (EXC-2094)-390783311. The work of FC is supported by GVA  Grant No. CDEIGENT/2020/003.


\begin{thebibliography}{138}%
\makeatletter
\providecommand \@ifxundefined [1]{%
 \@ifx{#1\undefined}
}%
\providecommand \@ifnum [1]{%
 \ifnum #1\expandafter \@firstoftwo
 \else \expandafter \@secondoftwo
 \fi
}%
\providecommand \@ifx [1]{%
 \ifx #1\expandafter \@firstoftwo
 \else \expandafter \@secondoftwo
 \fi
}%
\providecommand \natexlab [1]{#1}%
\providecommand \enquote  [1]{``#1''}%
\providecommand \bibnamefont  [1]{#1}%
\providecommand \bibfnamefont [1]{#1}%
\providecommand \citenamefont [1]{#1}%
\providecommand \href@noop [0]{\@secondoftwo}%
\providecommand \href [0]{\begingroup \@sanitize@url \@href}%
\providecommand \@href[1]{\@@startlink{#1}\@@href}%
\providecommand \@@href[1]{\endgroup#1\@@endlink}%
\providecommand \@sanitize@url [0]{\catcode `\\12\catcode `\$12\catcode
  `\&12\catcode `\#12\catcode `\^12\catcode `\_12\catcode `\%12\relax}%
\providecommand \@@startlink[1]{}%
\providecommand \@@endlink[0]{}%
\providecommand \url  [0]{\begingroup\@sanitize@url \@url }%
\providecommand \@url [1]{\endgroup\@href {#1}{\urlprefix }}%
\providecommand \urlprefix  [0]{URL }%
\providecommand \Eprint [0]{\href }%
\providecommand \doibase [0]{https://doi.org/}%
\providecommand \selectlanguage [0]{\@gobble}%
\providecommand \bibinfo  [0]{\@secondoftwo}%
\providecommand \bibfield  [0]{\@secondoftwo}%
\providecommand \translation [1]{[#1]}%
\providecommand \BibitemOpen [0]{}%
\providecommand \bibitemStop [0]{}%
\providecommand \bibitemNoStop [0]{.\EOS\space}%
\providecommand \EOS [0]{\spacefactor3000\relax}%
\providecommand \BibitemShut  [1]{\csname bibitem#1\endcsname}%
\let\auto@bib@innerbib\@empty
\bibitem [{\citenamefont {{Ruffert}}\ \emph {et~al.}(1996)\citenamefont
  {{Ruffert}}, \citenamefont {{Janka}},\ and\ \citenamefont
  {{Schaefer}}}]{Ruffert1996a}%
  \BibitemOpen
  \bibfield  {author} {\bibinfo {author} {\bibfnamefont {M.}~\bibnamefont
  {{Ruffert}}}, \bibinfo {author} {\bibfnamefont {H.-T.}\ \bibnamefont
  {{Janka}}},\ and\ \bibinfo {author} {\bibfnamefont {G.}~\bibnamefont
  {{Schaefer}}},\ }\href@noop {} {\bibfield  {journal} {\bibinfo  {journal}
  {\aap}\ }\textbf {\bibinfo {volume} {311}},\ \bibinfo {pages} {532} (\bibinfo
  {year} {1996})},\ \Eprint {https://arxiv.org/abs/astro-ph/9509006}
  {astro-ph/9509006} \BibitemShut {NoStop}%
\bibitem [{\citenamefont {{Rosswog}}\ \emph {et~al.}(2003)\citenamefont
  {{Rosswog}}, \citenamefont {{Ramirez-Ruiz}},\ and\ \citenamefont
  {{Davies}}}]{Rosswog2003a}%
  \BibitemOpen
  \bibfield  {author} {\bibinfo {author} {\bibfnamefont {S.}~\bibnamefont
  {{Rosswog}}}, \bibinfo {author} {\bibfnamefont {E.}~\bibnamefont
  {{Ramirez-Ruiz}}},\ and\ \bibinfo {author} {\bibfnamefont {M.~B.}\
  \bibnamefont {{Davies}}},\ }\href@noop {} {\bibfield  {journal} {\bibinfo
  {journal} {\mnras}\ }\textbf {\bibinfo {volume} {345}},\ \bibinfo {pages}
  {1077} (\bibinfo {year} {2003})},\ \Eprint
  {https://arxiv.org/abs/astro-ph/0306418} {arXiv:astro-ph/0306418 [astro-ph]}
  \BibitemShut {NoStop}%
\bibitem [{\citenamefont {{Bauswein}}\ \emph {et~al.}(2013)\citenamefont
  {{Bauswein}}, \citenamefont {{Goriely}},\ and\ \citenamefont
  {{Janka}}}]{Bauswein2013}%
  \BibitemOpen
  \bibfield  {author} {\bibinfo {author} {\bibfnamefont {A.}~\bibnamefont
  {{Bauswein}}}, \bibinfo {author} {\bibfnamefont {S.}~\bibnamefont
  {{Goriely}}},\ and\ \bibinfo {author} {\bibfnamefont {H.-T.}\ \bibnamefont
  {{Janka}}},\ }\href@noop {} {\bibfield  {journal} {\bibinfo  {journal}
  {\apj}\ }\textbf {\bibinfo {volume} {773}},\ \bibinfo {eid} {78} (\bibinfo
  {year} {2013})},\ \Eprint {https://arxiv.org/abs/1302.6530} {arXiv:1302.6530
  [astro-ph.SR]} \BibitemShut {NoStop}%
\bibitem [{\citenamefont {{Baiotti}}\ and\ \citenamefont
  {{Rezzolla}}(2017)}]{Baiotti2017a}%
  \BibitemOpen
  \bibfield  {author} {\bibinfo {author} {\bibfnamefont {L.}~\bibnamefont
  {{Baiotti}}}\ and\ \bibinfo {author} {\bibfnamefont {L.}~\bibnamefont
  {{Rezzolla}}},\ }\href@noop {} {\bibfield  {journal} {\bibinfo  {journal}
  {Reports on Progress in Physics}\ }\textbf {\bibinfo {volume} {80}},\
  \bibinfo {eid} {096901} (\bibinfo {year} {2017})},\ \Eprint
  {https://arxiv.org/abs/1607.03540} {arXiv:1607.03540 [gr-qc]} \BibitemShut
  {NoStop}%
\bibitem [{\citenamefont {{Bernuzzi}}(2020)}]{Bernuzzi2020b}%
  \BibitemOpen
  \bibfield  {author} {\bibinfo {author} {\bibfnamefont {S.}~\bibnamefont
  {{Bernuzzi}}},\ }\href {https://doi.org/10.1007/s10714-020-02752-5}
  {\bibfield  {journal} {\bibinfo  {journal} {General Relativity and
  Gravitation}\ }\textbf {\bibinfo {volume} {52}},\ \bibinfo {eid} {108}
  (\bibinfo {year} {2020})},\ \Eprint {https://arxiv.org/abs/2004.06419}
  {arXiv:2004.06419 [astro-ph.HE]} \BibitemShut {NoStop}%
\bibitem [{\citenamefont {{Ruiz}}\ \emph {et~al.}(2021)\citenamefont {{Ruiz}},
  \citenamefont {{Shapiro}},\ and\ \citenamefont {{Tsokaros}}}]{Ruiz2021d}%
  \BibitemOpen
  \bibfield  {author} {\bibinfo {author} {\bibfnamefont {M.}~\bibnamefont
  {{Ruiz}}}, \bibinfo {author} {\bibfnamefont {S.~L.}\ \bibnamefont
  {{Shapiro}}},\ and\ \bibinfo {author} {\bibfnamefont {A.}~\bibnamefont
  {{Tsokaros}}},\ }\href {https://doi.org/10.3389/fspas.2021.656907} {\bibfield
   {journal} {\bibinfo  {journal} {Frontiers in Astronomy and Space Sciences}\
  }\textbf {\bibinfo {volume} {8}},\ \bibinfo {eid} {39} (\bibinfo {year}
  {2021})},\ \Eprint {https://arxiv.org/abs/2102.03366} {arXiv:2102.03366
  [astro-ph.HE]} \BibitemShut {NoStop}%
\bibitem [{\citenamefont {{Radice}}\ \emph {et~al.}(2018)\citenamefont
  {{Radice}}, \citenamefont {{Perego}}, \citenamefont {{Hotokezaka}},
  \citenamefont {{Fromm}}, \citenamefont {{Bernuzzi}},\ and\ \citenamefont
  {{Roberts}}}]{Radice2018b}%
  \BibitemOpen
  \bibfield  {author} {\bibinfo {author} {\bibfnamefont {D.}~\bibnamefont
  {{Radice}}}, \bibinfo {author} {\bibfnamefont {A.}~\bibnamefont {{Perego}}},
  \bibinfo {author} {\bibfnamefont {K.}~\bibnamefont {{Hotokezaka}}}, \bibinfo
  {author} {\bibfnamefont {S.~A.}\ \bibnamefont {{Fromm}}}, \bibinfo {author}
  {\bibfnamefont {S.}~\bibnamefont {{Bernuzzi}}},\ and\ \bibinfo {author}
  {\bibfnamefont {L.~F.}\ \bibnamefont {{Roberts}}},\ }\href@noop {} {\bibfield
   {journal} {\bibinfo  {journal} {\apj}\ }\textbf {\bibinfo {volume} {869}},\
  \bibinfo {eid} {130} (\bibinfo {year} {2018})},\ \Eprint
  {https://arxiv.org/abs/1809.11161} {arXiv:1809.11161 [astro-ph.HE]}
  \BibitemShut {NoStop}%
\bibitem [{\citenamefont {{MacFadyen}}\ and\ \citenamefont
  {{Woosley}}(1999)}]{MacFadyen1999}%
  \BibitemOpen
  \bibfield  {author} {\bibinfo {author} {\bibfnamefont {A.~I.}\ \bibnamefont
  {{MacFadyen}}}\ and\ \bibinfo {author} {\bibfnamefont {S.~E.}\ \bibnamefont
  {{Woosley}}},\ }\href@noop {} {\bibfield  {journal} {\bibinfo  {journal}
  {\apj}\ }\textbf {\bibinfo {volume} {524}},\ \bibinfo {pages} {262} (\bibinfo
  {year} {1999})},\ \Eprint {https://arxiv.org/abs/arXiv:astro-ph/9810274}
  {arXiv:astro-ph/9810274} \BibitemShut {NoStop}%
\bibitem [{\citenamefont {{Surman}}\ \emph {et~al.}(2006)\citenamefont
  {{Surman}}, \citenamefont {{McLaughlin}},\ and\ \citenamefont
  {{Hix}}}]{Surman2006}%
  \BibitemOpen
  \bibfield  {author} {\bibinfo {author} {\bibfnamefont {R.}~\bibnamefont
  {{Surman}}}, \bibinfo {author} {\bibfnamefont {G.~C.}\ \bibnamefont
  {{McLaughlin}}},\ and\ \bibinfo {author} {\bibfnamefont {W.~R.}\ \bibnamefont
  {{Hix}}},\ }\href@noop {} {\bibfield  {journal} {\bibinfo  {journal} {\apj}\
  }\textbf {\bibinfo {volume} {643}},\ \bibinfo {pages} {1057} (\bibinfo {year}
  {2006})},\ \Eprint {https://arxiv.org/abs/arXiv:astro-ph/0509365}
  {arXiv:astro-ph/0509365} \BibitemShut {NoStop}%
\bibitem [{\citenamefont {{Nagataki}}\ \emph {et~al.}(2007)\citenamefont
  {{Nagataki}}, \citenamefont {{Takahashi}}, \citenamefont {{Mizuta}},\ and\
  \citenamefont {{Takiwaki}}}]{Nagataki2007}%
  \BibitemOpen
  \bibfield  {author} {\bibinfo {author} {\bibfnamefont {S.}~\bibnamefont
  {{Nagataki}}}, \bibinfo {author} {\bibfnamefont {R.}~\bibnamefont
  {{Takahashi}}}, \bibinfo {author} {\bibfnamefont {A.}~\bibnamefont
  {{Mizuta}}},\ and\ \bibinfo {author} {\bibfnamefont {T.}~\bibnamefont
  {{Takiwaki}}},\ }\href@noop {} {\bibfield  {journal} {\bibinfo  {journal}
  {\apj}\ }\textbf {\bibinfo {volume} {659}},\ \bibinfo {pages} {512} (\bibinfo
  {year} {2007})},\ \Eprint {https://arxiv.org/abs/arXiv:astro-ph/0608233}
  {arXiv:astro-ph/0608233} \BibitemShut {NoStop}%
\bibitem [{\citenamefont {{Siegel}}\ \emph {et~al.}(2019)\citenamefont
  {{Siegel}}, \citenamefont {{Barnes}},\ and\ \citenamefont
  {{Metzger}}}]{Siegel2019b}%
  \BibitemOpen
  \bibfield  {author} {\bibinfo {author} {\bibfnamefont {D.~M.}\ \bibnamefont
  {{Siegel}}}, \bibinfo {author} {\bibfnamefont {J.}~\bibnamefont {{Barnes}}},\
  and\ \bibinfo {author} {\bibfnamefont {B.~D.}\ \bibnamefont {{Metzger}}},\
  }\href@noop {} {\bibfield  {journal} {\bibinfo  {journal} {\nat}\ }\textbf
  {\bibinfo {volume} {569}},\ \bibinfo {pages} {241} (\bibinfo {year}
  {2019})},\ \Eprint {https://arxiv.org/abs/1810.00098} {arXiv:1810.00098
  [astro-ph.HE]} \BibitemShut {NoStop}%
\bibitem [{\citenamefont {{Obergaulinger}}\ and\ \citenamefont
  {{Aloy}}(2021)}]{Obergaulinger2021e}%
  \BibitemOpen
  \bibfield  {author} {\bibinfo {author} {\bibfnamefont {M.}~\bibnamefont
  {{Obergaulinger}}}\ and\ \bibinfo {author} {\bibfnamefont {M.~{\'A}.}\
  \bibnamefont {{Aloy}}},\ }\href@noop {} {\bibfield  {journal} {\bibinfo
  {journal} {arXiv e-prints}\ ,\ \bibinfo {eid} {arXiv:2108.13864}} (\bibinfo
  {year} {2021})},\ \Eprint {https://arxiv.org/abs/2108.13864}
  {arXiv:2108.13864 [astro-ph.HE]} \BibitemShut {NoStop}%
\bibitem [{\citenamefont {{Ruffert}}\ \emph {et~al.}(1997)\citenamefont
  {{Ruffert}}, \citenamefont {{Janka}}, \citenamefont {{Takahashi}},\ and\
  \citenamefont {{Schaefer}}}]{Ruffert1997}%
  \BibitemOpen
  \bibfield  {author} {\bibinfo {author} {\bibfnamefont {M.}~\bibnamefont
  {{Ruffert}}}, \bibinfo {author} {\bibfnamefont {H.-T.}\ \bibnamefont
  {{Janka}}}, \bibinfo {author} {\bibfnamefont {K.}~\bibnamefont
  {{Takahashi}}},\ and\ \bibinfo {author} {\bibfnamefont {G.}~\bibnamefont
  {{Schaefer}}},\ }\href@noop {} {\bibfield  {journal} {\bibinfo  {journal}
  {\aap}\ }\textbf {\bibinfo {volume} {319}},\ \bibinfo {pages} {122} (\bibinfo
  {year} {1997})},\ \Eprint {https://arxiv.org/abs/astro-ph/9606181}
  {astro-ph/9606181} \BibitemShut {NoStop}%
\bibitem [{\citenamefont {{Popham}}\ \emph {et~al.}(1999)\citenamefont
  {{Popham}}, \citenamefont {{Woosley}},\ and\ \citenamefont
  {{Fryer}}}]{Popham1999}%
  \BibitemOpen
  \bibfield  {author} {\bibinfo {author} {\bibfnamefont {R.}~\bibnamefont
  {{Popham}}}, \bibinfo {author} {\bibfnamefont {S.~E.}\ \bibnamefont
  {{Woosley}}},\ and\ \bibinfo {author} {\bibfnamefont {C.}~\bibnamefont
  {{Fryer}}},\ }\href@noop {} {\bibfield  {journal} {\bibinfo  {journal}
  {\apj}\ }\textbf {\bibinfo {volume} {518}},\ \bibinfo {pages} {356} (\bibinfo
  {year} {1999})},\ \Eprint {https://arxiv.org/abs/arXiv:astro-ph/9807028}
  {arXiv:astro-ph/9807028} \BibitemShut {NoStop}%
\bibitem [{\citenamefont {{Kohri}}\ and\ \citenamefont
  {{Mineshige}}(2002)}]{Kohri2002}%
  \BibitemOpen
  \bibfield  {author} {\bibinfo {author} {\bibfnamefont {K.}~\bibnamefont
  {{Kohri}}}\ and\ \bibinfo {author} {\bibfnamefont {S.}~\bibnamefont
  {{Mineshige}}},\ }\href@noop {} {\bibfield  {journal} {\bibinfo  {journal}
  {\apj}\ }\textbf {\bibinfo {volume} {577}},\ \bibinfo {pages} {311} (\bibinfo
  {year} {2002})},\ \Eprint {https://arxiv.org/abs/arXiv:astro-ph/0203177}
  {arXiv:astro-ph/0203177} \BibitemShut {NoStop}%
\bibitem [{\citenamefont {{Beloborodov}}(2003)}]{Beloborodov2003}%
  \BibitemOpen
  \bibfield  {author} {\bibinfo {author} {\bibfnamefont {A.~M.}\ \bibnamefont
  {{Beloborodov}}},\ }\href@noop {} {\bibfield  {journal} {\bibinfo  {journal}
  {\apj}\ }\textbf {\bibinfo {volume} {588}},\ \bibinfo {pages} {931} (\bibinfo
  {year} {2003})},\ \Eprint {https://arxiv.org/abs/arXiv:astro-ph/0210522}
  {arXiv:astro-ph/0210522} \BibitemShut {NoStop}%
\bibitem [{\citenamefont {{Chen}}\ and\ \citenamefont
  {{Beloborodov}}(2007)}]{Chen2007}%
  \BibitemOpen
  \bibfield  {author} {\bibinfo {author} {\bibfnamefont {W.}~\bibnamefont
  {{Chen}}}\ and\ \bibinfo {author} {\bibfnamefont {A.~M.}\ \bibnamefont
  {{Beloborodov}}},\ }\href@noop {} {\bibfield  {journal} {\bibinfo  {journal}
  {\apj}\ }\textbf {\bibinfo {volume} {657}},\ \bibinfo {pages} {383} (\bibinfo
  {year} {2007})},\ \Eprint {https://arxiv.org/abs/arXiv:astro-ph/0607145}
  {arXiv:astro-ph/0607145} \BibitemShut {NoStop}%
\bibitem [{\citenamefont {{Metzger}}\ \emph {et~al.}(2009)\citenamefont
  {{Metzger}}, \citenamefont {{Piro}},\ and\ \citenamefont
  {{Quataert}}}]{Metzger2009b}%
  \BibitemOpen
  \bibfield  {author} {\bibinfo {author} {\bibfnamefont {B.~D.}\ \bibnamefont
  {{Metzger}}}, \bibinfo {author} {\bibfnamefont {A.~L.}\ \bibnamefont
  {{Piro}}},\ and\ \bibinfo {author} {\bibfnamefont {E.}~\bibnamefont
  {{Quataert}}},\ }\href@noop {} {\bibfield  {journal} {\bibinfo  {journal}
  {\mnras}\ }\textbf {\bibinfo {volume} {396}},\ \bibinfo {pages} {304}
  (\bibinfo {year} {2009})},\ \Eprint {https://arxiv.org/abs/0810.2535}
  {arXiv:0810.2535} \BibitemShut {NoStop}%
\bibitem [{\citenamefont {{Fern{\'a}ndez}}\ and\ \citenamefont
  {{Metzger}}(2013)}]{Fernandez2013b}%
  \BibitemOpen
  \bibfield  {author} {\bibinfo {author} {\bibfnamefont {R.}~\bibnamefont
  {{Fern{\'a}ndez}}}\ and\ \bibinfo {author} {\bibfnamefont {B.~D.}\
  \bibnamefont {{Metzger}}},\ }\href@noop {} {\bibfield  {journal} {\bibinfo
  {journal} {\mnras}\ }\textbf {\bibinfo {volume} {435}},\ \bibinfo {pages}
  {502} (\bibinfo {year} {2013})},\ \Eprint {https://arxiv.org/abs/1304.6720}
  {arXiv:1304.6720 [astro-ph.HE]} \BibitemShut {NoStop}%
\bibitem [{\citenamefont {{Just}}\ \emph
  {et~al.}(2015{\natexlab{a}})\citenamefont {{Just}}, \citenamefont
  {{Bauswein}}, \citenamefont {{Pulpillo}}, \citenamefont {{Goriely}},\ and\
  \citenamefont {{Janka}}}]{Just2015a}%
  \BibitemOpen
  \bibfield  {author} {\bibinfo {author} {\bibfnamefont {O.}~\bibnamefont
  {{Just}}}, \bibinfo {author} {\bibfnamefont {A.}~\bibnamefont {{Bauswein}}},
  \bibinfo {author} {\bibfnamefont {R.~A.}\ \bibnamefont {{Pulpillo}}},
  \bibinfo {author} {\bibfnamefont {S.}~\bibnamefont {{Goriely}}},\ and\
  \bibinfo {author} {\bibfnamefont {H.-T.}\ \bibnamefont {{Janka}}},\
  }\href@noop {} {\bibfield  {journal} {\bibinfo  {journal} {\mnras}\ }\textbf
  {\bibinfo {volume} {448}},\ \bibinfo {pages} {541} (\bibinfo {year}
  {2015}{\natexlab{a}})},\ \Eprint {https://arxiv.org/abs/1406.2687}
  {arXiv:1406.2687 [astro-ph.SR]} \BibitemShut {NoStop}%
\bibitem [{\citenamefont {{Siegel}}\ \emph {et~al.}(2018)\citenamefont
  {{Siegel}}, \citenamefont {{Barnes}},\ and\ \citenamefont
  {{Metzger}}}]{Siegel2018a}%
  \BibitemOpen
  \bibfield  {author} {\bibinfo {author} {\bibfnamefont {D.~M.}\ \bibnamefont
  {{Siegel}}}, \bibinfo {author} {\bibfnamefont {J.}~\bibnamefont {{Barnes}}},\
  and\ \bibinfo {author} {\bibfnamefont {B.~D.}\ \bibnamefont {{Metzger}}},\
  }\href@noop {} {\bibfield  {journal} {\bibinfo  {journal} {ArXiv e-prints}\ }
  (\bibinfo {year} {2018})},\ \Eprint {https://arxiv.org/abs/1810.00098}
  {arXiv:1810.00098 [astro-ph.HE]} \BibitemShut {NoStop}%
\bibitem [{\citenamefont {{Janiuk}}(2019)}]{Janiuk2019a}%
  \BibitemOpen
  \bibfield  {author} {\bibinfo {author} {\bibfnamefont {A.}~\bibnamefont
  {{Janiuk}}},\ }\href@noop {} {\bibfield  {journal} {\bibinfo  {journal}
  {\apj}\ }\textbf {\bibinfo {volume} {882}},\ \bibinfo {eid} {163} (\bibinfo
  {year} {2019})},\ \Eprint {https://arxiv.org/abs/1907.00809}
  {arXiv:1907.00809 [astro-ph.HE]} \BibitemShut {NoStop}%
\bibitem [{\citenamefont {{Fujibayashi}}\ \emph {et~al.}(2020)\citenamefont
  {{Fujibayashi}}, \citenamefont {{Shibata}}, \citenamefont {{Wanajo}},
  \citenamefont {{Kiuchi}}, \citenamefont {{Kyutoku}},\ and\ \citenamefont
  {{Sekiguchi}}}]{Fujibayashi2020a}%
  \BibitemOpen
  \bibfield  {author} {\bibinfo {author} {\bibfnamefont {S.}~\bibnamefont
  {{Fujibayashi}}}, \bibinfo {author} {\bibfnamefont {M.}~\bibnamefont
  {{Shibata}}}, \bibinfo {author} {\bibfnamefont {S.}~\bibnamefont {{Wanajo}}},
  \bibinfo {author} {\bibfnamefont {K.}~\bibnamefont {{Kiuchi}}}, \bibinfo
  {author} {\bibfnamefont {K.}~\bibnamefont {{Kyutoku}}},\ and\ \bibinfo
  {author} {\bibfnamefont {Y.}~\bibnamefont {{Sekiguchi}}},\ }\href@noop {}
  {\bibfield  {journal} {\bibinfo  {journal} {\prd}\ }\textbf {\bibinfo
  {volume} {101}},\ \bibinfo {eid} {083029} (\bibinfo {year} {2020})},\ \Eprint
  {https://arxiv.org/abs/2001.04467} {arXiv:2001.04467 [astro-ph.HE]}
  \BibitemShut {NoStop}%
\bibitem [{\citenamefont {{Hayashi}}\ \emph {et~al.}(2021)\citenamefont
  {{Hayashi}}, \citenamefont {{Fujibayashi}}, \citenamefont {{Kiuchi}},
  \citenamefont {{Kyutoku}}, \citenamefont {{Sekiguchi}},\ and\ \citenamefont
  {{Shibata}}}]{Hayashi2021a}%
  \BibitemOpen
  \bibfield  {author} {\bibinfo {author} {\bibfnamefont {K.}~\bibnamefont
  {{Hayashi}}}, \bibinfo {author} {\bibfnamefont {S.}~\bibnamefont
  {{Fujibayashi}}}, \bibinfo {author} {\bibfnamefont {K.}~\bibnamefont
  {{Kiuchi}}}, \bibinfo {author} {\bibfnamefont {K.}~\bibnamefont {{Kyutoku}}},
  \bibinfo {author} {\bibfnamefont {Y.}~\bibnamefont {{Sekiguchi}}},\ and\
  \bibinfo {author} {\bibfnamefont {M.}~\bibnamefont {{Shibata}}},\ }\href@noop
  {} {\bibfield  {journal} {\bibinfo  {journal} {arXiv e-prints}\ ,\ \bibinfo
  {eid} {arXiv:2111.04621}} (\bibinfo {year} {2021})},\ \Eprint
  {https://arxiv.org/abs/2111.04621} {arXiv:2111.04621 [astro-ph.HE]}
  \BibitemShut {NoStop}%
\bibitem [{\citenamefont {{Most}}\ \emph {et~al.}(2021)\citenamefont {{Most}},
  \citenamefont {{Papenfort}}, \citenamefont {{Tootle}},\ and\ \citenamefont
  {{Rezzolla}}}]{Most2021q}%
  \BibitemOpen
  \bibfield  {author} {\bibinfo {author} {\bibfnamefont {E.~R.}\ \bibnamefont
  {{Most}}}, \bibinfo {author} {\bibfnamefont {L.~J.}\ \bibnamefont
  {{Papenfort}}}, \bibinfo {author} {\bibfnamefont {S.~D.}\ \bibnamefont
  {{Tootle}}},\ and\ \bibinfo {author} {\bibfnamefont {L.}~\bibnamefont
  {{Rezzolla}}},\ }\href@noop {} {\bibfield  {journal} {\bibinfo  {journal}
  {\mnras}\ }\textbf {\bibinfo {volume} {506}},\ \bibinfo {pages} {3511}
  (\bibinfo {year} {2021})},\ \Eprint {https://arxiv.org/abs/2106.06391}
  {arXiv:2106.06391 [astro-ph.HE]} \BibitemShut {NoStop}%
\bibitem [{\citenamefont {{Murguia-Berthier}}\ \emph
  {et~al.}(2021)\citenamefont {{Murguia-Berthier}}, \citenamefont {{Noble}},
  \citenamefont {{Roberts}}, \citenamefont {{Ramirez-Ruiz}}, \citenamefont
  {{Werneck}}, \citenamefont {{Kolacki}}, \citenamefont {{Etienne}},
  \citenamefont {{Avara}}, \citenamefont {{Campanelli}}, \citenamefont
  {{Ciolfi}}, \citenamefont {{Cipolletta}}, \citenamefont {{Drachler}},
  \citenamefont {{Ennoggi}}, \citenamefont {{Faber}}, \citenamefont {{Fiacco}},
  \citenamefont {{Giacomazzo}}, \citenamefont {{Gupte}}, \citenamefont {{Ha}},
  \citenamefont {{Kelly}}, \citenamefont {{Krolik}}, \citenamefont {{Lopez
  Armengol}}, \citenamefont {{Margalit}}, \citenamefont {{Moon}}, \citenamefont
  {{O'Shaughnessy}}, \citenamefont {{Rueda-Becerril}}, \citenamefont
  {{Schnittman}}, \citenamefont {{Zenati}},\ and\ \citenamefont
  {{Zlochower}}}]{Murguia-Berthier2021u}%
  \BibitemOpen
  \bibfield  {author} {\bibinfo {author} {\bibfnamefont {A.}~\bibnamefont
  {{Murguia-Berthier}}}, \bibinfo {author} {\bibfnamefont {S.~C.}\ \bibnamefont
  {{Noble}}}, \bibinfo {author} {\bibfnamefont {L.~F.}\ \bibnamefont
  {{Roberts}}}, \bibinfo {author} {\bibfnamefont {E.}~\bibnamefont
  {{Ramirez-Ruiz}}}, \bibinfo {author} {\bibfnamefont {L.~R.}\ \bibnamefont
  {{Werneck}}}, \bibinfo {author} {\bibfnamefont {M.}~\bibnamefont
  {{Kolacki}}}, \bibinfo {author} {\bibfnamefont {Z.~B.}\ \bibnamefont
  {{Etienne}}}, \bibinfo {author} {\bibfnamefont {M.}~\bibnamefont {{Avara}}},
  \bibinfo {author} {\bibfnamefont {M.}~\bibnamefont {{Campanelli}}}, \bibinfo
  {author} {\bibfnamefont {R.}~\bibnamefont {{Ciolfi}}}, \bibinfo {author}
  {\bibfnamefont {F.}~\bibnamefont {{Cipolletta}}}, \bibinfo {author}
  {\bibfnamefont {B.}~\bibnamefont {{Drachler}}}, \bibinfo {author}
  {\bibfnamefont {L.}~\bibnamefont {{Ennoggi}}}, \bibinfo {author}
  {\bibfnamefont {J.}~\bibnamefont {{Faber}}}, \bibinfo {author} {\bibfnamefont
  {G.}~\bibnamefont {{Fiacco}}}, \bibinfo {author} {\bibfnamefont
  {B.}~\bibnamefont {{Giacomazzo}}}, \bibinfo {author} {\bibfnamefont
  {T.}~\bibnamefont {{Gupte}}}, \bibinfo {author} {\bibfnamefont
  {T.}~\bibnamefont {{Ha}}}, \bibinfo {author} {\bibfnamefont {B.~J.}\
  \bibnamefont {{Kelly}}}, \bibinfo {author} {\bibfnamefont {J.~H.}\
  \bibnamefont {{Krolik}}}, \bibinfo {author} {\bibfnamefont {F.~G.}\
  \bibnamefont {{Lopez Armengol}}}, \bibinfo {author} {\bibfnamefont
  {B.}~\bibnamefont {{Margalit}}}, \bibinfo {author} {\bibfnamefont
  {T.}~\bibnamefont {{Moon}}}, \bibinfo {author} {\bibfnamefont
  {R.}~\bibnamefont {{O'Shaughnessy}}}, \bibinfo {author} {\bibfnamefont
  {J.~M.}\ \bibnamefont {{Rueda-Becerril}}}, \bibinfo {author} {\bibfnamefont
  {J.}~\bibnamefont {{Schnittman}}}, \bibinfo {author} {\bibfnamefont
  {Y.}~\bibnamefont {{Zenati}}},\ and\ \bibinfo {author} {\bibfnamefont
  {Y.}~\bibnamefont {{Zlochower}}},\ }\href
  {https://doi.org/10.3847/1538-4357/ac1119} {\bibfield  {journal} {\bibinfo
  {journal} {\apj}\ }\textbf {\bibinfo {volume} {919}},\ \bibinfo {eid} {95}
  (\bibinfo {year} {2021})},\ \Eprint {https://arxiv.org/abs/2106.05356}
  {arXiv:2106.05356 [astro-ph.HE]} \BibitemShut {NoStop}%
\bibitem [{\citenamefont {{Eichler}}\ \emph {et~al.}(1989)\citenamefont
  {{Eichler}}, \citenamefont {{Livio}}, \citenamefont {{Piran}},\ and\
  \citenamefont {{Schramm}}}]{Eichler1989}%
  \BibitemOpen
  \bibfield  {author} {\bibinfo {author} {\bibfnamefont {D.}~\bibnamefont
  {{Eichler}}}, \bibinfo {author} {\bibfnamefont {M.}~\bibnamefont {{Livio}}},
  \bibinfo {author} {\bibfnamefont {T.}~\bibnamefont {{Piran}}},\ and\ \bibinfo
  {author} {\bibfnamefont {D.~N.}\ \bibnamefont {{Schramm}}},\ }\href@noop {}
  {\bibfield  {journal} {\bibinfo  {journal} {\nat}\ }\textbf {\bibinfo
  {volume} {340}},\ \bibinfo {pages} {126} (\bibinfo {year}
  {1989})}\BibitemShut {NoStop}%
\bibitem [{\citenamefont {{Ruffert}}\ and\ \citenamefont
  {{Janka}}(1999)}]{Ruffert1999a}%
  \BibitemOpen
  \bibfield  {author} {\bibinfo {author} {\bibfnamefont {M.}~\bibnamefont
  {{Ruffert}}}\ and\ \bibinfo {author} {\bibfnamefont {H.-T.}\ \bibnamefont
  {{Janka}}},\ }\href@noop {} {\bibfield  {journal} {\bibinfo  {journal}
  {\aap}\ }\textbf {\bibinfo {volume} {344}},\ \bibinfo {pages} {573} (\bibinfo
  {year} {1999})},\ \Eprint {https://arxiv.org/abs/astro-ph/9809280}
  {astro-ph/9809280} \BibitemShut {NoStop}%
\bibitem [{\citenamefont {{Just}}\ \emph {et~al.}(2016)\citenamefont {{Just}},
  \citenamefont {{Obergaulinger}}, \citenamefont {{Janka}}, \citenamefont
  {{Bauswein}},\ and\ \citenamefont {{Schwarz}}}]{Just2016}%
  \BibitemOpen
  \bibfield  {author} {\bibinfo {author} {\bibfnamefont {O.}~\bibnamefont
  {{Just}}}, \bibinfo {author} {\bibfnamefont {M.}~\bibnamefont
  {{Obergaulinger}}}, \bibinfo {author} {\bibfnamefont {H.-T.}\ \bibnamefont
  {{Janka}}}, \bibinfo {author} {\bibfnamefont {A.}~\bibnamefont
  {{Bauswein}}},\ and\ \bibinfo {author} {\bibfnamefont {N.}~\bibnamefont
  {{Schwarz}}},\ }\href@noop {} {\bibfield  {journal} {\bibinfo  {journal}
  {\apjl}\ }\textbf {\bibinfo {volume} {816}},\ \bibinfo {eid} {L30} (\bibinfo
  {year} {2016})},\ \Eprint {https://arxiv.org/abs/1510.04288}
  {arXiv:1510.04288 [astro-ph.HE]} \BibitemShut {NoStop}%
\bibitem [{\citenamefont {{Surman}}\ \emph {et~al.}(2008)\citenamefont
  {{Surman}}, \citenamefont {{McLaughlin}}, \citenamefont {{Ruffert}},
  \citenamefont {{Janka}},\ and\ \citenamefont {{Hix}}}]{Surman2008b}%
  \BibitemOpen
  \bibfield  {author} {\bibinfo {author} {\bibfnamefont {R.}~\bibnamefont
  {{Surman}}}, \bibinfo {author} {\bibfnamefont {G.~C.}\ \bibnamefont
  {{McLaughlin}}}, \bibinfo {author} {\bibfnamefont {M.}~\bibnamefont
  {{Ruffert}}}, \bibinfo {author} {\bibfnamefont {H.-T.}\ \bibnamefont
  {{Janka}}},\ and\ \bibinfo {author} {\bibfnamefont {W.~R.}\ \bibnamefont
  {{Hix}}},\ }\href@noop {} {\bibfield  {journal} {\bibinfo  {journal} {\apjl}\
  }\textbf {\bibinfo {volume} {679}},\ \bibinfo {pages} {L117} (\bibinfo {year}
  {2008})},\ \Eprint {https://arxiv.org/abs/0803.1785} {arXiv:0803.1785}
  \BibitemShut {NoStop}%
\bibitem [{\citenamefont {{Wanajo}}\ and\ \citenamefont
  {{Janka}}(2012)}]{Wanajo2012}%
  \BibitemOpen
  \bibfield  {author} {\bibinfo {author} {\bibfnamefont {S.}~\bibnamefont
  {{Wanajo}}}\ and\ \bibinfo {author} {\bibfnamefont {H.-T.}\ \bibnamefont
  {{Janka}}},\ }\href@noop {} {\bibfield  {journal} {\bibinfo  {journal}
  {\apj}\ }\textbf {\bibinfo {volume} {746}},\ \bibinfo {eid} {180} (\bibinfo
  {year} {2012})},\ \Eprint {https://arxiv.org/abs/1106.6142} {arXiv:1106.6142
  [astro-ph.SR]} \BibitemShut {NoStop}%
\bibitem [{\citenamefont {{Wu}}\ \emph {et~al.}(2016)\citenamefont {{Wu}},
  \citenamefont {{Fern{\'a}ndez}}, \citenamefont {{Mart{\'{\i}}nez-Pinedo}},\
  and\ \citenamefont {{Metzger}}}]{Wu2016a}%
  \BibitemOpen
  \bibfield  {author} {\bibinfo {author} {\bibfnamefont {M.-R.}\ \bibnamefont
  {{Wu}}}, \bibinfo {author} {\bibfnamefont {R.}~\bibnamefont
  {{Fern{\'a}ndez}}}, \bibinfo {author} {\bibfnamefont {G.}~\bibnamefont
  {{Mart{\'{\i}}nez-Pinedo}}},\ and\ \bibinfo {author} {\bibfnamefont {B.~D.}\
  \bibnamefont {{Metzger}}},\ }\href@noop {} {\bibfield  {journal} {\bibinfo
  {journal} {\mnras}\ }\textbf {\bibinfo {volume} {463}},\ \bibinfo {pages}
  {2323} (\bibinfo {year} {2016})},\ \Eprint {https://arxiv.org/abs/1607.05290}
  {arXiv:1607.05290 [astro-ph.HE]} \BibitemShut {NoStop}%
\bibitem [{\citenamefont {{Kasen}}\ \emph {et~al.}(2017)\citenamefont
  {{Kasen}}, \citenamefont {{Metzger}}, \citenamefont {{Barnes}}, \citenamefont
  {{Quataert}},\ and\ \citenamefont {{Ramirez-Ruiz}}}]{Kasen2017a}%
  \BibitemOpen
  \bibfield  {author} {\bibinfo {author} {\bibfnamefont {D.}~\bibnamefont
  {{Kasen}}}, \bibinfo {author} {\bibfnamefont {B.}~\bibnamefont {{Metzger}}},
  \bibinfo {author} {\bibfnamefont {J.}~\bibnamefont {{Barnes}}}, \bibinfo
  {author} {\bibfnamefont {E.}~\bibnamefont {{Quataert}}},\ and\ \bibinfo
  {author} {\bibfnamefont {E.}~\bibnamefont {{Ramirez-Ruiz}}},\ }\href@noop {}
  {\bibfield  {journal} {\bibinfo  {journal} {\nat}\ }\textbf {\bibinfo
  {volume} {551}},\ \bibinfo {pages} {80} (\bibinfo {year} {2017})},\ \Eprint
  {https://arxiv.org/abs/1710.05463} {arXiv:1710.05463 [astro-ph.HE]}
  \BibitemShut {NoStop}%
\bibitem [{\citenamefont {{Miller}}\ \emph {et~al.}(2019)\citenamefont
  {{Miller}}, \citenamefont {{Ryan}}, \citenamefont {{Dolence}}, \citenamefont
  {{Burrows}}, \citenamefont {{Fontes}}, \citenamefont {{Fryer}}, \citenamefont
  {{Korobkin}}, \citenamefont {{Lippuner}}, \citenamefont {{Mumpower}},\ and\
  \citenamefont {{Wollaeger}}}]{Miller2019a}%
  \BibitemOpen
  \bibfield  {author} {\bibinfo {author} {\bibfnamefont {J.~M.}\ \bibnamefont
  {{Miller}}}, \bibinfo {author} {\bibfnamefont {B.~R.}\ \bibnamefont
  {{Ryan}}}, \bibinfo {author} {\bibfnamefont {J.~C.}\ \bibnamefont
  {{Dolence}}}, \bibinfo {author} {\bibfnamefont {A.}~\bibnamefont
  {{Burrows}}}, \bibinfo {author} {\bibfnamefont {C.~J.}\ \bibnamefont
  {{Fontes}}}, \bibinfo {author} {\bibfnamefont {C.~L.}\ \bibnamefont
  {{Fryer}}}, \bibinfo {author} {\bibfnamefont {O.}~\bibnamefont {{Korobkin}}},
  \bibinfo {author} {\bibfnamefont {J.}~\bibnamefont {{Lippuner}}}, \bibinfo
  {author} {\bibfnamefont {M.~R.}\ \bibnamefont {{Mumpower}}},\ and\ \bibinfo
  {author} {\bibfnamefont {R.~T.}\ \bibnamefont {{Wollaeger}}},\ }\href@noop {}
  {\bibfield  {journal} {\bibinfo  {journal} {\prd}\ }\textbf {\bibinfo
  {volume} {100}},\ \bibinfo {eid} {023008} (\bibinfo {year} {2019})},\ \Eprint
  {https://arxiv.org/abs/1905.07477} {arXiv:1905.07477 [astro-ph.HE]}
  \BibitemShut {NoStop}%
\bibitem [{\citenamefont {{Arnould}}\ and\ \citenamefont
  {{Goriely}}(2020)}]{Arnould2020f}%
  \BibitemOpen
  \bibfield  {author} {\bibinfo {author} {\bibfnamefont {M.}~\bibnamefont
  {{Arnould}}}\ and\ \bibinfo {author} {\bibfnamefont {S.}~\bibnamefont
  {{Goriely}}},\ }\href@noop {} {\bibfield  {journal} {\bibinfo  {journal}
  {Progress in Particle and Nuclear Physics}\ }\textbf {\bibinfo {volume}
  {112}},\ \bibinfo {eid} {103766} (\bibinfo {year} {2020})},\ \Eprint
  {https://arxiv.org/abs/2001.11228} {arXiv:2001.11228 [astro-ph.SR]}
  \BibitemShut {NoStop}%
\bibitem [{\citenamefont {{Cowan}}\ \emph {et~al.}(2021)\citenamefont
  {{Cowan}}, \citenamefont {{Sneden}}, \citenamefont {{Lawler}}, \citenamefont
  {{Aprahamian}}, \citenamefont {{Wiescher}}, \citenamefont {{Langanke}},
  \citenamefont {{Mart{\'\i}nez-Pinedo}},\ and\ \citenamefont
  {{Thielemann}}}]{Cowan2021g}%
  \BibitemOpen
  \bibfield  {author} {\bibinfo {author} {\bibfnamefont {J.~J.}\ \bibnamefont
  {{Cowan}}}, \bibinfo {author} {\bibfnamefont {C.}~\bibnamefont {{Sneden}}},
  \bibinfo {author} {\bibfnamefont {J.~E.}\ \bibnamefont {{Lawler}}}, \bibinfo
  {author} {\bibfnamefont {A.}~\bibnamefont {{Aprahamian}}}, \bibinfo {author}
  {\bibfnamefont {M.}~\bibnamefont {{Wiescher}}}, \bibinfo {author}
  {\bibfnamefont {K.}~\bibnamefont {{Langanke}}}, \bibinfo {author}
  {\bibfnamefont {G.}~\bibnamefont {{Mart{\'\i}nez-Pinedo}}},\ and\ \bibinfo
  {author} {\bibfnamefont {F.-K.}\ \bibnamefont {{Thielemann}}},\ }\href@noop
  {} {\bibfield  {journal} {\bibinfo  {journal} {Reviews of Modern Physics}\
  }\textbf {\bibinfo {volume} {93}},\ \bibinfo {eid} {015002} (\bibinfo {year}
  {2021})},\ \Eprint {https://arxiv.org/abs/1901.01410} {arXiv:1901.01410
  [astro-ph.HE]} \BibitemShut {NoStop}%
\bibitem [{\citenamefont {{Siegel}}\ and\ \citenamefont
  {{Metzger}}(2018)}]{Siegel2018c}%
  \BibitemOpen
  \bibfield  {author} {\bibinfo {author} {\bibfnamefont {D.~M.}\ \bibnamefont
  {{Siegel}}}\ and\ \bibinfo {author} {\bibfnamefont {B.~D.}\ \bibnamefont
  {{Metzger}}},\ }\href@noop {} {\bibfield  {journal} {\bibinfo  {journal}
  {\apj}\ }\textbf {\bibinfo {volume} {858}},\ \bibinfo {eid} {52} (\bibinfo
  {year} {2018})},\ \Eprint {https://arxiv.org/abs/1711.00868}
  {arXiv:1711.00868 [astro-ph.HE]} \BibitemShut {NoStop}%
\bibitem [{\citenamefont {Just}\ \emph {et~al.}(2021)\citenamefont {Just},
  \citenamefont {Goriely}, \citenamefont {Janka}, \citenamefont {Nagataki},\
  and\ \citenamefont {Bauswein}}]{Just2021i}%
  \BibitemOpen
  \bibfield  {author} {\bibinfo {author} {\bibfnamefont {O.}~\bibnamefont
  {Just}}, \bibinfo {author} {\bibfnamefont {S.}~\bibnamefont {Goriely}},
  \bibinfo {author} {\bibfnamefont {H.-T.}\ \bibnamefont {Janka}}, \bibinfo
  {author} {\bibfnamefont {S.}~\bibnamefont {Nagataki}},\ and\ \bibinfo
  {author} {\bibfnamefont {A.}~\bibnamefont {Bauswein}},\ }\href@noop {}
  {\bibfield  {journal} {\bibinfo  {journal} {Monthly Notices of the Royal
  Astronomical Society}\ }\textbf {\bibinfo {volume} {509}},\ \bibinfo {pages}
  {1377} (\bibinfo {year} {2021})},\ \Eprint
  {https://arxiv.org/abs/https://academic.oup.com/mnras/article-pdf/509/1/1377/41143986/stab2861.pdf}
  {https://academic.oup.com/mnras/article-pdf/509/1/1377/41143986/stab2861.pdf}
  \BibitemShut {NoStop}%
\bibitem [{\citenamefont {{C{\^o}t{\'e}}}\ \emph {et~al.}(2019)\citenamefont
  {{C{\^o}t{\'e}}}, \citenamefont {{Eichler}}, \citenamefont {{Arcones}},
  \citenamefont {{Hansen}}, \citenamefont {{Simonetti}}, \citenamefont
  {{Frebel}}, \citenamefont {{Fryer}}, \citenamefont {{Pignatari}},
  \citenamefont {{Reichert}}, \citenamefont {{Belczynski}},\ and\ \citenamefont
  {{Matteucci}}}]{Cote2019m}%
  \BibitemOpen
  \bibfield  {author} {\bibinfo {author} {\bibfnamefont {B.}~\bibnamefont
  {{C{\^o}t{\'e}}}}, \bibinfo {author} {\bibfnamefont {M.}~\bibnamefont
  {{Eichler}}}, \bibinfo {author} {\bibfnamefont {A.}~\bibnamefont
  {{Arcones}}}, \bibinfo {author} {\bibfnamefont {C.~J.}\ \bibnamefont
  {{Hansen}}}, \bibinfo {author} {\bibfnamefont {P.}~\bibnamefont
  {{Simonetti}}}, \bibinfo {author} {\bibfnamefont {A.}~\bibnamefont
  {{Frebel}}}, \bibinfo {author} {\bibfnamefont {C.~L.}\ \bibnamefont
  {{Fryer}}}, \bibinfo {author} {\bibfnamefont {M.}~\bibnamefont
  {{Pignatari}}}, \bibinfo {author} {\bibfnamefont {M.}~\bibnamefont
  {{Reichert}}}, \bibinfo {author} {\bibfnamefont {K.}~\bibnamefont
  {{Belczynski}}},\ and\ \bibinfo {author} {\bibfnamefont {F.}~\bibnamefont
  {{Matteucci}}},\ }\href {https://doi.org/10.3847/1538-4357/ab10db} {\bibfield
   {journal} {\bibinfo  {journal} {\apj}\ }\textbf {\bibinfo {volume} {875}},\
  \bibinfo {eid} {106} (\bibinfo {year} {2019})},\ \Eprint
  {https://arxiv.org/abs/1809.03525} {arXiv:1809.03525 [astro-ph.HE]}
  \BibitemShut {NoStop}%
\bibitem [{\citenamefont {Banerjee}\ \emph {et~al.}(2020)\citenamefont
  {Banerjee}, \citenamefont {Wu},\ and\ \citenamefont
  {Yuan}}]{Banerjee:2020eak}%
  \BibitemOpen
  \bibfield  {author} {\bibinfo {author} {\bibfnamefont {P.}~\bibnamefont
  {Banerjee}}, \bibinfo {author} {\bibfnamefont {M.-R.}\ \bibnamefont {Wu}},\
  and\ \bibinfo {author} {\bibfnamefont {Z.}~\bibnamefont {Yuan}},\ }\href
  {https://doi.org/10.3847/2041-8213/abbc0d} {\bibfield  {journal} {\bibinfo
  {journal} {Astrophys. J. Lett.}\ }\textbf {\bibinfo {volume} {902}},\
  \bibinfo {pages} {L34} (\bibinfo {year} {2020})},\ \Eprint
  {https://arxiv.org/abs/2007.04442} {arXiv:2007.04442 [astro-ph.GA]}
  \BibitemShut {NoStop}%
\bibitem [{\citenamefont {{van de Voort}}\ \emph {et~al.}(2020)\citenamefont
  {{van de Voort}}, \citenamefont {{Pakmor}}, \citenamefont {{Grand}},
  \citenamefont {{Springel}}, \citenamefont {{G{\'o}mez}},\ and\ \citenamefont
  {{Marinacci}}}]{van-de-Voort2020r}%
  \BibitemOpen
  \bibfield  {author} {\bibinfo {author} {\bibfnamefont {F.}~\bibnamefont {{van
  de Voort}}}, \bibinfo {author} {\bibfnamefont {R.}~\bibnamefont {{Pakmor}}},
  \bibinfo {author} {\bibfnamefont {R.~J.~J.}\ \bibnamefont {{Grand}}},
  \bibinfo {author} {\bibfnamefont {V.}~\bibnamefont {{Springel}}}, \bibinfo
  {author} {\bibfnamefont {F.~A.}\ \bibnamefont {{G{\'o}mez}}},\ and\ \bibinfo
  {author} {\bibfnamefont {F.}~\bibnamefont {{Marinacci}}},\ }\href@noop {}
  {\bibfield  {journal} {\bibinfo  {journal} {\mnras}\ }\textbf {\bibinfo
  {volume} {494}},\ \bibinfo {pages} {4867} (\bibinfo {year} {2020})},\ \Eprint
  {https://arxiv.org/abs/1907.01557} {arXiv:1907.01557 [astro-ph.GA]}
  \BibitemShut {NoStop}%
\bibitem [{\citenamefont {{Siegel}}(2019)}]{Siegel2019r}%
  \BibitemOpen
  \bibfield  {author} {\bibinfo {author} {\bibfnamefont {D.~M.}\ \bibnamefont
  {{Siegel}}},\ }\href@noop {} {\bibfield  {journal} {\bibinfo  {journal}
  {European Physical Journal A}\ }\textbf {\bibinfo {volume} {55}},\ \bibinfo
  {eid} {203} (\bibinfo {year} {2019})},\ \Eprint
  {https://arxiv.org/abs/1901.09044} {arXiv:1901.09044 [astro-ph.HE]}
  \BibitemShut {NoStop}%
\bibitem [{\citenamefont {{Fern{\'a}ndez}}\ \emph {et~al.}(2019)\citenamefont
  {{Fern{\'a}ndez}}, \citenamefont {{Tchekhovskoy}}, \citenamefont
  {{Quataert}}, \citenamefont {{Foucart}},\ and\ \citenamefont
  {{Kasen}}}]{Fernandez2019b}%
  \BibitemOpen
  \bibfield  {author} {\bibinfo {author} {\bibfnamefont {R.}~\bibnamefont
  {{Fern{\'a}ndez}}}, \bibinfo {author} {\bibfnamefont {A.}~\bibnamefont
  {{Tchekhovskoy}}}, \bibinfo {author} {\bibfnamefont {E.}~\bibnamefont
  {{Quataert}}}, \bibinfo {author} {\bibfnamefont {F.}~\bibnamefont
  {{Foucart}}},\ and\ \bibinfo {author} {\bibfnamefont {D.}~\bibnamefont
  {{Kasen}}},\ }\href@noop {} {\bibfield  {journal} {\bibinfo  {journal}
  {\mnras}\ }\textbf {\bibinfo {volume} {482}},\ \bibinfo {pages} {3373}
  (\bibinfo {year} {2019})},\ \Eprint {https://arxiv.org/abs/1808.00461}
  {arXiv:1808.00461 [astro-ph.HE]} \BibitemShut {NoStop}%
\bibitem [{\citenamefont {{Miller}}\ \emph {et~al.}(2020)\citenamefont
  {{Miller}}, \citenamefont {{Sprouse}}, \citenamefont {{Fryer}}, \citenamefont
  {{Ryan}}, \citenamefont {{Dolence}}, \citenamefont {{Mumpower}},\ and\
  \citenamefont {{Surman}}}]{Miller2020a}%
  \BibitemOpen
  \bibfield  {author} {\bibinfo {author} {\bibfnamefont {J.~M.}\ \bibnamefont
  {{Miller}}}, \bibinfo {author} {\bibfnamefont {T.~M.}\ \bibnamefont
  {{Sprouse}}}, \bibinfo {author} {\bibfnamefont {C.~L.}\ \bibnamefont
  {{Fryer}}}, \bibinfo {author} {\bibfnamefont {B.~R.}\ \bibnamefont {{Ryan}}},
  \bibinfo {author} {\bibfnamefont {J.~C.}\ \bibnamefont {{Dolence}}}, \bibinfo
  {author} {\bibfnamefont {M.~R.}\ \bibnamefont {{Mumpower}}},\ and\ \bibinfo
  {author} {\bibfnamefont {R.}~\bibnamefont {{Surman}}},\ }\href
  {https://doi.org/10.3847/1538-4357/abb4e3} {\bibfield  {journal} {\bibinfo
  {journal} {\apj}\ }\textbf {\bibinfo {volume} {902}},\ \bibinfo {eid} {66}
  (\bibinfo {year} {2020})},\ \Eprint {https://arxiv.org/abs/1912.03378}
  {arXiv:1912.03378 [astro-ph.HE]} \BibitemShut {NoStop}%
\bibitem [{\citenamefont {{Metzger}}\ \emph {et~al.}(2010)\citenamefont
  {{Metzger}}, \citenamefont {{Mart{\'{\i}}nez-Pinedo}}, \citenamefont
  {{Darbha}}, \citenamefont {{Quataert}}, \citenamefont {{Arcones}},
  \citenamefont {{Kasen}}, \citenamefont {{Thomas}}, \citenamefont {{Nugent}},
  \citenamefont {{Panov}},\ and\ \citenamefont {{Zinner}}}]{Metzger2010c}%
  \BibitemOpen
  \bibfield  {author} {\bibinfo {author} {\bibfnamefont {B.~D.}\ \bibnamefont
  {{Metzger}}}, \bibinfo {author} {\bibfnamefont {G.}~\bibnamefont
  {{Mart{\'{\i}}nez-Pinedo}}}, \bibinfo {author} {\bibfnamefont
  {S.}~\bibnamefont {{Darbha}}}, \bibinfo {author} {\bibfnamefont
  {E.}~\bibnamefont {{Quataert}}}, \bibinfo {author} {\bibfnamefont
  {A.}~\bibnamefont {{Arcones}}}, \bibinfo {author} {\bibfnamefont
  {D.}~\bibnamefont {{Kasen}}}, \bibinfo {author} {\bibfnamefont
  {R.}~\bibnamefont {{Thomas}}}, \bibinfo {author} {\bibfnamefont
  {P.}~\bibnamefont {{Nugent}}}, \bibinfo {author} {\bibfnamefont {I.~V.}\
  \bibnamefont {{Panov}}},\ and\ \bibinfo {author} {\bibfnamefont {N.~T.}\
  \bibnamefont {{Zinner}}},\ }\href@noop {} {\bibfield  {journal} {\bibinfo
  {journal} {\mnras}\ }\textbf {\bibinfo {volume} {406}},\ \bibinfo {pages}
  {2650} (\bibinfo {year} {2010})},\ \Eprint {https://arxiv.org/abs/1001.5029}
  {arXiv:1001.5029 [astro-ph.HE]} \BibitemShut {NoStop}%
\bibitem [{\citenamefont {{Roberts}}\ \emph {et~al.}(2011)\citenamefont
  {{Roberts}}, \citenamefont {{Kasen}}, \citenamefont {{Lee}},\ and\
  \citenamefont {{Ramirez-Ruiz}}}]{Roberts2011}%
  \BibitemOpen
  \bibfield  {author} {\bibinfo {author} {\bibfnamefont {L.~F.}\ \bibnamefont
  {{Roberts}}}, \bibinfo {author} {\bibfnamefont {D.}~\bibnamefont {{Kasen}}},
  \bibinfo {author} {\bibfnamefont {W.~H.}\ \bibnamefont {{Lee}}},\ and\
  \bibinfo {author} {\bibfnamefont {E.}~\bibnamefont {{Ramirez-Ruiz}}},\
  }\href@noop {} {\bibfield  {journal} {\bibinfo  {journal} {\apjl}\ }\textbf
  {\bibinfo {volume} {736}},\ \bibinfo {pages} {L21+} (\bibinfo {year}
  {2011})},\ \Eprint {https://arxiv.org/abs/1104.5504} {arXiv:1104.5504
  [astro-ph.HE]} \BibitemShut {NoStop}%
\bibitem [{\citenamefont {{Goriely}}\ \emph {et~al.}(2011)\citenamefont
  {{Goriely}}, \citenamefont {{Bauswein}},\ and\ \citenamefont
  {{Janka}}}]{Goriely2011}%
  \BibitemOpen
  \bibfield  {author} {\bibinfo {author} {\bibfnamefont {S.}~\bibnamefont
  {{Goriely}}}, \bibinfo {author} {\bibfnamefont {A.}~\bibnamefont
  {{Bauswein}}},\ and\ \bibinfo {author} {\bibfnamefont {H.-T.}\ \bibnamefont
  {{Janka}}},\ }\href@noop {} {\bibfield  {journal} {\bibinfo  {journal}
  {\apjl}\ }\textbf {\bibinfo {volume} {738}},\ \bibinfo {pages} {L32+}
  (\bibinfo {year} {2011})},\ \Eprint {https://arxiv.org/abs/1107.0899}
  {arXiv:1107.0899 [astro-ph.SR]} \BibitemShut {NoStop}%
\bibitem [{\citenamefont {{Tanaka}}\ \emph {et~al.}(2017)\citenamefont
  {{Tanaka}}, \citenamefont {{Utsumi}}, \citenamefont {{Mazzali}},
  \citenamefont {{Tominaga}}, \citenamefont {{Yoshida}}, \citenamefont
  {{Sekiguchi}}, \citenamefont {{Morokuma}}, \citenamefont {{Motohara}},
  \citenamefont {{Ohta}}, \citenamefont {{Kawabata}}, \citenamefont {{Abe}},
  \citenamefont {{Aoki}}, \citenamefont {{Asakura}}, \citenamefont {{Baar}},
  \citenamefont {{Barway}}, \citenamefont {{Bond}}, \citenamefont {{Doi}},
  \citenamefont {{Fujiyoshi}}, \citenamefont {{Furusawa}}, \citenamefont
  {{Honda}}, \citenamefont {{Itoh}}, \citenamefont {{Kawabata}}, \citenamefont
  {{Kawai}}, \citenamefont {{Kim}}, \citenamefont {{Lee}}, \citenamefont
  {{Miyazaki}}, \citenamefont {{Morihana}}, \citenamefont {{Nagashima}},
  \citenamefont {{Nagayama}}, \citenamefont {{Nakaoka}}, \citenamefont
  {{Nakata}}, \citenamefont {{Ohsawa}}, \citenamefont {{Ohshima}},
  \citenamefont {{Okita}}, \citenamefont {{Saito}}, \citenamefont {{Sumi}},
  \citenamefont {{Tajitsu}}, \citenamefont {{Takahashi}}, \citenamefont
  {{Takayama}}, \citenamefont {{Tamura}}, \citenamefont {{Tanaka}},
  \citenamefont {{Terai}}, \citenamefont {{Tristram}}, \citenamefont
  {{Yasuda}},\ and\ \citenamefont {{Zenko}}}]{Tanaka2017t}%
  \BibitemOpen
  \bibfield  {author} {\bibinfo {author} {\bibfnamefont {M.}~\bibnamefont
  {{Tanaka}}}, \bibinfo {author} {\bibfnamefont {Y.}~\bibnamefont {{Utsumi}}},
  \bibinfo {author} {\bibfnamefont {P.~A.}\ \bibnamefont {{Mazzali}}}, \bibinfo
  {author} {\bibfnamefont {N.}~\bibnamefont {{Tominaga}}}, \bibinfo {author}
  {\bibfnamefont {M.}~\bibnamefont {{Yoshida}}}, \bibinfo {author}
  {\bibfnamefont {Y.}~\bibnamefont {{Sekiguchi}}}, \bibinfo {author}
  {\bibfnamefont {T.}~\bibnamefont {{Morokuma}}}, \bibinfo {author}
  {\bibfnamefont {K.}~\bibnamefont {{Motohara}}}, \bibinfo {author}
  {\bibfnamefont {K.}~\bibnamefont {{Ohta}}}, \bibinfo {author} {\bibfnamefont
  {K.~S.}\ \bibnamefont {{Kawabata}}}, \bibinfo {author} {\bibfnamefont
  {F.}~\bibnamefont {{Abe}}}, \bibinfo {author} {\bibfnamefont
  {K.}~\bibnamefont {{Aoki}}}, \bibinfo {author} {\bibfnamefont
  {Y.}~\bibnamefont {{Asakura}}}, \bibinfo {author} {\bibfnamefont
  {S.}~\bibnamefont {{Baar}}}, \bibinfo {author} {\bibfnamefont
  {S.}~\bibnamefont {{Barway}}}, \bibinfo {author} {\bibfnamefont {I.~A.}\
  \bibnamefont {{Bond}}}, \bibinfo {author} {\bibfnamefont {M.}~\bibnamefont
  {{Doi}}}, \bibinfo {author} {\bibfnamefont {T.}~\bibnamefont {{Fujiyoshi}}},
  \bibinfo {author} {\bibfnamefont {H.}~\bibnamefont {{Furusawa}}}, \bibinfo
  {author} {\bibfnamefont {S.}~\bibnamefont {{Honda}}}, \bibinfo {author}
  {\bibfnamefont {Y.}~\bibnamefont {{Itoh}}}, \bibinfo {author} {\bibfnamefont
  {M.}~\bibnamefont {{Kawabata}}}, \bibinfo {author} {\bibfnamefont
  {N.}~\bibnamefont {{Kawai}}}, \bibinfo {author} {\bibfnamefont {J.~H.}\
  \bibnamefont {{Kim}}}, \bibinfo {author} {\bibfnamefont {C.-H.}\ \bibnamefont
  {{Lee}}}, \bibinfo {author} {\bibfnamefont {S.}~\bibnamefont {{Miyazaki}}},
  \bibinfo {author} {\bibfnamefont {K.}~\bibnamefont {{Morihana}}}, \bibinfo
  {author} {\bibfnamefont {H.}~\bibnamefont {{Nagashima}}}, \bibinfo {author}
  {\bibfnamefont {T.}~\bibnamefont {{Nagayama}}}, \bibinfo {author}
  {\bibfnamefont {T.}~\bibnamefont {{Nakaoka}}}, \bibinfo {author}
  {\bibfnamefont {F.}~\bibnamefont {{Nakata}}}, \bibinfo {author}
  {\bibfnamefont {R.}~\bibnamefont {{Ohsawa}}}, \bibinfo {author}
  {\bibfnamefont {T.}~\bibnamefont {{Ohshima}}}, \bibinfo {author}
  {\bibfnamefont {H.}~\bibnamefont {{Okita}}}, \bibinfo {author} {\bibfnamefont
  {T.}~\bibnamefont {{Saito}}}, \bibinfo {author} {\bibfnamefont
  {T.}~\bibnamefont {{Sumi}}}, \bibinfo {author} {\bibfnamefont
  {A.}~\bibnamefont {{Tajitsu}}}, \bibinfo {author} {\bibfnamefont
  {J.}~\bibnamefont {{Takahashi}}}, \bibinfo {author} {\bibfnamefont
  {M.}~\bibnamefont {{Takayama}}}, \bibinfo {author} {\bibfnamefont
  {Y.}~\bibnamefont {{Tamura}}}, \bibinfo {author} {\bibfnamefont
  {I.}~\bibnamefont {{Tanaka}}}, \bibinfo {author} {\bibfnamefont
  {T.}~\bibnamefont {{Terai}}}, \bibinfo {author} {\bibfnamefont {P.~J.}\
  \bibnamefont {{Tristram}}}, \bibinfo {author} {\bibfnamefont
  {N.}~\bibnamefont {{Yasuda}}},\ and\ \bibinfo {author} {\bibfnamefont
  {T.}~\bibnamefont {{Zenko}}},\ }\href@noop {} {\bibfield  {journal} {\bibinfo
   {journal} {\pasj}\ }\textbf {\bibinfo {volume} {69}},\ \bibinfo {eid} {102}
  (\bibinfo {year} {2017})},\ \Eprint {https://arxiv.org/abs/1710.05850}
  {arXiv:1710.05850 [astro-ph.HE]} \BibitemShut {NoStop}%
\bibitem [{\citenamefont {{Metzger}}(2019)}]{Metzger2019a}%
  \BibitemOpen
  \bibfield  {author} {\bibinfo {author} {\bibfnamefont {B.~D.}\ \bibnamefont
  {{Metzger}}},\ }\href@noop {} {\bibfield  {journal} {\bibinfo  {journal}
  {Living Reviews in Relativity}\ }\textbf {\bibinfo {volume} {23}},\ \bibinfo
  {eid} {1} (\bibinfo {year} {2019})},\ \Eprint
  {https://arxiv.org/abs/1910.01617} {arXiv:1910.01617 [astro-ph.HE]}
  \BibitemShut {NoStop}%
\bibitem [{\citenamefont {{Hoffman}}\ \emph {et~al.}(1997)\citenamefont
  {{Hoffman}}, \citenamefont {{Woosley}},\ and\ \citenamefont
  {{Qian}}}]{Hoffman1997}%
  \BibitemOpen
  \bibfield  {author} {\bibinfo {author} {\bibfnamefont {R.~D.}\ \bibnamefont
  {{Hoffman}}}, \bibinfo {author} {\bibfnamefont {S.~E.}\ \bibnamefont
  {{Woosley}}},\ and\ \bibinfo {author} {\bibfnamefont {Y.-Z.}\ \bibnamefont
  {{Qian}}},\ }\href@noop {} {\bibfield  {journal} {\bibinfo  {journal} {\apj}\
  }\textbf {\bibinfo {volume} {482}},\ \bibinfo {pages} {951} (\bibinfo {year}
  {1997})},\ \Eprint {https://arxiv.org/abs/arXiv:astro-ph/9611097}
  {arXiv:astro-ph/9611097} \BibitemShut {NoStop}%
\bibitem [{\citenamefont {{Kasen}}\ \emph {et~al.}(2015)\citenamefont
  {{Kasen}}, \citenamefont {{Fern{\'a}ndez}},\ and\ \citenamefont
  {{Metzger}}}]{Kasen2015}%
  \BibitemOpen
  \bibfield  {author} {\bibinfo {author} {\bibfnamefont {D.}~\bibnamefont
  {{Kasen}}}, \bibinfo {author} {\bibfnamefont {R.}~\bibnamefont
  {{Fern{\'a}ndez}}},\ and\ \bibinfo {author} {\bibfnamefont {B.~D.}\
  \bibnamefont {{Metzger}}},\ }\href@noop {} {\bibfield  {journal} {\bibinfo
  {journal} {\mnras}\ }\textbf {\bibinfo {volume} {450}},\ \bibinfo {pages}
  {1777} (\bibinfo {year} {2015})},\ \Eprint {https://arxiv.org/abs/1411.3726}
  {arXiv:1411.3726 [astro-ph.HE]} \BibitemShut {NoStop}%
\bibitem [{\citenamefont {{Lema{\^\i}tre}}\ \emph {et~al.}(2021)\citenamefont
  {{Lema{\^\i}tre}}, \citenamefont {{Goriely}}, \citenamefont {{Bauswein}},\
  and\ \citenamefont {{Janka}}}]{Lemaitre2021m}%
  \BibitemOpen
  \bibfield  {author} {\bibinfo {author} {\bibfnamefont {J.~F.}\ \bibnamefont
  {{Lema{\^\i}tre}}}, \bibinfo {author} {\bibfnamefont {S.}~\bibnamefont
  {{Goriely}}}, \bibinfo {author} {\bibfnamefont {A.}~\bibnamefont
  {{Bauswein}}},\ and\ \bibinfo {author} {\bibfnamefont {H.~T.}\ \bibnamefont
  {{Janka}}},\ }\href@noop {} {\bibfield  {journal} {\bibinfo  {journal}
  {\prc}\ }\textbf {\bibinfo {volume} {103}},\ \bibinfo {eid} {025806}
  (\bibinfo {year} {2021})},\ \Eprint {https://arxiv.org/abs/2102.03686}
  {arXiv:2102.03686 [nucl-th]} \BibitemShut {NoStop}%
\bibitem [{\citenamefont {{Kawaguchi}}\ \emph {et~al.}(2018)\citenamefont
  {{Kawaguchi}}, \citenamefont {{Shibata}},\ and\ \citenamefont
  {{Tanaka}}}]{Kawaguchi2018a}%
  \BibitemOpen
  \bibfield  {author} {\bibinfo {author} {\bibfnamefont {K.}~\bibnamefont
  {{Kawaguchi}}}, \bibinfo {author} {\bibfnamefont {M.}~\bibnamefont
  {{Shibata}}},\ and\ \bibinfo {author} {\bibfnamefont {M.}~\bibnamefont
  {{Tanaka}}},\ }\href@noop {} {\bibfield  {journal} {\bibinfo  {journal}
  {\apjl}\ }\textbf {\bibinfo {volume} {865}},\ \bibinfo {eid} {L21} (\bibinfo
  {year} {2018})},\ \Eprint {https://arxiv.org/abs/1806.04088}
  {arXiv:1806.04088 [astro-ph.HE]} \BibitemShut {NoStop}%
\bibitem [{\citenamefont {{Qian}}\ and\ \citenamefont
  {{Woosley}}(1996)}]{Qian1996}%
  \BibitemOpen
  \bibfield  {author} {\bibinfo {author} {\bibfnamefont {Y.}~\bibnamefont
  {{Qian}}}\ and\ \bibinfo {author} {\bibfnamefont {S.~E.}\ \bibnamefont
  {{Woosley}}},\ }\href@noop {} {\bibfield  {journal} {\bibinfo  {journal}
  {\apj}\ }\textbf {\bibinfo {volume} {471}},\ \bibinfo {pages} {331} (\bibinfo
  {year} {1996})},\ \Eprint {https://arxiv.org/abs/arXiv:astro-ph/9611094}
  {arXiv:astro-ph/9611094} \BibitemShut {NoStop}%
\bibitem [{\citenamefont {Duan}\ \emph {et~al.}(2011)\citenamefont {Duan},
  \citenamefont {Friedland}, \citenamefont {McLaughlin},\ and\ \citenamefont
  {Surman}}]{Duan:2010af}%
  \BibitemOpen
  \bibfield  {author} {\bibinfo {author} {\bibfnamefont {H.}~\bibnamefont
  {Duan}}, \bibinfo {author} {\bibfnamefont {A.}~\bibnamefont {Friedland}},
  \bibinfo {author} {\bibfnamefont {G.}~\bibnamefont {McLaughlin}},\ and\
  \bibinfo {author} {\bibfnamefont {R.}~\bibnamefont {Surman}},\ }\href
  {https://doi.org/10.1088/0954-3899/38/3/035201} {\bibfield  {journal}
  {\bibinfo  {journal} {J. Phys. G}\ }\textbf {\bibinfo {volume} {38}},\
  \bibinfo {pages} {035201} (\bibinfo {year} {2011})},\ \Eprint
  {https://arxiv.org/abs/1012.0532} {arXiv:1012.0532 [astro-ph.SR]}
  \BibitemShut {NoStop}%
\bibitem [{\citenamefont {Malkus}\ \emph {et~al.}(2012)\citenamefont {Malkus},
  \citenamefont {Kneller}, \citenamefont {McLaughlin},\ and\ \citenamefont
  {Surman}}]{Malkus:2012ts}%
  \BibitemOpen
  \bibfield  {author} {\bibinfo {author} {\bibfnamefont {A.}~\bibnamefont
  {Malkus}}, \bibinfo {author} {\bibfnamefont {J.~P.}\ \bibnamefont {Kneller}},
  \bibinfo {author} {\bibfnamefont {G.~C.}\ \bibnamefont {McLaughlin}},\ and\
  \bibinfo {author} {\bibfnamefont {R.}~\bibnamefont {Surman}},\ }\href@noop {}
  {\bibfield  {journal} {\bibinfo  {journal} {Phys. Rev.}\ }\textbf {\bibinfo
  {volume} {D86}},\ \bibinfo {pages} {085015} (\bibinfo {year} {2012})},\
  \Eprint {https://arxiv.org/abs/1207.6648} {arXiv:1207.6648 [hep-ph]}
  \BibitemShut {NoStop}%
\bibitem [{\citenamefont {Wu}\ \emph {et~al.}(2015)\citenamefont {Wu},
  \citenamefont {Qian}, \citenamefont {Martinez-Pinedo}, \citenamefont
  {Fischer},\ and\ \citenamefont {Huther}}]{Wu:2014kaa}%
  \BibitemOpen
  \bibfield  {author} {\bibinfo {author} {\bibfnamefont {M.-R.}\ \bibnamefont
  {Wu}}, \bibinfo {author} {\bibfnamefont {Y.-Z.}\ \bibnamefont {Qian}},
  \bibinfo {author} {\bibfnamefont {G.}~\bibnamefont {Martinez-Pinedo}},
  \bibinfo {author} {\bibfnamefont {T.}~\bibnamefont {Fischer}},\ and\ \bibinfo
  {author} {\bibfnamefont {L.}~\bibnamefont {Huther}},\ }\href
  {https://doi.org/10.1103/PhysRevD.91.065016} {\bibfield  {journal} {\bibinfo
  {journal} {Phys. Rev. D}\ }\textbf {\bibinfo {volume} {91}},\ \bibinfo
  {pages} {065016} (\bibinfo {year} {2015})},\ \Eprint
  {https://arxiv.org/abs/1412.8587} {arXiv:1412.8587 [astro-ph.HE]}
  \BibitemShut {NoStop}%
\bibitem [{\citenamefont {{Pllumbi}}\ \emph {et~al.}(2015)\citenamefont
  {{Pllumbi}}, \citenamefont {{Tamborra}}, \citenamefont {{Wanajo}},
  \citenamefont {{Janka}},\ and\ \citenamefont
  {{H{\"u}depohl}}}]{Pllumbi2015a}%
  \BibitemOpen
  \bibfield  {author} {\bibinfo {author} {\bibfnamefont {E.}~\bibnamefont
  {{Pllumbi}}}, \bibinfo {author} {\bibfnamefont {I.}~\bibnamefont
  {{Tamborra}}}, \bibinfo {author} {\bibfnamefont {S.}~\bibnamefont
  {{Wanajo}}}, \bibinfo {author} {\bibfnamefont {H.-T.}\ \bibnamefont
  {{Janka}}},\ and\ \bibinfo {author} {\bibfnamefont {L.}~\bibnamefont
  {{H{\"u}depohl}}},\ }\href@noop {} {\bibfield  {journal} {\bibinfo  {journal}
  {\apj}\ }\textbf {\bibinfo {volume} {808}},\ \bibinfo {eid} {188} (\bibinfo
  {year} {2015})},\ \Eprint {https://arxiv.org/abs/1406.2596} {arXiv:1406.2596
  [astro-ph.SR]} \BibitemShut {NoStop}%
\bibitem [{\citenamefont {Frensel}\ \emph {et~al.}(2017)\citenamefont
  {Frensel}, \citenamefont {Wu}, \citenamefont {Volpe},\ and\ \citenamefont
  {Perego}}]{Frensel:2016fge}%
  \BibitemOpen
  \bibfield  {author} {\bibinfo {author} {\bibfnamefont {M.}~\bibnamefont
  {Frensel}}, \bibinfo {author} {\bibfnamefont {M.-R.}\ \bibnamefont {Wu}},
  \bibinfo {author} {\bibfnamefont {C.}~\bibnamefont {Volpe}},\ and\ \bibinfo
  {author} {\bibfnamefont {A.}~\bibnamefont {Perego}},\ }\href@noop {}
  {\bibfield  {journal} {\bibinfo  {journal} {Phys. Rev.}\ }\textbf {\bibinfo
  {volume} {D95}},\ \bibinfo {pages} {023011} (\bibinfo {year} {2017})},\
  \Eprint {https://arxiv.org/abs/1607.05938} {arXiv:1607.05938 [astro-ph.HE]}
  \BibitemShut {NoStop}%
\bibitem [{\citenamefont {Sasaki}\ \emph {et~al.}(2017)\citenamefont {Sasaki},
  \citenamefont {Kajino}, \citenamefont {Takiwaki}, \citenamefont {Hayakawa},
  \citenamefont {Balantekin},\ and\ \citenamefont {Pehlivan}}]{Sasaki:2017jry}%
  \BibitemOpen
  \bibfield  {author} {\bibinfo {author} {\bibfnamefont {H.}~\bibnamefont
  {Sasaki}}, \bibinfo {author} {\bibfnamefont {T.}~\bibnamefont {Kajino}},
  \bibinfo {author} {\bibfnamefont {T.}~\bibnamefont {Takiwaki}}, \bibinfo
  {author} {\bibfnamefont {T.}~\bibnamefont {Hayakawa}}, \bibinfo {author}
  {\bibfnamefont {A.~B.}\ \bibnamefont {Balantekin}},\ and\ \bibinfo {author}
  {\bibfnamefont {Y.}~\bibnamefont {Pehlivan}},\ }\href
  {https://doi.org/10.1103/PhysRevD.96.043013} {\bibfield  {journal} {\bibinfo
  {journal} {Phys. Rev. D}\ }\textbf {\bibinfo {volume} {96}},\ \bibinfo
  {pages} {043013} (\bibinfo {year} {2017})},\ \Eprint
  {https://arxiv.org/abs/1707.09111} {arXiv:1707.09111 [astro-ph.HE]}
  \BibitemShut {NoStop}%
\bibitem [{\citenamefont {Zhu}\ \emph {et~al.}(2016)\citenamefont {Zhu},
  \citenamefont {Perego},\ and\ \citenamefont {McLaughlin}}]{Zhu:2016mwa}%
  \BibitemOpen
  \bibfield  {author} {\bibinfo {author} {\bibfnamefont {Y.-L.}\ \bibnamefont
  {Zhu}}, \bibinfo {author} {\bibfnamefont {A.}~\bibnamefont {Perego}},\ and\
  \bibinfo {author} {\bibfnamefont {G.~C.}\ \bibnamefont {McLaughlin}},\
  }\href@noop {} {\bibfield  {journal} {\bibinfo  {journal} {Phys. Rev.}\
  }\textbf {\bibinfo {volume} {D94}},\ \bibinfo {pages} {105006} (\bibinfo
  {year} {2016})},\ \Eprint {https://arxiv.org/abs/1607.04671}
  {arXiv:1607.04671 [hep-ph]} \BibitemShut {NoStop}%
\bibitem [{\citenamefont {Wu}\ \emph {et~al.}(2017)\citenamefont {Wu},
  \citenamefont {Tamborra}, \citenamefont {Just},\ and\ \citenamefont
  {Janka}}]{Wu:2017drk}%
  \BibitemOpen
  \bibfield  {author} {\bibinfo {author} {\bibfnamefont {M.-R.}\ \bibnamefont
  {Wu}}, \bibinfo {author} {\bibfnamefont {I.}~\bibnamefont {Tamborra}},
  \bibinfo {author} {\bibfnamefont {O.}~\bibnamefont {Just}},\ and\ \bibinfo
  {author} {\bibfnamefont {H.-T.}\ \bibnamefont {Janka}},\ }\href@noop {}
  {\bibfield  {journal} {\bibinfo  {journal} {Phys. Rev.}\ }\textbf {\bibinfo
  {volume} {D96}},\ \bibinfo {pages} {123015} (\bibinfo {year} {2017})},\
  \Eprint {https://arxiv.org/abs/1711.00477} {arXiv:1711.00477 [astro-ph.HE]}
  \BibitemShut {NoStop}%
\bibitem [{\citenamefont {{Deaton}}\ \emph {et~al.}(2018)\citenamefont
  {{Deaton}}, \citenamefont {{O'Connor}}, \citenamefont {{Zhu}}, \citenamefont
  {{Bohn}}, \citenamefont {{Jesse}}, \citenamefont {{Foucart}}, \citenamefont
  {{Duez}},\ and\ \citenamefont {{McLaughlin}}}]{Deaton2018a}%
  \BibitemOpen
  \bibfield  {author} {\bibinfo {author} {\bibfnamefont {M.~B.}\ \bibnamefont
  {{Deaton}}}, \bibinfo {author} {\bibfnamefont {E.}~\bibnamefont
  {{O'Connor}}}, \bibinfo {author} {\bibfnamefont {Y.~L.}\ \bibnamefont
  {{Zhu}}}, \bibinfo {author} {\bibfnamefont {A.}~\bibnamefont {{Bohn}}},
  \bibinfo {author} {\bibfnamefont {J.}~\bibnamefont {{Jesse}}}, \bibinfo
  {author} {\bibfnamefont {F.}~\bibnamefont {{Foucart}}}, \bibinfo {author}
  {\bibfnamefont {M.~D.}\ \bibnamefont {{Duez}}},\ and\ \bibinfo {author}
  {\bibfnamefont {G.~C.}\ \bibnamefont {{McLaughlin}}},\ }\href@noop {}
  {\bibfield  {journal} {\bibinfo  {journal} {\prd}\ }\textbf {\bibinfo
  {volume} {98}},\ \bibinfo {eid} {103014} (\bibinfo {year} {2018})},\ \Eprint
  {https://arxiv.org/abs/1806.10255} {arXiv:1806.10255 [astro-ph.HE]}
  \BibitemShut {NoStop}%
\bibitem [{\citenamefont {{George}}\ \emph {et~al.}(2020)\citenamefont
  {{George}}, \citenamefont {{Wu}}, \citenamefont {{Tamborra}}, \citenamefont
  {{Ardevol-Pulpillo}},\ and\ \citenamefont {{Janka}}}]{George2020a}%
  \BibitemOpen
  \bibfield  {author} {\bibinfo {author} {\bibfnamefont {M.}~\bibnamefont
  {{George}}}, \bibinfo {author} {\bibfnamefont {M.-R.}\ \bibnamefont {{Wu}}},
  \bibinfo {author} {\bibfnamefont {I.}~\bibnamefont {{Tamborra}}}, \bibinfo
  {author} {\bibfnamefont {R.}~\bibnamefont {{Ardevol-Pulpillo}}},\ and\
  \bibinfo {author} {\bibfnamefont {H.-T.}\ \bibnamefont {{Janka}}},\
  }\href@noop {} {\bibfield  {journal} {\bibinfo  {journal} {\prd}\ }\textbf
  {\bibinfo {volume} {102}},\ \bibinfo {eid} {103015} (\bibinfo {year}
  {2020})},\ \Eprint {https://arxiv.org/abs/2009.04046} {arXiv:2009.04046
  [astro-ph.HE]} \BibitemShut {NoStop}%
\bibitem [{\citenamefont {{Xiong}}\ \emph {et~al.}(2020)\citenamefont
  {{Xiong}}, \citenamefont {{Sieverding}}, \citenamefont {{Sen}},\ and\
  \citenamefont {{Qian}}}]{Xiong2020c}%
  \BibitemOpen
  \bibfield  {author} {\bibinfo {author} {\bibfnamefont {Z.}~\bibnamefont
  {{Xiong}}}, \bibinfo {author} {\bibfnamefont {A.}~\bibnamefont
  {{Sieverding}}}, \bibinfo {author} {\bibfnamefont {M.}~\bibnamefont
  {{Sen}}},\ and\ \bibinfo {author} {\bibfnamefont {Y.-Z.}\ \bibnamefont
  {{Qian}}},\ }\href@noop {} {\bibfield  {journal} {\bibinfo  {journal} {\apj}\
  }\textbf {\bibinfo {volume} {900}},\ \bibinfo {eid} {144} (\bibinfo {year}
  {2020})},\ \Eprint {https://arxiv.org/abs/2006.11414} {arXiv:2006.11414
  [astro-ph.HE]} \BibitemShut {NoStop}%
\bibitem [{\citenamefont {{Myers}}\ \emph {et~al.}(2021)\citenamefont
  {{Myers}}, \citenamefont {{Cooper}}, \citenamefont {{Warren}}, \citenamefont
  {{Kneller}}, \citenamefont {{McLaughlin}}, \citenamefont {{Richers}},
  \citenamefont {{Grohs}},\ and\ \citenamefont {{Frohlich}}}]{Myers2021w}%
  \BibitemOpen
  \bibfield  {author} {\bibinfo {author} {\bibfnamefont {M.}~\bibnamefont
  {{Myers}}}, \bibinfo {author} {\bibfnamefont {T.}~\bibnamefont {{Cooper}}},
  \bibinfo {author} {\bibfnamefont {M.}~\bibnamefont {{Warren}}}, \bibinfo
  {author} {\bibfnamefont {J.}~\bibnamefont {{Kneller}}}, \bibinfo {author}
  {\bibfnamefont {G.}~\bibnamefont {{McLaughlin}}}, \bibinfo {author}
  {\bibfnamefont {S.}~\bibnamefont {{Richers}}}, \bibinfo {author}
  {\bibfnamefont {E.}~\bibnamefont {{Grohs}}},\ and\ \bibinfo {author}
  {\bibfnamefont {C.}~\bibnamefont {{Frohlich}}},\ }\href@noop {} {\bibfield
  {journal} {\bibinfo  {journal} {arXiv e-prints}\ ,\ \bibinfo {eid}
  {arXiv:2111.13722}} (\bibinfo {year} {2021})},\ \Eprint
  {https://arxiv.org/abs/2111.13722} {arXiv:2111.13722 [hep-ph]} \BibitemShut
  {NoStop}%
\bibitem [{\citenamefont {Sawyer}(2005)}]{Sawyer:2005jk}%
  \BibitemOpen
  \bibfield  {author} {\bibinfo {author} {\bibfnamefont {R.~F.}\ \bibnamefont
  {Sawyer}},\ }\href@noop {} {\bibfield  {journal} {\bibinfo  {journal} {Phys.
  Rev.}\ }\textbf {\bibinfo {volume} {D72}},\ \bibinfo {pages} {045003}
  (\bibinfo {year} {2005})},\ \Eprint {https://arxiv.org/abs/hep-ph/0503013}
  {arXiv:hep-ph/0503013 [hep-ph]} \BibitemShut {NoStop}%
\bibitem [{\citenamefont {Sawyer}(2016)}]{Sawyer:2015dsa}%
  \BibitemOpen
  \bibfield  {author} {\bibinfo {author} {\bibfnamefont {R.~F.}\ \bibnamefont
  {Sawyer}},\ }\href@noop {} {\bibfield  {journal} {\bibinfo  {journal} {Phys.
  Rev. Lett.}\ }\textbf {\bibinfo {volume} {116}},\ \bibinfo {pages} {081101}
  (\bibinfo {year} {2016})},\ \Eprint {https://arxiv.org/abs/1509.03323}
  {arXiv:1509.03323 [astro-ph.HE]} \BibitemShut {NoStop}%
\bibitem [{\citenamefont {Capozzi}\ and\ \citenamefont
  {Saviano}(2022)}]{Capozzi:2022slf}%
  \BibitemOpen
  \bibfield  {author} {\bibinfo {author} {\bibfnamefont {F.}~\bibnamefont
  {Capozzi}}\ and\ \bibinfo {author} {\bibfnamefont {N.}~\bibnamefont
  {Saviano}},\ }\href {https://doi.org/10.3390/universe8020094} {\bibfield
  {journal} {\bibinfo  {journal} {Universe}\ }\textbf {\bibinfo {volume} {8}},\
  \bibinfo {pages} {94} (\bibinfo {year} {2022})},\ \Eprint
  {https://arxiv.org/abs/2202.02494} {arXiv:2202.02494 [hep-ph]} \BibitemShut
  {NoStop}%
\bibitem [{\citenamefont {Tamborra}\ and\ \citenamefont
  {Shalgar}(2021)}]{Tamborra:2020cul}%
  \BibitemOpen
  \bibfield  {author} {\bibinfo {author} {\bibfnamefont {I.}~\bibnamefont
  {Tamborra}}\ and\ \bibinfo {author} {\bibfnamefont {S.}~\bibnamefont
  {Shalgar}},\ }\href {https://doi.org/10.1146/annurev-nucl-102920-050505}
  {\bibfield  {journal} {\bibinfo  {journal} {Ann. Rev. Nucl. Part. Sci.}\
  }\textbf {\bibinfo {volume} {71}},\ \bibinfo {pages} {165} (\bibinfo {year}
  {2021})},\ \Eprint {https://arxiv.org/abs/2011.01948} {arXiv:2011.01948
  [astro-ph.HE]} \BibitemShut {NoStop}%
\bibitem [{\citenamefont {Duan}\ \emph {et~al.}(2010)\citenamefont {Duan},
  \citenamefont {Fuller},\ and\ \citenamefont {Qian}}]{Duan:2010bg}%
  \BibitemOpen
  \bibfield  {author} {\bibinfo {author} {\bibfnamefont {H.}~\bibnamefont
  {Duan}}, \bibinfo {author} {\bibfnamefont {G.~M.}\ \bibnamefont {Fuller}},\
  and\ \bibinfo {author} {\bibfnamefont {Y.-Z.}\ \bibnamefont {Qian}},\
  }\href@noop {} {\bibfield  {journal} {\bibinfo  {journal} {Ann. Rev. Nucl.
  Part. Sci.}\ }\textbf {\bibinfo {volume} {60}},\ \bibinfo {pages} {569}
  (\bibinfo {year} {2010})},\ \Eprint {https://arxiv.org/abs/1001.2799}
  {arXiv:1001.2799 [hep-ph]} \BibitemShut {NoStop}%
\bibitem [{\citenamefont {Martin}\ \emph {et~al.}(2021)\citenamefont {Martin},
  \citenamefont {Carlson}, \citenamefont {Cirigliano},\ and\ \citenamefont
  {Duan}}]{Martin:2021xyl}%
  \BibitemOpen
  \bibfield  {author} {\bibinfo {author} {\bibfnamefont {J.~D.}\ \bibnamefont
  {Martin}}, \bibinfo {author} {\bibfnamefont {J.}~\bibnamefont {Carlson}},
  \bibinfo {author} {\bibfnamefont {V.}~\bibnamefont {Cirigliano}},\ and\
  \bibinfo {author} {\bibfnamefont {H.}~\bibnamefont {Duan}},\ }\href
  {https://doi.org/10.1103/PhysRevD.103.063001} {\bibfield  {journal} {\bibinfo
   {journal} {Phys. Rev. D}\ }\textbf {\bibinfo {volume} {103}},\ \bibinfo
  {pages} {063001} (\bibinfo {year} {2021})},\ \Eprint
  {https://arxiv.org/abs/2101.01278} {arXiv:2101.01278 [hep-ph]} \BibitemShut
  {NoStop}%
\bibitem [{\citenamefont {Shalgar}\ and\ \citenamefont
  {Tamborra}(2021{\natexlab{a}})}]{Shalgar:2020wcx}%
  \BibitemOpen
  \bibfield  {author} {\bibinfo {author} {\bibfnamefont {S.}~\bibnamefont
  {Shalgar}}\ and\ \bibinfo {author} {\bibfnamefont {I.}~\bibnamefont
  {Tamborra}},\ }\href {https://doi.org/10.1103/PhysRevD.103.063002} {\bibfield
   {journal} {\bibinfo  {journal} {Phys. Rev. D}\ }\textbf {\bibinfo {volume}
  {103}},\ \bibinfo {pages} {063002} (\bibinfo {year} {2021}{\natexlab{a}})},\
  \Eprint {https://arxiv.org/abs/2011.00004} {arXiv:2011.00004 [astro-ph.HE]}
  \BibitemShut {NoStop}%
\bibitem [{\citenamefont {{Sasaki}}\ and\ \citenamefont
  {{Takiwaki}}(2021)}]{Sasaki:2021zld}%
  \BibitemOpen
  \bibfield  {author} {\bibinfo {author} {\bibfnamefont {H.}~\bibnamefont
  {{Sasaki}}}\ and\ \bibinfo {author} {\bibfnamefont {T.}~\bibnamefont
  {{Takiwaki}}},\ }\href@noop {} {\bibfield  {journal} {\bibinfo  {journal}
  {arXiv e-prints}\ ,\ \bibinfo {eid} {arXiv:2109.14011}} (\bibinfo {year}
  {2021})},\ \Eprint {https://arxiv.org/abs/2109.14011} {arXiv:2109.14011
  [hep-ph]} \BibitemShut {NoStop}%
\bibitem [{\citenamefont {Capozzi}\ \emph {et~al.}(2019)\citenamefont
  {Capozzi}, \citenamefont {Dasgupta}, \citenamefont {Mirizzi}, \citenamefont
  {Sen},\ and\ \citenamefont {Sigl}}]{Capozzi:2018clo}%
  \BibitemOpen
  \bibfield  {author} {\bibinfo {author} {\bibfnamefont {F.}~\bibnamefont
  {Capozzi}}, \bibinfo {author} {\bibfnamefont {B.}~\bibnamefont {Dasgupta}},
  \bibinfo {author} {\bibfnamefont {A.}~\bibnamefont {Mirizzi}}, \bibinfo
  {author} {\bibfnamefont {M.}~\bibnamefont {Sen}},\ and\ \bibinfo {author}
  {\bibfnamefont {G.}~\bibnamefont {Sigl}},\ }\href@noop {} {\bibfield
  {journal} {\bibinfo  {journal} {Phys. Rev. Lett.}\ }\textbf {\bibinfo
  {volume} {122}},\ \bibinfo {pages} {091101} (\bibinfo {year} {2019})},\
  \Eprint {https://arxiv.org/abs/1808.06618} {arXiv:1808.06618 [hep-ph]}
  \BibitemShut {NoStop}%
\bibitem [{\citenamefont {{Johns}}(2021)}]{Johns:2021qby}%
  \BibitemOpen
  \bibfield  {author} {\bibinfo {author} {\bibfnamefont {L.}~\bibnamefont
  {{Johns}}},\ }\href@noop {} {\bibfield  {journal} {\bibinfo  {journal} {arXiv
  e-prints}\ ,\ \bibinfo {eid} {arXiv:2104.11369}} (\bibinfo {year} {2021})},\
  \Eprint {https://arxiv.org/abs/2104.11369} {arXiv:2104.11369 [hep-ph]}
  \BibitemShut {NoStop}%
\bibitem [{\citenamefont {Sigl}(2022)}]{Sigl:2021tmj}%
  \BibitemOpen
  \bibfield  {author} {\bibinfo {author} {\bibfnamefont {G.}~\bibnamefont
  {Sigl}},\ }\href {https://doi.org/10.1103/PhysRevD.105.043005} {\bibfield
  {journal} {\bibinfo  {journal} {Phys. Rev. D}\ }\textbf {\bibinfo {volume}
  {105}},\ \bibinfo {pages} {043005} (\bibinfo {year} {2022})},\ \Eprint
  {https://arxiv.org/abs/2109.00091} {arXiv:2109.00091 [hep-ph]} \BibitemShut
  {NoStop}%
\bibitem [{\citenamefont {Izaguirre}\ \emph {et~al.}(2017)\citenamefont
  {Izaguirre}, \citenamefont {Raffelt},\ and\ \citenamefont
  {Tamborra}}]{Izaguirre:2016gsx}%
  \BibitemOpen
  \bibfield  {author} {\bibinfo {author} {\bibfnamefont {I.}~\bibnamefont
  {Izaguirre}}, \bibinfo {author} {\bibfnamefont {G.}~\bibnamefont {Raffelt}},\
  and\ \bibinfo {author} {\bibfnamefont {I.}~\bibnamefont {Tamborra}},\
  }\href@noop {} {\bibfield  {journal} {\bibinfo  {journal} {Phys. Rev. Lett.}\
  }\textbf {\bibinfo {volume} {118}},\ \bibinfo {pages} {021101} (\bibinfo
  {year} {2017})},\ \Eprint {https://arxiv.org/abs/1610.01612}
  {arXiv:1610.01612 [hep-ph]} \BibitemShut {NoStop}%
\bibitem [{\citenamefont {Capozzi}\ \emph {et~al.}(2017)\citenamefont
  {Capozzi}, \citenamefont {Dasgupta}, \citenamefont {Lisi}, \citenamefont
  {Marrone},\ and\ \citenamefont {Mirizzi}}]{Capozzi:2017gqd}%
  \BibitemOpen
  \bibfield  {author} {\bibinfo {author} {\bibfnamefont {F.}~\bibnamefont
  {Capozzi}}, \bibinfo {author} {\bibfnamefont {B.}~\bibnamefont {Dasgupta}},
  \bibinfo {author} {\bibfnamefont {E.}~\bibnamefont {Lisi}}, \bibinfo {author}
  {\bibfnamefont {A.}~\bibnamefont {Marrone}},\ and\ \bibinfo {author}
  {\bibfnamefont {A.}~\bibnamefont {Mirizzi}},\ }\href@noop {} {\bibfield
  {journal} {\bibinfo  {journal} {Phys. Rev. D}\ }\textbf {\bibinfo {volume}
  {96}},\ \bibinfo {pages} {043016} (\bibinfo {year} {2017})},\ \Eprint
  {https://arxiv.org/abs/1706.03360} {arXiv:1706.03360 [hep-ph]} \BibitemShut
  {NoStop}%
\bibitem [{\citenamefont {Yi}\ \emph {et~al.}(2019)\citenamefont {Yi},
  \citenamefont {Ma}, \citenamefont {Martin},\ and\ \citenamefont
  {Duan}}]{Yi:2019hrp}%
  \BibitemOpen
  \bibfield  {author} {\bibinfo {author} {\bibfnamefont {C.}~\bibnamefont
  {Yi}}, \bibinfo {author} {\bibfnamefont {L.}~\bibnamefont {Ma}}, \bibinfo
  {author} {\bibfnamefont {J.~D.}\ \bibnamefont {Martin}},\ and\ \bibinfo
  {author} {\bibfnamefont {H.}~\bibnamefont {Duan}},\ }\href@noop {} {\bibfield
   {journal} {\bibinfo  {journal} {Phys. Rev.}\ }\textbf {\bibinfo {volume}
  {D99}},\ \bibinfo {pages} {063005} (\bibinfo {year} {2019})},\ \Eprint
  {https://arxiv.org/abs/1901.01546} {arXiv:1901.01546 [hep-ph]} \BibitemShut
  {NoStop}%
\bibitem [{\citenamefont {{Morinaga}}(2021)}]{Morinaga2021f}%
  \BibitemOpen
  \bibfield  {author} {\bibinfo {author} {\bibfnamefont {T.}~\bibnamefont
  {{Morinaga}}},\ }\href@noop {} {\bibfield  {journal} {\bibinfo  {journal}
  {arXiv e-prints}\ ,\ \bibinfo {eid} {arXiv:2103.15267}} (\bibinfo {year}
  {2021})},\ \Eprint {https://arxiv.org/abs/2103.15267} {arXiv:2103.15267
  [hep-ph]} \BibitemShut {NoStop}%
\bibitem [{\citenamefont {Dasgupta}(2022)}]{Dasgupta:2021gfs}%
  \BibitemOpen
  \bibfield  {author} {\bibinfo {author} {\bibfnamefont {B.}~\bibnamefont
  {Dasgupta}},\ }\href {https://doi.org/10.1103/PhysRevLett.128.081102}
  {\bibfield  {journal} {\bibinfo  {journal} {Phys. Rev. Lett.}\ }\textbf
  {\bibinfo {volume} {128}},\ \bibinfo {pages} {081102} (\bibinfo {year}
  {2022})},\ \Eprint {https://arxiv.org/abs/2110.00192} {arXiv:2110.00192
  [hep-ph]} \BibitemShut {NoStop}%
\bibitem [{\citenamefont {{Wu}}\ and\ \citenamefont
  {{Tamborra}}(2017)}]{Wu2017l}%
  \BibitemOpen
  \bibfield  {author} {\bibinfo {author} {\bibfnamefont {M.-R.}\ \bibnamefont
  {{Wu}}}\ and\ \bibinfo {author} {\bibfnamefont {I.}~\bibnamefont
  {{Tamborra}}},\ }\href {https://doi.org/10.1103/PhysRevD.95.103007}
  {\bibfield  {journal} {\bibinfo  {journal} {\prd}\ }\textbf {\bibinfo
  {volume} {95}},\ \bibinfo {eid} {103007} (\bibinfo {year} {2017})},\ \Eprint
  {https://arxiv.org/abs/1701.06580} {arXiv:1701.06580 [astro-ph.HE]}
  \BibitemShut {NoStop}%
\bibitem [{\citenamefont {{Wu}}\ \emph {et~al.}(2017)\citenamefont {{Wu}},
  \citenamefont {{Tamborra}}, \citenamefont {{Just}},\ and\ \citenamefont
  {{Janka}}}]{Wu2017a}%
  \BibitemOpen
  \bibfield  {author} {\bibinfo {author} {\bibfnamefont {M.-R.}\ \bibnamefont
  {{Wu}}}, \bibinfo {author} {\bibfnamefont {I.}~\bibnamefont {{Tamborra}}},
  \bibinfo {author} {\bibfnamefont {O.}~\bibnamefont {{Just}}},\ and\ \bibinfo
  {author} {\bibfnamefont {H.-T.}\ \bibnamefont {{Janka}}},\ }\href@noop {}
  {\bibfield  {journal} {\bibinfo  {journal} {\prd}\ }\textbf {\bibinfo
  {volume} {96}},\ \bibinfo {eid} {123015} (\bibinfo {year} {2017})},\ \Eprint
  {https://arxiv.org/abs/1711.00477} {arXiv:1711.00477 [astro-ph.HE]}
  \BibitemShut {NoStop}%
\bibitem [{\citenamefont {Tamborra}\ \emph {et~al.}(2017)\citenamefont
  {Tamborra}, \citenamefont {Huedepohl}, \citenamefont {Raffelt},\ and\
  \citenamefont {Janka}}]{Tamborra:2017ubu}%
  \BibitemOpen
  \bibfield  {author} {\bibinfo {author} {\bibfnamefont {I.}~\bibnamefont
  {Tamborra}}, \bibinfo {author} {\bibfnamefont {L.}~\bibnamefont {Huedepohl}},
  \bibinfo {author} {\bibfnamefont {G.}~\bibnamefont {Raffelt}},\ and\ \bibinfo
  {author} {\bibfnamefont {H.-T.}\ \bibnamefont {Janka}},\ }\href
  {https://doi.org/10.3847/1538-4357/aa6a18} {\bibfield  {journal} {\bibinfo
  {journal} {Astrophys. J.}\ }\textbf {\bibinfo {volume} {839}},\ \bibinfo
  {pages} {132} (\bibinfo {year} {2017})},\ \Eprint
  {https://arxiv.org/abs/1702.00060} {arXiv:1702.00060 [astro-ph.HE]}
  \BibitemShut {NoStop}%
\bibitem [{\citenamefont {Abbar}\ \emph {et~al.}(2019)\citenamefont {Abbar},
  \citenamefont {Duan}, \citenamefont {Sumiyoshi}, \citenamefont {Takiwaki},\
  and\ \citenamefont {Volpe}}]{Abbar:2018shq}%
  \BibitemOpen
  \bibfield  {author} {\bibinfo {author} {\bibfnamefont {S.}~\bibnamefont
  {Abbar}}, \bibinfo {author} {\bibfnamefont {H.}~\bibnamefont {Duan}},
  \bibinfo {author} {\bibfnamefont {K.}~\bibnamefont {Sumiyoshi}}, \bibinfo
  {author} {\bibfnamefont {T.}~\bibnamefont {Takiwaki}},\ and\ \bibinfo
  {author} {\bibfnamefont {M.~C.}\ \bibnamefont {Volpe}},\ }\href
  {https://doi.org/10.1103/PhysRevD.100.043004} {\bibfield  {journal} {\bibinfo
   {journal} {Phys. Rev. D}\ }\textbf {\bibinfo {volume} {100}},\ \bibinfo
  {pages} {043004} (\bibinfo {year} {2019})},\ \Eprint
  {https://arxiv.org/abs/1812.06883} {arXiv:1812.06883 [astro-ph.HE]}
  \BibitemShut {NoStop}%
\bibitem [{\citenamefont {Delfan~Azari}\ \emph {et~al.}(2019)\citenamefont
  {Delfan~Azari}, \citenamefont {Yamada}, \citenamefont {Morinaga},
  \citenamefont {Iwakami}, \citenamefont {Okawa}, \citenamefont {Nagakura},\
  and\ \citenamefont {Sumiyoshi}}]{DelfanAzari:2019epo}%
  \BibitemOpen
  \bibfield  {author} {\bibinfo {author} {\bibfnamefont {M.}~\bibnamefont
  {Delfan~Azari}}, \bibinfo {author} {\bibfnamefont {S.}~\bibnamefont
  {Yamada}}, \bibinfo {author} {\bibfnamefont {T.}~\bibnamefont {Morinaga}},
  \bibinfo {author} {\bibfnamefont {W.}~\bibnamefont {Iwakami}}, \bibinfo
  {author} {\bibfnamefont {H.}~\bibnamefont {Okawa}}, \bibinfo {author}
  {\bibfnamefont {H.}~\bibnamefont {Nagakura}},\ and\ \bibinfo {author}
  {\bibfnamefont {K.}~\bibnamefont {Sumiyoshi}},\ }\href
  {https://doi.org/10.1103/PhysRevD.99.103011} {\bibfield  {journal} {\bibinfo
  {journal} {Phys. Rev. D}\ }\textbf {\bibinfo {volume} {99}},\ \bibinfo
  {pages} {103011} (\bibinfo {year} {2019})},\ \Eprint
  {https://arxiv.org/abs/1902.07467} {arXiv:1902.07467 [astro-ph.HE]}
  \BibitemShut {NoStop}%
\bibitem [{\citenamefont {Morinaga}\ \emph {et~al.}(2020)\citenamefont
  {Morinaga}, \citenamefont {Nagakura}, \citenamefont {Kato},\ and\
  \citenamefont {Yamada}}]{Morinaga:2019wsv}%
  \BibitemOpen
  \bibfield  {author} {\bibinfo {author} {\bibfnamefont {T.}~\bibnamefont
  {Morinaga}}, \bibinfo {author} {\bibfnamefont {H.}~\bibnamefont {Nagakura}},
  \bibinfo {author} {\bibfnamefont {C.}~\bibnamefont {Kato}},\ and\ \bibinfo
  {author} {\bibfnamefont {S.}~\bibnamefont {Yamada}},\ }\href
  {https://doi.org/10.1103/PhysRevResearch.2.012046} {\bibfield  {journal}
  {\bibinfo  {journal} {Phys. Rev. Res.}\ }\textbf {\bibinfo {volume} {2}},\
  \bibinfo {pages} {012046} (\bibinfo {year} {2020})},\ \Eprint
  {https://arxiv.org/abs/1909.13131} {arXiv:1909.13131 [astro-ph.HE]}
  \BibitemShut {NoStop}%
\bibitem [{\citenamefont {Delfan~Azari}\ \emph {et~al.}(2020)\citenamefont
  {Delfan~Azari}, \citenamefont {Yamada}, \citenamefont {Morinaga},
  \citenamefont {Nagakura}, \citenamefont {Furusawa}, \citenamefont {Harada},
  \citenamefont {Okawa}, \citenamefont {Iwakami},\ and\ \citenamefont
  {Sumiyoshi}}]{DelfanAzari:2019tez}%
  \BibitemOpen
  \bibfield  {author} {\bibinfo {author} {\bibfnamefont {M.}~\bibnamefont
  {Delfan~Azari}}, \bibinfo {author} {\bibfnamefont {S.}~\bibnamefont
  {Yamada}}, \bibinfo {author} {\bibfnamefont {T.}~\bibnamefont {Morinaga}},
  \bibinfo {author} {\bibfnamefont {H.}~\bibnamefont {Nagakura}}, \bibinfo
  {author} {\bibfnamefont {S.}~\bibnamefont {Furusawa}}, \bibinfo {author}
  {\bibfnamefont {A.}~\bibnamefont {Harada}}, \bibinfo {author} {\bibfnamefont
  {H.}~\bibnamefont {Okawa}}, \bibinfo {author} {\bibfnamefont
  {W.}~\bibnamefont {Iwakami}},\ and\ \bibinfo {author} {\bibfnamefont
  {K.}~\bibnamefont {Sumiyoshi}},\ }\href
  {https://doi.org/10.1103/PhysRevD.101.023018} {\bibfield  {journal} {\bibinfo
   {journal} {Phys. Rev. D}\ }\textbf {\bibinfo {volume} {101}},\ \bibinfo
  {pages} {023018} (\bibinfo {year} {2020})},\ \Eprint
  {https://arxiv.org/abs/1910.06176} {arXiv:1910.06176 [astro-ph.HE]}
  \BibitemShut {NoStop}%
\bibitem [{\citenamefont {Nagakura}\ \emph {et~al.}(2019)\citenamefont
  {Nagakura}, \citenamefont {Sumiyoshi},\ and\ \citenamefont
  {Yamada}}]{Nagakura:2019sig}%
  \BibitemOpen
  \bibfield  {author} {\bibinfo {author} {\bibfnamefont {H.}~\bibnamefont
  {Nagakura}}, \bibinfo {author} {\bibfnamefont {K.}~\bibnamefont
  {Sumiyoshi}},\ and\ \bibinfo {author} {\bibfnamefont {S.}~\bibnamefont
  {Yamada}},\ }\href@noop {} {\bibfield  {journal} {\bibinfo  {journal}
  {Astrophys. J. Lett.}\ }\textbf {\bibinfo {volume} {880}},\ \bibinfo {pages}
  {L28} (\bibinfo {year} {2019})},\ \Eprint {https://arxiv.org/abs/1907.04863}
  {arXiv:1907.04863 [astro-ph.HE]} \BibitemShut {NoStop}%
\bibitem [{\citenamefont {{Abbar}}\ \emph {et~al.}(2019)\citenamefont
  {{Abbar}}, \citenamefont {{Duan}}, \citenamefont {{Sumiyoshi}}, \citenamefont
  {{Takiwaki}},\ and\ \citenamefont {{Volpe}}}]{Abbar2019a}%
  \BibitemOpen
  \bibfield  {author} {\bibinfo {author} {\bibfnamefont {S.}~\bibnamefont
  {{Abbar}}}, \bibinfo {author} {\bibfnamefont {H.}~\bibnamefont {{Duan}}},
  \bibinfo {author} {\bibfnamefont {K.}~\bibnamefont {{Sumiyoshi}}}, \bibinfo
  {author} {\bibfnamefont {T.}~\bibnamefont {{Takiwaki}}},\ and\ \bibinfo
  {author} {\bibfnamefont {M.~C.}\ \bibnamefont {{Volpe}}},\ }\href@noop {}
  {\bibfield  {journal} {\bibinfo  {journal} {\prd}\ }\textbf {\bibinfo
  {volume} {100}},\ \bibinfo {eid} {043004} (\bibinfo {year} {2019})},\ \Eprint
  {https://arxiv.org/abs/1812.06883} {arXiv:1812.06883 [astro-ph.HE]}
  \BibitemShut {NoStop}%
\bibitem [{\citenamefont {Glas}\ \emph {et~al.}(2020)\citenamefont {Glas},
  \citenamefont {Janka}, \citenamefont {Capozzi}, \citenamefont {Sen},
  \citenamefont {Dasgupta}, \citenamefont {Mirizzi},\ and\ \citenamefont
  {Sigl}}]{Glas:2019ijo}%
  \BibitemOpen
  \bibfield  {author} {\bibinfo {author} {\bibfnamefont {R.}~\bibnamefont
  {Glas}}, \bibinfo {author} {\bibfnamefont {H.~T.}\ \bibnamefont {Janka}},
  \bibinfo {author} {\bibfnamefont {F.}~\bibnamefont {Capozzi}}, \bibinfo
  {author} {\bibfnamefont {M.}~\bibnamefont {Sen}}, \bibinfo {author}
  {\bibfnamefont {B.}~\bibnamefont {Dasgupta}}, \bibinfo {author}
  {\bibfnamefont {A.}~\bibnamefont {Mirizzi}},\ and\ \bibinfo {author}
  {\bibfnamefont {G.}~\bibnamefont {Sigl}},\ }\href
  {https://doi.org/10.1103/PhysRevD.101.063001} {\bibfield  {journal} {\bibinfo
   {journal} {Phys. Rev. D}\ }\textbf {\bibinfo {volume} {101}},\ \bibinfo
  {pages} {063001} (\bibinfo {year} {2020})},\ \Eprint
  {https://arxiv.org/abs/1912.00274} {arXiv:1912.00274 [astro-ph.HE]}
  \BibitemShut {NoStop}%
\bibitem [{\citenamefont {{Abbar}}\ \emph {et~al.}(2021)\citenamefont
  {{Abbar}}, \citenamefont {{Capozzi}}, \citenamefont {{Glas}}, \citenamefont
  {{Janka}},\ and\ \citenamefont {{Tamborra}}}]{Abbar2021r}%
  \BibitemOpen
  \bibfield  {author} {\bibinfo {author} {\bibfnamefont {S.}~\bibnamefont
  {{Abbar}}}, \bibinfo {author} {\bibfnamefont {F.}~\bibnamefont {{Capozzi}}},
  \bibinfo {author} {\bibfnamefont {R.}~\bibnamefont {{Glas}}}, \bibinfo
  {author} {\bibfnamefont {H.~T.}\ \bibnamefont {{Janka}}},\ and\ \bibinfo
  {author} {\bibfnamefont {I.}~\bibnamefont {{Tamborra}}},\ }\href@noop {}
  {\bibfield  {journal} {\bibinfo  {journal} {\prd}\ }\textbf {\bibinfo
  {volume} {103}},\ \bibinfo {eid} {063033} (\bibinfo {year} {2021})},\ \Eprint
  {https://arxiv.org/abs/2012.06594} {arXiv:2012.06594 [astro-ph.HE]}
  \BibitemShut {NoStop}%
\bibitem [{\citenamefont {Capozzi}\ \emph {et~al.}(2021)\citenamefont
  {Capozzi}, \citenamefont {Abbar}, \citenamefont {Bollig},\ and\ \citenamefont
  {Janka}}]{Capozzi2021b}%
  \BibitemOpen
  \bibfield  {author} {\bibinfo {author} {\bibfnamefont {F.}~\bibnamefont
  {Capozzi}}, \bibinfo {author} {\bibfnamefont {S.}~\bibnamefont {Abbar}},
  \bibinfo {author} {\bibfnamefont {R.}~\bibnamefont {Bollig}},\ and\ \bibinfo
  {author} {\bibfnamefont {H.-T.}\ \bibnamefont {Janka}},\ }\href
  {https://doi.org/10.1103/PhysRevD.103.063013} {\bibfield  {journal} {\bibinfo
   {journal} {Phys. Rev. D}\ }\textbf {\bibinfo {volume} {103}},\ \bibinfo
  {pages} {063013} (\bibinfo {year} {2021})}\BibitemShut {NoStop}%
\bibitem [{\citenamefont {{Nagakura}}\ \emph {et~al.}(2021)\citenamefont
  {{Nagakura}}, \citenamefont {{Burrows}}, \citenamefont {{Johns}},\ and\
  \citenamefont {{Fuller}}}]{Nagakura2021r}%
  \BibitemOpen
  \bibfield  {author} {\bibinfo {author} {\bibfnamefont {H.}~\bibnamefont
  {{Nagakura}}}, \bibinfo {author} {\bibfnamefont {A.}~\bibnamefont
  {{Burrows}}}, \bibinfo {author} {\bibfnamefont {L.}~\bibnamefont {{Johns}}},\
  and\ \bibinfo {author} {\bibfnamefont {G.~M.}\ \bibnamefont {{Fuller}}},\
  }\href@noop {} {\bibfield  {journal} {\bibinfo  {journal} {\prd}\ }\textbf
  {\bibinfo {volume} {104}},\ \bibinfo {eid} {083025} (\bibinfo {year}
  {2021})},\ \Eprint {https://arxiv.org/abs/2108.07281} {arXiv:2108.07281
  [astro-ph.HE]} \BibitemShut {NoStop}%
\bibitem [{\citenamefont {{Harada}}\ and\ \citenamefont
  {{Nagakura}}(2021)}]{Harada2021h}%
  \BibitemOpen
  \bibfield  {author} {\bibinfo {author} {\bibfnamefont {A.}~\bibnamefont
  {{Harada}}}\ and\ \bibinfo {author} {\bibfnamefont {H.}~\bibnamefont
  {{Nagakura}}},\ }\href@noop {} {\bibfield  {journal} {\bibinfo  {journal}
  {arXiv e-prints}\ ,\ \bibinfo {eid} {arXiv:2110.08291}} (\bibinfo {year}
  {2021})},\ \Eprint {https://arxiv.org/abs/2110.08291} {arXiv:2110.08291
  [astro-ph.HE]} \BibitemShut {NoStop}%
\bibitem [{\citenamefont {{Li}}\ and\ \citenamefont
  {{Siegel}}(2021)}]{Li2021g}%
  \BibitemOpen
  \bibfield  {author} {\bibinfo {author} {\bibfnamefont {X.}~\bibnamefont
  {{Li}}}\ and\ \bibinfo {author} {\bibfnamefont {D.~M.}\ \bibnamefont
  {{Siegel}}},\ }\href {https://doi.org/10.1103/PhysRevLett.126.251101}
  {\bibfield  {journal} {\bibinfo  {journal} {\prl}\ }\textbf {\bibinfo
  {volume} {126}},\ \bibinfo {eid} {251101} (\bibinfo {year} {2021})},\ \Eprint
  {https://arxiv.org/abs/2103.02616} {arXiv:2103.02616 [astro-ph.HE]}
  \BibitemShut {NoStop}%
\bibitem [{\citenamefont {George}\ \emph {et~al.}(2020)\citenamefont {George},
  \citenamefont {Wu}, \citenamefont {Tamborra}, \citenamefont
  {Ardevol-Pulpillo},\ and\ \citenamefont {Janka}}]{George:2020veu}%
  \BibitemOpen
  \bibfield  {author} {\bibinfo {author} {\bibfnamefont {M.}~\bibnamefont
  {George}}, \bibinfo {author} {\bibfnamefont {M.-R.}\ \bibnamefont {Wu}},
  \bibinfo {author} {\bibfnamefont {I.}~\bibnamefont {Tamborra}}, \bibinfo
  {author} {\bibfnamefont {R.}~\bibnamefont {Ardevol-Pulpillo}},\ and\ \bibinfo
  {author} {\bibfnamefont {H.-T.}\ \bibnamefont {Janka}},\ }\href@noop {}
  {\bibfield  {journal} {\bibinfo  {journal} {Phys. Rev. D}\ }\textbf {\bibinfo
  {volume} {102}},\ \bibinfo {pages} {103015} (\bibinfo {year} {2020})},\
  \Eprint {https://arxiv.org/abs/2009.04046} {arXiv:2009.04046 [astro-ph.HE]}
  \BibitemShut {NoStop}%
\bibitem [{\citenamefont {{Abbar}}(2020)}]{Abbar2020m}%
  \BibitemOpen
  \bibfield  {author} {\bibinfo {author} {\bibfnamefont {S.}~\bibnamefont
  {{Abbar}}},\ }\href@noop {} {\bibfield  {journal} {\bibinfo  {journal}
  {\jcap}\ }\textbf {\bibinfo {volume} {2020}},\ \bibinfo {eid} {027} (\bibinfo
  {year} {2020})},\ \Eprint {https://arxiv.org/abs/2003.00969}
  {arXiv:2003.00969 [astro-ph.HE]} \BibitemShut {NoStop}%
\bibitem [{\citenamefont {Obergaulinger}(2008)}]{Obergaulinger2008a}%
  \BibitemOpen
  \bibfield  {author} {\bibinfo {author} {\bibfnamefont {M.}~\bibnamefont
  {Obergaulinger}},\ }\emph {\bibinfo {title} {Astrophysical
  magnetohydrodynamics and radiative transfer}},\ \href@noop {} {\bibinfo
  {type} {Dissertation}},\ \bibinfo  {school} {Technische Universit{\"a}t
  M{\"u}nchen}, \bibinfo {address} {M{\"u}nchen} (\bibinfo {year}
  {2008})\BibitemShut {NoStop}%
\bibitem [{\citenamefont {{Just}}\ \emph
  {et~al.}(2015{\natexlab{b}})\citenamefont {{Just}}, \citenamefont
  {{Obergaulinger}},\ and\ \citenamefont {{Janka}}}]{Just2015b}%
  \BibitemOpen
  \bibfield  {author} {\bibinfo {author} {\bibfnamefont {O.}~\bibnamefont
  {{Just}}}, \bibinfo {author} {\bibfnamefont {M.}~\bibnamefont
  {{Obergaulinger}}},\ and\ \bibinfo {author} {\bibfnamefont {H.-T.}\
  \bibnamefont {{Janka}}},\ }\href@noop {} {\bibfield  {journal} {\bibinfo
  {journal} {\mnras}\ }\textbf {\bibinfo {volume} {453}},\ \bibinfo {pages}
  {3386} (\bibinfo {year} {2015}{\natexlab{b}})},\ \Eprint
  {https://arxiv.org/abs/1501.02999} {arXiv:1501.02999 [astro-ph.HE]}
  \BibitemShut {NoStop}%
\bibitem [{\citenamefont {{Chakraborty}}\ and\ \citenamefont
  {{Chakraborty}}(2020)}]{Chakraborty:2019wxe}%
  \BibitemOpen
  \bibfield  {author} {\bibinfo {author} {\bibfnamefont {M.}~\bibnamefont
  {{Chakraborty}}}\ and\ \bibinfo {author} {\bibfnamefont {S.}~\bibnamefont
  {{Chakraborty}}},\ }\href {https://doi.org/10.1088/1475-7516/2020/01/005}
  {\bibfield  {journal} {\bibinfo  {journal} {\jcap}\ }\textbf {\bibinfo
  {volume} {2020}},\ \bibinfo {eid} {005} (\bibinfo {year} {2020})},\ \Eprint
  {https://arxiv.org/abs/1909.10420} {arXiv:1909.10420 [hep-ph]} \BibitemShut
  {NoStop}%
\bibitem [{\citenamefont {Shalgar}\ and\ \citenamefont
  {Tamborra}(2021{\natexlab{b}})}]{Shalgar:2021wlj}%
  \BibitemOpen
  \bibfield  {author} {\bibinfo {author} {\bibfnamefont {S.}~\bibnamefont
  {Shalgar}}\ and\ \bibinfo {author} {\bibfnamefont {I.}~\bibnamefont
  {Tamborra}},\ }\href {https://doi.org/10.1103/PhysRevD.104.023011} {\bibfield
   {journal} {\bibinfo  {journal} {Phys. Rev. D}\ }\textbf {\bibinfo {volume}
  {104}},\ \bibinfo {pages} {023011} (\bibinfo {year} {2021}{\natexlab{b}})},\
  \Eprint {https://arxiv.org/abs/2103.12743} {arXiv:2103.12743 [hep-ph]}
  \BibitemShut {NoStop}%
\bibitem [{\citenamefont {Capozzi}\ \emph {et~al.}(2020)\citenamefont
  {Capozzi}, \citenamefont {Chakraborty}, \citenamefont {Chakraborty},\ and\
  \citenamefont {Sen}}]{Capozzi:2020kge}%
  \BibitemOpen
  \bibfield  {author} {\bibinfo {author} {\bibfnamefont {F.}~\bibnamefont
  {Capozzi}}, \bibinfo {author} {\bibfnamefont {M.}~\bibnamefont
  {Chakraborty}}, \bibinfo {author} {\bibfnamefont {S.}~\bibnamefont
  {Chakraborty}},\ and\ \bibinfo {author} {\bibfnamefont {M.}~\bibnamefont
  {Sen}},\ }\href {https://doi.org/10.1103/PhysRevLett.125.251801} {\bibfield
  {journal} {\bibinfo  {journal} {Phys. Rev. Lett.}\ }\textbf {\bibinfo
  {volume} {125}},\ \bibinfo {pages} {251801} (\bibinfo {year} {2020})},\
  \Eprint {https://arxiv.org/abs/2005.14204} {arXiv:2005.14204 [hep-ph]}
  \BibitemShut {NoStop}%
\bibitem [{\citenamefont {{Bollig}}\ \emph {et~al.}(2017)\citenamefont
  {{Bollig}}, \citenamefont {{Janka}}, \citenamefont {{Lohs}}, \citenamefont
  {{Mart{\'{\i}}nez-Pinedo}}, \citenamefont {{Horowitz}},\ and\ \citenamefont
  {{Melson}}}]{Bollig2017a}%
  \BibitemOpen
  \bibfield  {author} {\bibinfo {author} {\bibfnamefont {R.}~\bibnamefont
  {{Bollig}}}, \bibinfo {author} {\bibfnamefont {H.-T.}\ \bibnamefont
  {{Janka}}}, \bibinfo {author} {\bibfnamefont {A.}~\bibnamefont {{Lohs}}},
  \bibinfo {author} {\bibfnamefont {G.}~\bibnamefont
  {{Mart{\'{\i}}nez-Pinedo}}}, \bibinfo {author} {\bibfnamefont {C.~J.}\
  \bibnamefont {{Horowitz}}},\ and\ \bibinfo {author} {\bibfnamefont
  {T.}~\bibnamefont {{Melson}}},\ }\href@noop {} {\bibfield  {journal}
  {\bibinfo  {journal} {Physical Review Letters}\ }\textbf {\bibinfo {volume}
  {119}},\ \bibinfo {eid} {242702} (\bibinfo {year} {2017})},\ \Eprint
  {https://arxiv.org/abs/1706.04630} {arXiv:1706.04630 [astro-ph.HE]}
  \BibitemShut {NoStop}%
\bibitem [{\citenamefont {Dasgupta}\ \emph {et~al.}(2018)\citenamefont
  {Dasgupta}, \citenamefont {Mirizzi},\ and\ \citenamefont
  {Sen}}]{Dasgupta:2018ulw}%
  \BibitemOpen
  \bibfield  {author} {\bibinfo {author} {\bibfnamefont {B.}~\bibnamefont
  {Dasgupta}}, \bibinfo {author} {\bibfnamefont {A.}~\bibnamefont {Mirizzi}},\
  and\ \bibinfo {author} {\bibfnamefont {M.}~\bibnamefont {Sen}},\ }\href
  {https://doi.org/10.1103/PhysRevD.98.103001} {\bibfield  {journal} {\bibinfo
  {journal} {Phys. Rev. D}\ }\textbf {\bibinfo {volume} {98}},\ \bibinfo
  {pages} {103001} (\bibinfo {year} {2018})},\ \Eprint
  {https://arxiv.org/abs/1807.03322} {arXiv:1807.03322 [hep-ph]} \BibitemShut
  {NoStop}%
\bibitem [{\citenamefont {{Nagakura}}\ and\ \citenamefont
  {{Johns}}(2021)}]{Nagakura2021a}%
  \BibitemOpen
  \bibfield  {author} {\bibinfo {author} {\bibfnamefont {H.}~\bibnamefont
  {{Nagakura}}}\ and\ \bibinfo {author} {\bibfnamefont {L.}~\bibnamefont
  {{Johns}}},\ }\href@noop {} {\bibfield  {journal} {\bibinfo  {journal} {arXiv
  e-prints}\ ,\ \bibinfo {eid} {arXiv:2104.05729}} (\bibinfo {year} {2021})},\
  \Eprint {https://arxiv.org/abs/2104.05729} {arXiv:2104.05729 [astro-ph.HE]}
  \BibitemShut {NoStop}%
\bibitem [{\citenamefont {Nagakura}\ and\ \citenamefont
  {Johns}(2021)}]{Nagakura:2021suv}%
  \BibitemOpen
  \bibfield  {author} {\bibinfo {author} {\bibfnamefont {H.}~\bibnamefont
  {Nagakura}}\ and\ \bibinfo {author} {\bibfnamefont {L.}~\bibnamefont
  {Johns}},\ }\href {https://doi.org/10.1103/PhysRevD.104.063014} {\bibfield
  {journal} {\bibinfo  {journal} {Phys. Rev. D}\ }\textbf {\bibinfo {volume}
  {104}},\ \bibinfo {pages} {063014} (\bibinfo {year} {2021})},\ \Eprint
  {https://arxiv.org/abs/2106.02650} {arXiv:2106.02650 [astro-ph.HE]}
  \BibitemShut {NoStop}%
\bibitem [{\citenamefont {{Johns}}\ and\ \citenamefont
  {{Nagakura}}(2021)}]{Johns2021w}%
  \BibitemOpen
  \bibfield  {author} {\bibinfo {author} {\bibfnamefont {L.}~\bibnamefont
  {{Johns}}}\ and\ \bibinfo {author} {\bibfnamefont {H.}~\bibnamefont
  {{Nagakura}}},\ }\href@noop {} {\bibfield  {journal} {\bibinfo  {journal}
  {\prd}\ }\textbf {\bibinfo {volume} {103}},\ \bibinfo {eid} {123012}
  (\bibinfo {year} {2021})},\ \Eprint {https://arxiv.org/abs/2104.04106}
  {arXiv:2104.04106 [hep-ph]} \BibitemShut {NoStop}%
\bibitem [{\citenamefont {Bhattacharyya}\ and\ \citenamefont
  {Dasgupta}(2021)}]{Bhattacharyya:2020jpj}%
  \BibitemOpen
  \bibfield  {author} {\bibinfo {author} {\bibfnamefont {S.}~\bibnamefont
  {Bhattacharyya}}\ and\ \bibinfo {author} {\bibfnamefont {B.}~\bibnamefont
  {Dasgupta}},\ }\href {https://doi.org/10.1103/PhysRevLett.126.061302}
  {\bibfield  {journal} {\bibinfo  {journal} {Phys. Rev. Lett.}\ }\textbf
  {\bibinfo {volume} {126}},\ \bibinfo {pages} {061302} (\bibinfo {year}
  {2021})},\ \Eprint {https://arxiv.org/abs/2009.03337} {arXiv:2009.03337
  [hep-ph]} \BibitemShut {NoStop}%
\bibitem [{\citenamefont {{Wu}}\ \emph {et~al.}(2021)\citenamefont {{Wu}},
  \citenamefont {{George}}, \citenamefont {{Lin}},\ and\ \citenamefont
  {{Xiong}}}]{Wu2021q}%
  \BibitemOpen
  \bibfield  {author} {\bibinfo {author} {\bibfnamefont {M.-R.}\ \bibnamefont
  {{Wu}}}, \bibinfo {author} {\bibfnamefont {M.}~\bibnamefont {{George}}},
  \bibinfo {author} {\bibfnamefont {C.-Y.}\ \bibnamefont {{Lin}}},\ and\
  \bibinfo {author} {\bibfnamefont {Z.}~\bibnamefont {{Xiong}}},\ }\href@noop
  {} {\bibfield  {journal} {\bibinfo  {journal} {\prd}\ }\textbf {\bibinfo
  {volume} {104}},\ \bibinfo {eid} {103003} (\bibinfo {year} {2021})},\ \Eprint
  {https://arxiv.org/abs/2108.09886} {arXiv:2108.09886 [hep-ph]} \BibitemShut
  {NoStop}%
\bibitem [{\citenamefont {{Richers}}\ \emph {et~al.}(2021)\citenamefont
  {{Richers}}, \citenamefont {{Willcox}}, \citenamefont {{Ford}},\ and\
  \citenamefont {{Myers}}}]{Richers2021j}%
  \BibitemOpen
  \bibfield  {author} {\bibinfo {author} {\bibfnamefont {S.}~\bibnamefont
  {{Richers}}}, \bibinfo {author} {\bibfnamefont {D.~E.}\ \bibnamefont
  {{Willcox}}}, \bibinfo {author} {\bibfnamefont {N.~M.}\ \bibnamefont
  {{Ford}}},\ and\ \bibinfo {author} {\bibfnamefont {A.}~\bibnamefont
  {{Myers}}},\ }\href {https://doi.org/10.1103/PhysRevD.103.083013} {\bibfield
  {journal} {\bibinfo  {journal} {\prd}\ }\textbf {\bibinfo {volume} {103}},\
  \bibinfo {eid} {083013} (\bibinfo {year} {2021})},\ \Eprint
  {https://arxiv.org/abs/2101.02745} {arXiv:2101.02745 [astro-ph.HE]}
  \BibitemShut {NoStop}%
\bibitem [{\citenamefont {{Xiong}}\ and\ \citenamefont
  {{Qian}}(2021)}]{Xiong2021o}%
  \BibitemOpen
  \bibfield  {author} {\bibinfo {author} {\bibfnamefont {Z.}~\bibnamefont
  {{Xiong}}}\ and\ \bibinfo {author} {\bibfnamefont {Y.-Z.}\ \bibnamefont
  {{Qian}}},\ }\href@noop {} {\bibfield  {journal} {\bibinfo  {journal}
  {Physics Letters B}\ }\textbf {\bibinfo {volume} {820}},\ \bibinfo {eid}
  {136550} (\bibinfo {year} {2021})},\ \Eprint
  {https://arxiv.org/abs/2104.05618} {arXiv:2104.05618 [astro-ph.HE]}
  \BibitemShut {NoStop}%
\bibitem [{\citenamefont {Padilla-Gay}\ \emph {et~al.}(2022)\citenamefont
  {Padilla-Gay}, \citenamefont {Tamborra},\ and\ \citenamefont
  {Raffelt}}]{Padilla-Gay:2021haz}%
  \BibitemOpen
  \bibfield  {author} {\bibinfo {author} {\bibfnamefont {I.}~\bibnamefont
  {Padilla-Gay}}, \bibinfo {author} {\bibfnamefont {I.}~\bibnamefont
  {Tamborra}},\ and\ \bibinfo {author} {\bibfnamefont {G.~G.}\ \bibnamefont
  {Raffelt}},\ }\href {https://doi.org/10.1103/PhysRevLett.128.121102}
  {\bibfield  {journal} {\bibinfo  {journal} {Phys. Rev. Lett.}\ }\textbf
  {\bibinfo {volume} {128}},\ \bibinfo {pages} {121102} (\bibinfo {year}
  {2022})},\ \Eprint {https://arxiv.org/abs/2109.14627} {arXiv:2109.14627
  [astro-ph.HE]} \BibitemShut {NoStop}%
\bibitem [{\citenamefont {Padilla-Gay}\ \emph {et~al.}(2021)\citenamefont
  {Padilla-Gay}, \citenamefont {Shalgar},\ and\ \citenamefont
  {Tamborra}}]{Padilla-Gay2021y}%
  \BibitemOpen
  \bibfield  {author} {\bibinfo {author} {\bibfnamefont {I.}~\bibnamefont
  {Padilla-Gay}}, \bibinfo {author} {\bibfnamefont {S.}~\bibnamefont
  {Shalgar}},\ and\ \bibinfo {author} {\bibfnamefont {I.}~\bibnamefont
  {Tamborra}},\ }\href@noop {} {\bibfield  {journal} {\bibinfo  {journal}
  {Journal of Cosmology and Astroparticle Physics}\ }\textbf {\bibinfo {volume}
  {2021}}\bibinfo  {number} { (01)},\ \bibinfo {pages} {017}}\BibitemShut
  {NoStop}%
\bibitem [{\citenamefont {Duan}\ \emph {et~al.}(2021)\citenamefont {Duan},
  \citenamefont {Martin},\ and\ \citenamefont {Omanakuttan}}]{Duan:2021woc}%
  \BibitemOpen
\bibfield  {number} {  }\bibfield  {author} {\bibinfo {author} {\bibfnamefont
  {H.}~\bibnamefont {Duan}}, \bibinfo {author} {\bibfnamefont {J.~D.}\
  \bibnamefont {Martin}},\ and\ \bibinfo {author} {\bibfnamefont
  {S.}~\bibnamefont {Omanakuttan}},\ }\href
  {https://doi.org/10.1103/PhysRevD.104.123026} {\bibfield  {journal} {\bibinfo
   {journal} {Phys. Rev. D}\ }\textbf {\bibinfo {volume} {104}},\ \bibinfo
  {pages} {123026} (\bibinfo {year} {2021})},\ \Eprint
  {https://arxiv.org/abs/2110.02286} {arXiv:2110.02286 [hep-ph]} \BibitemShut
  {NoStop}%
\bibitem [{\citenamefont {{Horowitz}}(2002)}]{Horowitz2002a}%
  \BibitemOpen
  \bibfield  {author} {\bibinfo {author} {\bibfnamefont {C.~J.}\ \bibnamefont
  {{Horowitz}}},\ }\href@noop {} {\bibfield  {journal} {\bibinfo  {journal}
  {\prd}\ }\textbf {\bibinfo {volume} {65}},\ \bibinfo {eid} {043001} (\bibinfo
  {year} {2002})},\ \Eprint {https://arxiv.org/abs/astro-ph/0109209}
  {astro-ph/0109209} \BibitemShut {NoStop}%
\bibitem [{\citenamefont {{Bruenn}}(1985)}]{Bruenn1985}%
  \BibitemOpen
  \bibfield  {author} {\bibinfo {author} {\bibfnamefont {S.~W.}\ \bibnamefont
  {{Bruenn}}},\ }\href@noop {} {\bibfield  {journal} {\bibinfo  {journal}
  {\apjs}\ }\textbf {\bibinfo {volume} {58}},\ \bibinfo {pages} {771} (\bibinfo
  {year} {1985})}\BibitemShut {NoStop}%
\bibitem [{\citenamefont {{Pons}}\ \emph {et~al.}(1998)\citenamefont {{Pons}},
  \citenamefont {{Miralles}},\ and\ \citenamefont {{Ibanez}}}]{Pons1998}%
  \BibitemOpen
  \bibfield  {author} {\bibinfo {author} {\bibfnamefont {J.~A.}\ \bibnamefont
  {{Pons}}}, \bibinfo {author} {\bibfnamefont {J.~A.}\ \bibnamefont
  {{Miralles}}},\ and\ \bibinfo {author} {\bibfnamefont {J.~M.~A.}\
  \bibnamefont {{Ibanez}}},\ }\href@noop {} {\bibfield  {journal} {\bibinfo
  {journal} {\aaps}\ }\textbf {\bibinfo {volume} {129}},\ \bibinfo {pages}
  {343} (\bibinfo {year} {1998})},\ \Eprint
  {https://arxiv.org/abs/arXiv:astro-ph/9802333} {arXiv:astro-ph/9802333}
  \BibitemShut {NoStop}%
\bibitem [{\citenamefont {{Hannestad}}\ and\ \citenamefont
  {{Raffelt}}(1998)}]{Hannestad1998}%
  \BibitemOpen
  \bibfield  {author} {\bibinfo {author} {\bibfnamefont {S.}~\bibnamefont
  {{Hannestad}}}\ and\ \bibinfo {author} {\bibfnamefont {G.}~\bibnamefont
  {{Raffelt}}},\ }\href@noop {} {\bibfield  {journal} {\bibinfo  {journal}
  {\apj}\ }\textbf {\bibinfo {volume} {507}},\ \bibinfo {pages} {339} (\bibinfo
  {year} {1998})},\ \Eprint {https://arxiv.org/abs/astro-ph/9711132}
  {astro-ph/9711132} \BibitemShut {NoStop}%
\bibitem [{\citenamefont {O'Connor}(2015)}]{OConnor2015a}%
  \BibitemOpen
  \bibfield  {author} {\bibinfo {author} {\bibfnamefont {E.}~\bibnamefont
  {O'Connor}},\ }\href@noop {} {\bibfield  {journal} {\bibinfo  {journal}
  {\apjs}\ }\textbf {\bibinfo {volume} {219}},\ \bibinfo {eid} {24} (\bibinfo
  {year} {2015})},\ \Eprint {https://arxiv.org/abs/1411.7058} {arXiv:1411.7058
  [astro-ph.HE]} \BibitemShut {NoStop}%
\bibitem [{\citenamefont {{Shalgar}}\ \emph {et~al.}(2020)\citenamefont
  {{Shalgar}}, \citenamefont {{Padilla-Gay}},\ and\ \citenamefont
  {{Tamborra}}}]{Shalgar:2019qwg}%
  \BibitemOpen
  \bibfield  {author} {\bibinfo {author} {\bibfnamefont {S.}~\bibnamefont
  {{Shalgar}}}, \bibinfo {author} {\bibfnamefont {I.}~\bibnamefont
  {{Padilla-Gay}}},\ and\ \bibinfo {author} {\bibfnamefont {I.}~\bibnamefont
  {{Tamborra}}},\ }\href {https://doi.org/10.1088/1475-7516/2020/06/048}
  {\bibfield  {journal} {\bibinfo  {journal} {\jcap}\ }\textbf {\bibinfo
  {volume} {2020}},\ \bibinfo {eid} {048} (\bibinfo {year} {2020})},\ \Eprint
  {https://arxiv.org/abs/1911.09110} {arXiv:1911.09110 [astro-ph.HE]}
  \BibitemShut {NoStop}%
\bibitem [{\citenamefont {Vlasenko}\ \emph {et~al.}(2014)\citenamefont
  {Vlasenko}, \citenamefont {Fuller},\ and\ \citenamefont
  {Cirigliano}}]{Vlasenko:2013fja}%
  \BibitemOpen
  \bibfield  {author} {\bibinfo {author} {\bibfnamefont {A.}~\bibnamefont
  {Vlasenko}}, \bibinfo {author} {\bibfnamefont {G.~M.}\ \bibnamefont
  {Fuller}},\ and\ \bibinfo {author} {\bibfnamefont {V.}~\bibnamefont
  {Cirigliano}},\ }\href@noop {} {\bibfield  {journal} {\bibinfo  {journal}
  {Phys. Rev.}\ }\textbf {\bibinfo {volume} {D89}},\ \bibinfo {pages} {105004}
  (\bibinfo {year} {2014})},\ \Eprint {https://arxiv.org/abs/1309.2628}
  {arXiv:1309.2628 [hep-ph]} \BibitemShut {NoStop}%
\bibitem [{\citenamefont {Volpe}(2015)}]{Volpe:2015rla}%
  \BibitemOpen
  \bibfield  {author} {\bibinfo {author} {\bibfnamefont {C.}~\bibnamefont
  {Volpe}},\ }\href@noop {} {\bibfield  {journal} {\bibinfo  {journal} {Int. J.
  Mod. Phys.}\ }\textbf {\bibinfo {volume} {E24}},\ \bibinfo {pages} {1541009}
  (\bibinfo {year} {2015})},\ \Eprint {https://arxiv.org/abs/1506.06222}
  {arXiv:1506.06222 [astro-ph.SR]} \BibitemShut {NoStop}%
\bibitem [{\citenamefont {Abbar}(2020)}]{Abbar:2020ggq}%
  \BibitemOpen
  \bibfield  {author} {\bibinfo {author} {\bibfnamefont {S.}~\bibnamefont
  {Abbar}},\ }\href {https://doi.org/10.1103/PhysRevD.101.103032} {\bibfield
  {journal} {\bibinfo  {journal} {Phys. Rev. D}\ }\textbf {\bibinfo {volume}
  {101}},\ \bibinfo {pages} {103032} (\bibinfo {year} {2020})},\ \Eprint
  {https://arxiv.org/abs/2001.04876} {arXiv:2001.04876 [astro-ph.HE]}
  \BibitemShut {NoStop}%
\bibitem [{\citenamefont {Sasaki}\ and\ \citenamefont
  {Takiwaki}(2021)}]{Sasaki:2021bvu}%
  \BibitemOpen
  \bibfield  {author} {\bibinfo {author} {\bibfnamefont {H.}~\bibnamefont
  {Sasaki}}\ and\ \bibinfo {author} {\bibfnamefont {T.}~\bibnamefont
  {Takiwaki}},\ }\href {https://doi.org/10.1103/PhysRevD.104.023018} {\bibfield
   {journal} {\bibinfo  {journal} {Phys. Rev. D}\ }\textbf {\bibinfo {volume}
  {104}},\ \bibinfo {pages} {023018} (\bibinfo {year} {2021})},\ \Eprint
  {https://arxiv.org/abs/2106.02181} {arXiv:2106.02181 [hep-ph]} \BibitemShut
  {NoStop}%
\bibitem [{\citenamefont {Kharlanov}\ and\ \citenamefont
  {Shustov}(2021)}]{Kharlanov:2020cti}%
  \BibitemOpen
  \bibfield  {author} {\bibinfo {author} {\bibfnamefont {O.~G.}\ \bibnamefont
  {Kharlanov}}\ and\ \bibinfo {author} {\bibfnamefont {P.~I.}\ \bibnamefont
  {Shustov}},\ }\href {https://doi.org/10.1103/PhysRevD.103.095004} {\bibfield
  {journal} {\bibinfo  {journal} {Phys. Rev. D}\ }\textbf {\bibinfo {volume}
  {103}},\ \bibinfo {pages} {095004} (\bibinfo {year} {2021})},\ \Eprint
  {https://arxiv.org/abs/2010.05329} {arXiv:2010.05329 [hep-ph]} \BibitemShut
  {NoStop}%
\bibitem [{\citenamefont {{Shakura}}\ and\ \citenamefont
  {{Sunyaev}}(1973)}]{Shakura1973}%
  \BibitemOpen
  \bibfield  {author} {\bibinfo {author} {\bibfnamefont {N.~I.}\ \bibnamefont
  {{Shakura}}}\ and\ \bibinfo {author} {\bibfnamefont {R.~A.}\ \bibnamefont
  {{Sunyaev}}},\ }\href@noop {} {\bibfield  {journal} {\bibinfo  {journal}
  {A\&A}\ }\textbf {\bibinfo {volume} {24}},\ \bibinfo {pages} {337} (\bibinfo
  {year} {1973})}\BibitemShut {NoStop}%
\bibitem [{\citenamefont {{Zhou}}\ \emph {et~al.}(2021)\citenamefont {{Zhou}},
  \citenamefont {{Kiuchi}}, \citenamefont {{Shibata}}, \citenamefont
  {{Tsokaros}},\ and\ \citenamefont {{Uryu}}}]{Zhou2021j}%
  \BibitemOpen
  \bibfield  {author} {\bibinfo {author} {\bibfnamefont {E.}~\bibnamefont
  {{Zhou}}}, \bibinfo {author} {\bibfnamefont {K.}~\bibnamefont {{Kiuchi}}},
  \bibinfo {author} {\bibfnamefont {M.}~\bibnamefont {{Shibata}}}, \bibinfo
  {author} {\bibfnamefont {A.}~\bibnamefont {{Tsokaros}}},\ and\ \bibinfo
  {author} {\bibfnamefont {K.}~\bibnamefont {{Uryu}}},\ }\href@noop {}
  {\bibfield  {journal} {\bibinfo  {journal} {arXiv e-prints}\ ,\ \bibinfo
  {eid} {arXiv:2111.00958}} (\bibinfo {year} {2021})},\ \Eprint
  {https://arxiv.org/abs/2111.00958} {arXiv:2111.00958 [astro-ph.HE]}
  \BibitemShut {NoStop}%
\bibitem [{\citenamefont {Papaloizou}\ and\ \citenamefont
  {Pringle}(1984)}]{Papaloizou1984d}%
  \BibitemOpen
  \bibfield  {author} {\bibinfo {author} {\bibfnamefont {J.~C.~B.}\
  \bibnamefont {Papaloizou}}\ and\ \bibinfo {author} {\bibfnamefont {J.~E.}\
  \bibnamefont {Pringle}},\ }\href {https://doi.org/10.1093/mnras/208.4.721}
  {\bibfield  {journal} {\bibinfo  {journal} {Monthly Notices of the Royal
  Astronomical Society}\ }\textbf {\bibinfo {volume} {208}},\ \bibinfo {pages}
  {721} (\bibinfo {year} {1984})},\ \Eprint
  {https://arxiv.org/abs/https://academic.oup.com/mnras/article-pdf/208/4/721/2896279/mnras208-0721.pdf}
  {https://academic.oup.com/mnras/article-pdf/208/4/721/2896279/mnras208-0721.pdf}
  \BibitemShut {NoStop}%
\bibitem [{\citenamefont {{Xie}}\ \emph {et~al.}(2020)\citenamefont {{Xie}},
  \citenamefont {{Hawke}}, \citenamefont {{Passamonti}},\ and\ \citenamefont
  {{Andersson}}}]{Xie2020j}%
  \BibitemOpen
  \bibfield  {author} {\bibinfo {author} {\bibfnamefont {X.}~\bibnamefont
  {{Xie}}}, \bibinfo {author} {\bibfnamefont {I.}~\bibnamefont {{Hawke}}},
  \bibinfo {author} {\bibfnamefont {A.}~\bibnamefont {{Passamonti}}},\ and\
  \bibinfo {author} {\bibfnamefont {N.}~\bibnamefont {{Andersson}}},\ }\href
  {https://doi.org/10.1103/PhysRevD.102.044040} {\bibfield  {journal} {\bibinfo
   {journal} {\prd}\ }\textbf {\bibinfo {volume} {102}},\ \bibinfo {eid}
  {044040} (\bibinfo {year} {2020})},\ \Eprint
  {https://arxiv.org/abs/2005.13696} {arXiv:2005.13696 [astro-ph.HE]}
  \BibitemShut {NoStop}%
\bibitem [{\citenamefont {Korobkin}\ \emph {et~al.}(2012)\citenamefont
  {Korobkin}, \citenamefont {Abdikamalov}, \citenamefont {Stergioulas},
  \citenamefont {Schnetter}, \citenamefont {Zink}, \citenamefont {Rosswog},\
  and\ \citenamefont {Ott}}]{Korobkin2012l}%
  \BibitemOpen
  \bibfield  {author} {\bibinfo {author} {\bibfnamefont {O.}~\bibnamefont
  {Korobkin}}, \bibinfo {author} {\bibfnamefont {E.}~\bibnamefont
  {Abdikamalov}}, \bibinfo {author} {\bibfnamefont {N.}~\bibnamefont
  {Stergioulas}}, \bibinfo {author} {\bibfnamefont {E.}~\bibnamefont
  {Schnetter}}, \bibinfo {author} {\bibfnamefont {B.}~\bibnamefont {Zink}},
  \bibinfo {author} {\bibfnamefont {S.}~\bibnamefont {Rosswog}},\ and\ \bibinfo
  {author} {\bibfnamefont {C.}~\bibnamefont {Ott}},\ }\bibfield  {journal}
  {\bibinfo  {journal} {Monthly Notices of the Royal Astronomical Society}\
  }\textbf {\bibinfo {volume} {431}},\ \href
  {https://doi.org/10.1093/mnras/stt166} {10.1093/mnras/stt166} (\bibinfo
  {year} {2012})\BibitemShut {NoStop}%
\bibitem [{\citenamefont {{Moffatt}}(1978)}]{Moffatt1978}%
  \BibitemOpen
  \bibfield  {author} {\bibinfo {author} {\bibfnamefont {H.~K.}\ \bibnamefont
  {{Moffatt}}},\ }\href@noop {} {\emph {\bibinfo {title} {{Magnetic field
  generation in electrically conducting fluids}}}}\ (\bibinfo  {publisher}
  {Cambridge, England, Cambridge University Press, 1978.~353 p.},\ \bibinfo
  {year} {1978})\BibitemShut {NoStop}%
\bibitem [{\citenamefont {{Setiawan}}\ \emph {et~al.}(2006)\citenamefont
  {{Setiawan}}, \citenamefont {{Ruffert}},\ and\ \citenamefont
  {{Janka}}}]{Setiawan2006}%
  \BibitemOpen
  \bibfield  {author} {\bibinfo {author} {\bibfnamefont {S.}~\bibnamefont
  {{Setiawan}}}, \bibinfo {author} {\bibfnamefont {M.}~\bibnamefont
  {{Ruffert}}},\ and\ \bibinfo {author} {\bibfnamefont {H.-T.}\ \bibnamefont
  {{Janka}}},\ }\href@noop {} {\bibfield  {journal} {\bibinfo  {journal}
  {\aap}\ }\textbf {\bibinfo {volume} {458}},\ \bibinfo {pages} {553} (\bibinfo
  {year} {2006})},\ \Eprint {https://arxiv.org/abs/astro-ph/0509300}
  {astro-ph/0509300} \BibitemShut {NoStop}%
\bibitem [{\citenamefont {{Siegel}}\ and\ \citenamefont
  {{Metzger}}(2017)}]{Siegel2017b}%
  \BibitemOpen
  \bibfield  {author} {\bibinfo {author} {\bibfnamefont {D.~M.}\ \bibnamefont
  {{Siegel}}}\ and\ \bibinfo {author} {\bibfnamefont {B.~D.}\ \bibnamefont
  {{Metzger}}},\ }\href@noop {} {\bibfield  {journal} {\bibinfo  {journal}
  {Physical Review Letters}\ }\textbf {\bibinfo {volume} {119}},\ \bibinfo
  {eid} {231102} (\bibinfo {year} {2017})},\ \Eprint
  {https://arxiv.org/abs/1705.05473} {arXiv:1705.05473 [astro-ph.HE]}
  \BibitemShut {NoStop}%
\bibitem [{\citenamefont {{Metzger}}\ \emph {et~al.}(2008)\citenamefont
  {{Metzger}}, \citenamefont {{Piro}},\ and\ \citenamefont
  {{Quataert}}}]{Metzger2008c}%
  \BibitemOpen
  \bibfield  {author} {\bibinfo {author} {\bibfnamefont {B.~D.}\ \bibnamefont
  {{Metzger}}}, \bibinfo {author} {\bibfnamefont {A.~L.}\ \bibnamefont
  {{Piro}}},\ and\ \bibinfo {author} {\bibfnamefont {E.}~\bibnamefont
  {{Quataert}}},\ }\href@noop {} {\bibfield  {journal} {\bibinfo  {journal}
  {\mnras}\ }\textbf {\bibinfo {volume} {390}},\ \bibinfo {pages} {781}
  (\bibinfo {year} {2008})},\ \Eprint {https://arxiv.org/abs/0805.4415}
  {arXiv:0805.4415} \BibitemShut {NoStop}%
\bibitem [{\citenamefont {Just}\ \emph {et~al.}(2022)\citenamefont {Just},
  \citenamefont {Kullmann}, \citenamefont {Goriely}, \citenamefont {Bauswein},
  \citenamefont {Janka},\ and\ \citenamefont {Collins}}]{Just2022a}%
  \BibitemOpen
  \bibfield  {author} {\bibinfo {author} {\bibfnamefont {O.}~\bibnamefont
  {Just}}, \bibinfo {author} {\bibfnamefont {I.}~\bibnamefont {Kullmann}},
  \bibinfo {author} {\bibfnamefont {S.}~\bibnamefont {Goriely}}, \bibinfo
  {author} {\bibfnamefont {A.}~\bibnamefont {Bauswein}}, \bibinfo {author}
  {\bibfnamefont {H.-T.}\ \bibnamefont {Janka}},\ and\ \bibinfo {author}
  {\bibfnamefont {C.~E.}\ \bibnamefont {Collins}},\ }\href
  {https://doi.org/10.1093/mnras/stab3327} {\bibfield  {journal} {\bibinfo
  {journal} {Monthly Notices of the Royal Astronomical Society}\ }\textbf
  {\bibinfo {volume} {510}},\ \bibinfo {pages} {2820} (\bibinfo {year}
  {2022})},\ \Eprint
  {https://arxiv.org/abs/https://academic.oup.com/mnras/article-pdf/510/2/2820/42084363/stab3327.pdf}
  {https://academic.oup.com/mnras/article-pdf/510/2/2820/42084363/stab3327.pdf}
  \BibitemShut {NoStop}%
\bibitem [{\citenamefont {{De}}\ and\ \citenamefont
  {{Siegel}}(2020)}]{De2020a}%
  \BibitemOpen
  \bibfield  {author} {\bibinfo {author} {\bibfnamefont {S.}~\bibnamefont
  {{De}}}\ and\ \bibinfo {author} {\bibfnamefont {D.}~\bibnamefont
  {{Siegel}}},\ }\href@noop {} {\bibfield  {journal} {\bibinfo  {journal}
  {arXiv e-prints}\ ,\ \bibinfo {eid} {arXiv:2011.07176}} (\bibinfo {year}
  {2020})},\ \Eprint {https://arxiv.org/abs/2011.07176} {arXiv:2011.07176
  [astro-ph.HE]} \BibitemShut {NoStop}%
\bibitem [{\citenamefont {{Christie}}\ \emph {et~al.}(2019)\citenamefont
  {{Christie}}, \citenamefont {{Lalakos}}, \citenamefont {{Tchekhovskoy}},
  \citenamefont {{Fern{\'a}ndez}}, \citenamefont {{Foucart}}, \citenamefont
  {{Quataert}},\ and\ \citenamefont {{Kasen}}}]{Christie2019a}%
  \BibitemOpen
  \bibfield  {author} {\bibinfo {author} {\bibfnamefont {I.~M.}\ \bibnamefont
  {{Christie}}}, \bibinfo {author} {\bibfnamefont {A.}~\bibnamefont
  {{Lalakos}}}, \bibinfo {author} {\bibfnamefont {A.}~\bibnamefont
  {{Tchekhovskoy}}}, \bibinfo {author} {\bibfnamefont {R.}~\bibnamefont
  {{Fern{\'a}ndez}}}, \bibinfo {author} {\bibfnamefont {F.}~\bibnamefont
  {{Foucart}}}, \bibinfo {author} {\bibfnamefont {E.}~\bibnamefont
  {{Quataert}}},\ and\ \bibinfo {author} {\bibfnamefont {D.}~\bibnamefont
  {{Kasen}}},\ }\href@noop {} {\bibfield  {journal} {\bibinfo  {journal}
  {\mnras}\ }\textbf {\bibinfo {volume} {490}},\ \bibinfo {pages} {4811}
  (\bibinfo {year} {2019})},\ \Eprint {https://arxiv.org/abs/1907.02079}
  {arXiv:1907.02079 [astro-ph.HE]} \BibitemShut {NoStop}%
\end{thebibliography}

%



\end{document}